\documentclass{aa}  

\usepackage{graphicx}
\usepackage{natbib}
\bibpunct{(}{)}{;}{a}{}{,} 
\usepackage{txfonts}
\usepackage{lipsum}
\usepackage{tikz}
\usetikzlibrary{patterns}
\usepackage{subcaption}         
\usepackage{lscape}            
\usepackage{placeins}

\begin{document}

   \title{Higher-order methods of radiative transfer in simulations of the epoch of reionisation: $\rm P_n$ versus $\rm M_1$}

     \author{M. Palanque \inst{1} \inst{2}
          \and
          P. Ocvirk\inst{1}
          \and
          E. Franck\inst{2}
          \and
          P. Gerhard\inst{3}
          \and
          D. Aubert\inst{1}
          \and
          O. Marchal\inst{1}
          \fnmsep
          }

   \institute{Observatoire Astronomique de Strasbourg, Universite de Strasbourg, CNRS UMR               7550,
                11 rue de l’Universite, F-67000 Strasbourg, France
         \and
             Institut de Recherche Mathématique Avancée (IRMA), University of Strasbourg
        \and
            Direction du Numérique - Pôle CESAR, University of Strasbourg
        }
   \date{Received 23/07/2025; accepted 06/07/2026}
 
   \abstract{In current cosmological simulations, radiative transfer modules generally rely on the $\rm M_1$ approximation, which has some glaring flaws related to its fluid-like behaviour, such as spurious pseudo-sources and loss of directionality when radiation fronts from different directions collide. $\rm P_n$, another moment-based model, may correct these issues.}
   {We aim to test $\rm P_n$ in an astrophysical setting and compare it to $\rm M_1$, in order to see if it can indeed correct $\rm M_1$'s issues. Also, we want to use $\rm P_n$'s solutions to better pinpoint and quantify $\rm M_1$ errors.}
   {We implemented a $\rm P_n$ radiation transport method, coupled it to a photo-thermo-chemistry module to account for the interaction of ionising radiation with the hydrogen gas, and benchmarked it using tests for radiative transfer models comparison in astrophysics.}
   {We implemented a $\rm P_n$ radiation transport method, coupled it to a photo-thermo-chemistry module to account for the interaction of ionising radiation with the hydrogen gas, and benchmarked it using tests for radiative transfer models comparison in astrophysics.}
   {We find that high-order $\rm P_n$ (e.g. $\rm P_9$) indeed corrects $\rm M_1$'s flaws, while faring as well or even better in some aspects in the tests, in particular when directionality is important or colliding radiation fronts occur. By comparing $\rm P_9$ and $\rm M_1$ radiation fields in an idealised cosmological test case, we highlight a new, thus far unreported artefact of $\rm M_1$, the ‘dark sombrero’. A dark sombrero appears as a spherical photon-deficit shell around the source, at typically one third to one quarter of the distance to the next source. The photon density in dark sombreros can be underestimated by a factor of up to 2-3.
   These artefacts occur in regions where a source's radiation field connects with that of another source or group of sources. These basic properties (the position and amplitude) of the dark sombreros may depend on the sources' relative intensities, positions, and spatial resolution, although we have not been able to test this in detail in this study.
   Moreover, the larger-scale $\rm M_1$ photon density also exhibits spurious features, enhancing or reducing photon density in various regions. We use a small reionisation-like test simulation to characterise the  relative error in hydrogen neutral fractions between $\rm M_1$ and $\rm P_9$. The relative error is well represented by a Gaussian with a dispersion of 0.27 dex in $\rm \log_{10}(\rm x_{HI})$.  Both aspects are likely related to the photons' collisional behaviour in $\rm M_1$. }


   \maketitle

\section{Introduction}

   The epoch of reionisation (EoR) takes place starting 150 Myr after the Big Bang, and ends approximately around 1 Gyr after the Big Bang (redshift 20 to 6) \citep{Barkana&Loeb2001}. During this early period of our Universe's history, its entire gas content gets ionised by the recently formed first stars and galaxies. This is an important process of the history of our Universe. However, details about how this reionisation took place are still debated, mainly in relation to the mass range of the galaxies driving it and the nature of ionising sources (stars, compact objects, etc.). This tension originates in the difficult task of deriving the properties of galaxies in such a remote epoch, as well as those of the intergalactic medium (IGM) and its evolution, even though progress is being made thanks to, for example, JWST \citep{Harikane_2025, papovich_2025} and new high-redshift spectroscopic quasar samples such as XQR-30 \citep{Dodorico_2023}. Further complementary constraints will be brought by the Square Kilometre Array telescope SKA \citep{SKA}. In the meantime, numerical simulations are a crucial tool to interpret existing observations of the EoR and help us prepare future observational campaigns. Some data hints at a major role played by the more numerous small- and medium-sized galaxies \citep{Lewis_2020, Katz_2019}, while some other data favour higher-mass galaxies \citep{Naidu_2020}. There are even proposed scenarios in which both the lower and higher ends of the galaxy mass spectrum drive reionisation at different epochs \citep{Lymanalpha_opacities}.\\
   
   In all of these simulations, radiative transfer is a central component. With the ionisation of the IGM being the main subject of study of this epoch, the need for a reliable radiative transfer model is central to better describe this process. One of the most commonly used models nowadays is the moment-derived model $\rm M_1$ \citep{LEVERMORE_1984, Aubert_2008}. However, $\rm M_1$ is not without flaws, as, for example, it tends to create pseudo-sources at the colliding point of two wave fronts. This stems from the fact that $\rm M_1$ approximates photons as a fluid, making colliding photon flux add up. It is even more glaringly obvious when observing two intersecting photon beams cross each other. In a physical setting, the two light beams should not interact with each other, but with the $\rm M_1$ approximation, they merge into one single beam whose direction is the sum of the direction vectors of the two initial beams \citep{Rosdahl_2013}. $\rm M_1$'s collisionality could be one of the reasons why the model tends to under estimate the photo-ionisation rate at small scales \citep{Wu_2021} and potentially impact the results of previous simulations.\\

    There are many other radiative transfer models in the literature despite the omnipresence of $\rm M_1$ in the field of cosmology simulations \citep{osti, Garett_2014}. One of the easiest ways to simulate light in a simulation is to follow numerical particles representing the path of light in the medium. This method is the basis for the Monte Carlo ray tracing models, such as CRASH \citep{Maselli_2003}, $\rm C^2$-ray \citep{C2ray}, or the hybrid characteristics (HC) method \citep{HC_method}. However, these methods suffer from a scaling issue with the number of sources in a simulation, as well as from stochastic noise, with improvements proposed by \citet{Pawlik_2008} in the form of adaptative ray tracing to mitigate the cost of radiative transfer. Alternatives to Monte Carlo methods are discrete ordinate methods, or $\rm S_n$ \citep{Larsen_2010}. Despite their simplicity, $\rm S_n$ methods may become costly because of the tracking of fluxes in a large enough number of directions to avoid angular artefacts (above 80 according to \citet{AREPO-IDORT}). Alternatively, many numerical methods using moments of the equation of radiative transfer have been proposed and studied in the literature (see \citet{Garett_2014} for review). Among them, the $\rm M_n$ methods are derived from the Boltzmann entropy \citep{alldredge_2012}, which guarantees the positivity of the photon density. However, only $\rm M_1$, its first order, has an analytical closure, and solving higher orders of $\rm M_n$ involves numerical root finding, with a potentially significant computing cost \citep{Dubroca_1999}. Another moment derived model is $\rm P_n$ \citep{osti,Meltz_2015}, which makes use of a simple spectral description of the fluxes and is expected to describe well the transport for isotropic sources even at low orders, while anisotropic radiation fields will require higher orders. \\
    
    In this paper, we focused on Pn due to the simplicity of its closure and formalism. The objective is determining if increasing the order of moment derived models through $\rm P_n$ can correct some of the issues that can appear in $\rm M_1$. We also aim to characterise $\rm M_1$'s artefacts in the observable space relevant to the EoR (Lyman-$\rm \alpha$ forest transmissions and neutral hydrogen fraction).\\

    In this paper, we first focus on the methodology and implementation of the $\rm P_n$ model and the photo-chemistry kernel in section \ref{section:methodology}. Then, in section \ref{section:test_cases}, we show the results of a first set of simple radiation-only test cases to compare the two models. Finally, in \ref{section:comparisonproject}, we perform a set of benchmarked cosmological test cases on $\rm P_n$ and $\rm M_1$ to gauge the ability of the models to be used in a physical setting.\\

\section{Methodology}\label{section:methodology}

In this section, we describe in more depth how the $\rm P_n$ model works and its implementation for our test cases, as well as the photo-thermo-chemistry kernel used for these tests and the code we used to run all of them on GPUs. 

\subsection{The $\rm P_n$ model}\label{subsection:Pn}

$\rm P_n$ is a moment-derived model of order $\rm n$ of the equation of radiative transfer. Just like $\rm M_n$ or any other moment-derived model, it approximates the equation of radiative transfer in vacuum (Eq. \ref{equ:Radiative_Transfer}) by deriving its moments until reaching the $\rm n- th$ order. At this point, a need arises for a closure equation to close the system. This is where the approximations for $\rm M_n$, from which $\rm M_1$ is the first order, and $\rm P_n$ diverge. The grey approximation of the radiative transfer equation in vacuum, with $\rm I$ the radiative intensity, $\rm r$ the spatial position, $\rm \Omega(\theta,\phi)$ the angular unit vector, $\rm S$ the ionising source terms, and $\rm t$ the time, reads:\\

\begin{equation}\label{equ:Radiative_Transfer}
    \frac{1}{c}\partial_{t}\rm{I}(r,\Omega,t)+\mathbf{\Omega} \cdot \nabla \rm{I}(r,\Omega,t) = S.
\end{equation}

$\rm M_1$ is a moment-derived model of order 1, which means that the closure equation consists of writing the moment of order 2 as a combination of lower-order moments. A detailed explanation of the $\rm M_1$ model's closure equation can be found in \citet{Aubert_2008}. The equation can be projected on a simple $\rm (1, \Omega)$ basis. However, since $\rm P_n$ can technically be derived up to any moment, $\rm n$, we need to project our equation on an infinite basis. We used the spherical harmonics basis $\rm Y_{l,m}$ (see Appendix \ref{Appendix:SH}) as was done in \citet{Meltz_2015}.\\

For the sake of subsequent calculations, we replaced the energy intensity, $\rm{I}$, with the number intensity, $\rm{N}$, through a normalisation by a single photon energy, $\rm h\nu$. We applied the $\rm P_n$ closure equation, which consists of truncating our equation projected on $\rm Y_{l,m}$ at the order $\rm n$. Let us call $\rm w=(w_{0,0},...,w_{l,m},...w_{n,n})$ 
our solution vector. In other terms, we approximated our intensity, $\rm{N}$, as
\begin{equation}
    \rm N(t,r,\mathbf{\Omega}) = \sum\limits_{l=0}^n \sum\limits_{m=-l}^{l} \rm w_{l,m}Y_{l,m} 
\end{equation}
or
\begin{equation}
    \forall l>n, \forall m \in [\![-l;l]\!], \rm w_{l,m}(t,r) = 0.
\end{equation}

 With our convention of considering number intensities, the first term of our solution vector, $\rm w_0$, is a photon density. Recurrence formulas developed in \citet{Meltz_2015} show that this now-closed system can be simplified and written as

\begin{equation}\label{equ:projectedRT}
    \frac{1}{c}\partial_t w + J^x \partial_x w + J^y \partial_y w + J^z \partial_z w = S,
\end{equation}

where $\rm J^x,J^y,J^z$ are constant matrices of size $\rm (n+1)^2 \times (n+1)^2$ defined in appendix \ref{appendixA}. This is a very convenient notation, since these matrices are constant and only need to be computed once at the beginning of a simulation run. However, they are sparse despite being quite large, which can take a lot of memory space. The simplicity of the closure equation being a simple truncation means this method may be sensitive to strong anisotropy in the radiation field as well as fast time-varying sources, which can create oscillations. On top of that, it is not based on a physical entropy as $\rm M_1$ is, and thus does not guarantee the positivity of the photon density, which we delve deeper into in subsections \ref{subsection:chemistry} and \ref{subsection:Non-iso}.

\subsection{Numerical Scheme}

Our code makes use of two separate kernels running successively to perform a complete time step, p. The first one, referred to as the transport kernel, is responsible for solving the RT equation in a vacuum using the $\rm P_n$ closure, while the second one, referred to as the chemistry kernel or photo-thermo-chemistry kernel, updates physical fields such as the temperature of the gas, the fraction of ionised hydrogen, and the re-emission and absorption of photons by the medium. As such, we call $\rm p$, $\rm p+1/2$, and $\rm p+1$ the state of our variables at, respectively, the beginning of the transport kernel, the interface between the transport and chemistry kernels, and the beginning of the next time step $\rm p+1$. This notation is used throughout this subsection.\\

With this two-kernel system, we can compute multiple adimensional tests of section \ref{section:test_cases} by turning off the chemistry kernel. Let us point out once again that neither of those two kernels applies any transformation to the hydrogen density itself, as our simplified model does not take into account hydrodynamics and gravity. As for the boundary conditions used throughout this work, they are detailed in Appendix \ref{Appendix:boundary_conditions}.

\subsubsection{Transport kernel}\label{subsection:transport}

We used a classical finite-volume solver for our model, which is described on a three-dimensional Cartesian grid of fixed size. Each cell has a grid position $\rm (i,j,k)$ and dimensions ($\rm \Delta x$, $\rm \Delta y$, $\rm \Delta z$). For all our tests, we used cubic cells; hence, $\rm \Delta x = \Delta y = \Delta z$. Our time step, $\rm \Delta t$, was controlled by a Courant condition (CFL) defined as $\rm CFL = \frac{\Delta t c}{h_{min}}$, with $\rm c$ the speed of light and $\rm h_{min}=\frac{Volume_{cell}}{Surface_{cell}}=\frac{\Delta x^3}{6\Delta x^2}$, and remained constant throughout a run. We had to satisfy that $\rm CFL<1$ to ensure stability of our scheme for an explicit integration such as ours, determining the value of $\rm \Delta t$. In this paper, the time step was consistently chosen to satisfy $\rm CFL=0.8$, except for a few exceptions detailed at the end of subsection \ref{subsection:chemistry}.\\

To compute our explicit transport time step using the $\rm P_n$ model, we used a Rusanov scheme  \citep{RUSANOV1962} as advised in \citet{Meltz_2015} and \citet{sahmim_2005} for its good stability properties. For comparison purposes, and just as was done in \citet{Aubert_2008}, which we are referencing, our own implementation of $\rm M_1$ also used that scheme. We focus on the Rusanov scheme despite its diffusivity as it is the numerical scheme implemented in large simulation codes along $\rm M_1$ such as RAMSES or DYABLO. A diffusivity study of the $\rm M_1$ closure using an HLL scheme can be found in \citet{Berthon_2010}.\\

Let us call $\rm w_l$ the factor corresponding to the moment of order l in our solution vector, $\rm w$. Eq. \ref{equ:projectedRT} can be discretised as follows, for a time step, p, and positions $\rm (i,j,k)$:

\begin{equation}\label{equ:RusanovScheme}
    \begin{aligned}
    \forall l\in[\![1,n]\!]\\
    \frac{w_l^{p+1/2,i,j,k}-w_l^{p,i,j,k}}{\Delta t} + c [&\frac{w_{l+1}^{p,i+1/2,j,k}-  w_{l+1}^{p,i-1/2,j,k}}{\Delta x}  +\\ 
    &\frac{w_{l+1}^{p,i,j+1/2,k}-w_{l+1}^{p,i,j-1/2,k}}{\Delta y} +\\ 
    &\frac{w_{l+1}^{p,i,j,k+1/2}-w_{l+1}^{p,i,j,k-1/2}}{\Delta z}] = S,
    \end{aligned}
\end{equation}

with $\rm i+1/2$ the flux at the interface between the cells $\rm i$ and $\rm i+1$ of our grid, in the x direction. This flux can be computed with the Rusanov scheme as follows:\\
\begin{equation}
    w^{p,i+1/2,j,k} = \frac{1}{2} J^x (w^{p,i+1,j,k}+w^{p,i,j,k}) - \frac{B}{2}(w^{p,i+1,j,k}-w^{p,i,j,k}).
\end{equation}

In the same way, we can compute the two other directions:\\
\begin{equation}
    w^{p,i,j+1/2,k} = \frac{1}{2} J^y (w^{p,i,j+1,k}+w^{p,i,j,k}) - \frac{B}{2}(w^{p,i,j+1,k}-w^{p,i,j,k}),
\end{equation}
\begin{equation}
    w^{p,i,j,k+1/2} = \frac{1}{2} J^z (w^{p,i,j,k+1}+w^{p,i,j,k}) - \frac{B}{2}(w^{p,i,j,k+1}-w^{p,i,j,k}),
\end{equation}

where B is an upper bound of the spectral radius of the matrices, $\rm J$. Said spectral radius being $\rm ]-1;1[$  \citep{Meltz_2015}, we chose $\rm B=1$ for our implementation. \\

\subsubsection{Chemistry kernel}\label{subsection:chemistry}

During a time step, $\rm p$, the chemistry kernel takes in the photon density, $\rm w_0^{p+1/2}$, from the transport kernel, defined as our coefficient of order 0, and outputs the updated temperature, $\rm T^p(r)$, ionised fraction, $\rm x^p(r)$, and photon density variation, $\rm dw_0^p(r)$. This last parameter is then used to update all moments in the transport kernel to take into account the photon recombination in the photon budget. As a reminder, the only chemical compound we take into account is hydrogen, and its density, $\rm n_H$, remains constant in time as a result of the absence of a hydrodynamical solver in our tests.\\

Following the methodology showcased in \citet{Aubert_2008}, we derived an implicit equation for $\rm x$ and $\rm dw_0$ and an explicit equation for $\rm T$. With $\rm n_{HI}$ and $\rm n_{HII}$, respectively, the density of neutral and ionised hydrogen, $\rm n_e$ the density of free electrons equal to $\rm n_{HII}$, and $\rm x$ the fraction of ionised hydrogen, the hydrogen density, $\rm n_H$, can be written as follows:\\
\begin{equation}
    n_H = n_{HI} + n_{HII} = (1-x) n_H + x n_H .
\end{equation}

With that in mind, we aim to solve the following equations to obtain the updated $\rm w_0$, $\rm w_k$, $\rm x$, and $\rm T$:\\
\begin{equation}\label{equ:photodens}
    \frac{\partial w_0}{\partial t} = -(1-x)n_H c\sigma w_0 + x^2 n_H^2 (\alpha_A - \alpha_B),
\end{equation}
\begin{equation}
    \frac{\partial w_k}{\partial t} = - (1-x)n_H c \sigma w_k, \forall k \in [1,n],
\end{equation}
\begin{equation}\label{equ:ionisingEqu}
    \frac{\partial x}{\partial t} = (1-x)c\sigma w_0 - x^2 n_H^2 \alpha_A - x(1-x)n_H \beta,
\end{equation}
\begin{equation}\label{equ:energy}
    \frac{\partial E}{\partial t} = \mathcal{H}-\mathcal{L},
\end{equation}

where $\rm \alpha_A(T)$, $\rm \alpha_B(T)$, and $\rm \beta(T)$ correspond, respectively, to the case A and case B recombination rates, and the HI collisional ionisation coefficient as defined in \citet{Hui_Gnedin_97}, $\rm \sigma$ is the effective HI cross section at photon energy $\rm e_{HI}$, $\rm E(T,t)$ the thermal energy of the gas, and $\rm \mathcal{H}(x,w_0,n_H)$ and $\rm \mathcal{L}(T,n_H)$ are the photoionisation heating rate and cooling rate. In all of our test cases of section \ref{section:comparisonproject}, we considered $\rm 10^5\:\rm K$ black body sources to comply with literature test cases \citep{Iliev_2006} implying a cross-section of $\rm \sigma = 1.63\times 10^{-22}\:\rm m^2$ at $\rm e_{HI}$= 29.61  \citep{Aubert_2008}, except for test 1 in which we considered $\rm 3.10^4\:\rm K$ black body sources and $\rm \sigma = 6.3\times 10^{-22}\:\rm m^2$ for a photon energy equal to the ionisation energy, $\rm e_{HI}$= 13.6 eV \citep{Osterbrock_1974}.\\

As we said previously in subsection \ref{subsection:Pn}, and as we delve deeper into in subsection \ref{subsection:Non-iso}, $\rm P_n$, in essence, does not guarantee that the photon density, $\rm w_0^{p+1/2}$, at the end of a transport step is positive. Our way around this possible non-positivity of the density is to approximate it as zero in negative regions when passing it to the chemistry kernel, as is shown in Eq. \ref{equ:positivation}. This might seem like a strong choice, but as is shown in Appendix \ref{Appendix:order_negativity}, the negative photon density output is mostly negligible as long as $\rm P_n$'s order is sufficiently high in spherical cases.

\begin{equation}\label{equ:positivation}
    \Bar{w_0}^{p+1/2}=max(w_0^{p+1/2},0).
\end{equation}

In the chemistry kernel, we only ever use this truncated version $\rm \Bar{w_0}$ of the photon density. To follow the rate of reionisation, the neutral fraction of gas is a key observable that can be derived from the 21cm emission of neutral hydrogen \citep{Zaroubi_2012}. This parameter can be defined as $\rm x_{HI} = 1 - x$, where $\rm x$ is the fraction of ionised hydrogen in our simulation.\\

After discretising Eq.\ref{equ:photodens} to \ref{equ:energy}, and with $\rm x^p = x^{p+1/2}=x$ and $\rm x^{p+1}=X$, we derived the updated photon density, $\rm w_0^{p+1}$, the updated moments of order higher than 0 $\rm w_l^{p+1}$ and the updated temperature $\rm T^{p+1}$:

\begin{subequations}
    \begin{equation}
        \begin{aligned}
        &w_0^{p+1}=w_0^{p+1/2}+dw_0^{p+1/2},\\
        &with\:dw_0^{p+1/2} = \beta n_H^2 (1-X)X \Delta t - \alpha_B n_H^2 X^2 \Delta t - n_H(X-x),
        \end{aligned}
    \end{equation}
    \begin{equation}
        w_l^{p+1} = \frac{w_l^{p+1/2}}{1+ n_H \sigma c \Delta t (1-X)},
    \end{equation}
    \begin{equation}
        \begin{aligned}
        \frac{T^{p+1}-T^{p+\frac{1}{2}}}{\Delta t} &= \frac{1}{X+1} \times \\
            &\left[\frac{2[\mathcal{H}(x,X,\Bar{w_0}^{p+1/2})-\mathcal{L}(X,T^{p+\frac{1}{2}})]}{3k_B n_H} - \frac{(X-x)T^{p+1}}{\Delta t}\right],
        \end{aligned}
    \end{equation}
\end{subequations}

with $\rm w_l^{p+1/2}= w_l^p - \frac{dw_{l+1}^p}{dr} \Delta t$ the value of the moment at the end of the transport step and $\rm \mathcal{H}$ and $\rm \mathcal{L}$, respectively, the heating and cooling rate of the hydrogen gas. Details about the steps to derive these equations can be found in  Appendix \ref{Appendix:chemistry}. Since computing the updated temperature, $\rm T^{p+1}$, requires prior knowledge of the updated ionised fraction, $\rm X$, it was computed last during a chemistry time step. \\

By this point, we had updated all chemistry variables for the time step $\rm p+1$. Updated values of $\rm w^{p+1}$ were sent back to the transport kernel as new initial conditions for the next transport step. We note that we did not implement any sub-cycling for our chemistry kernel in this paper, despite the possibility of $\rm \Delta t$ being too large to accurately describe the short-timescale behaviour of our chemistry. In cases in which the time step would have been too great to be handled by the kernel, we opted for a diminished CFL, which increased our computational cost. However, sub-cycling our chemistry kernel would make our implementation more optimised and should be implemented during future use of the model.\\ 

\begin{figure}[!ht]
   \centering
   \includegraphics[width=1.0\linewidth,trim={1.6cm 0 2.5cm 1cm},clip]{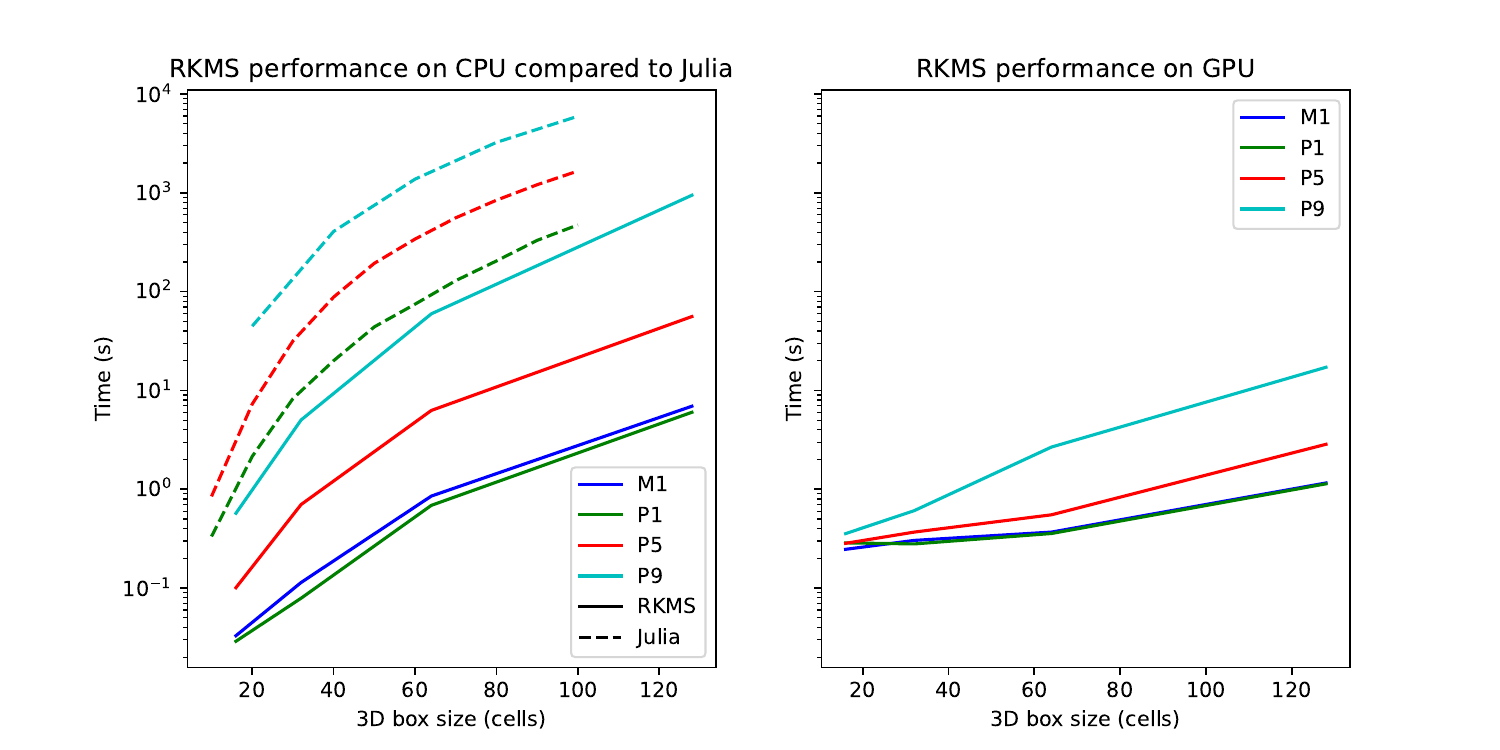}
      \caption{Execution time comparison of RKMS compared to our Julia implementation on mono-CPU and mono-GPU depending on the box size. This was done using an impulse response test described in Appendix \ref{Appendix:order_negativity} for a duration of 200 time steps.}
    \label{fig:acceleration}
\end{figure}

\subsection{RKMS}
 For a given order, $\rm n$, of a moment-based model, the number of coefficients to be computed at each time step, in each cell, is $\rm (n+1)^2$. This means that, while $\rm M_1$ only requires the computation of four coefficients at each time step, $\rm P_9$ will need a hundred. This quickly became a problem for our Python-based and then Julia-based non-parallel CPU implementations, which struggled to compute our tests in a reasonable time frame.\\

Reduced kinetic model solver (RKMS) is an openCL based, Python-wrapped, mono-GPU implementation of $\rm P_n$ developed by Pierre Gerhard during his PHD thesis\footnote{\url{https://github.com/p-gerhard/rkms}} \citep{gerhard2020}. Its fully optimised GPU core makes it a lot faster than our previous implementations. Extensive work on RKMS was necessary to implement the chemistry kernel and the dimensionalisation required for the test cases in section \ref{section:comparisonproject}, but allowed us to run a lot of tests far more quickly than we expected, as is shown in Fig. \ref{fig:acceleration}. However, one of the limits of RKMS is that it is not parallelised on several GPUs, and thus can still take a long time to compute higher orders of $\rm P_n$. As a result, the biggest box size used in this paper is a $\rm 128^3$ box with $\rm P_9$ implementation.\\

\begin{figure*}[!ht]
   \centering
   \includegraphics[width=1\linewidth,trim={2cm 0.4cm 1.5cm 1.5cm},clip]{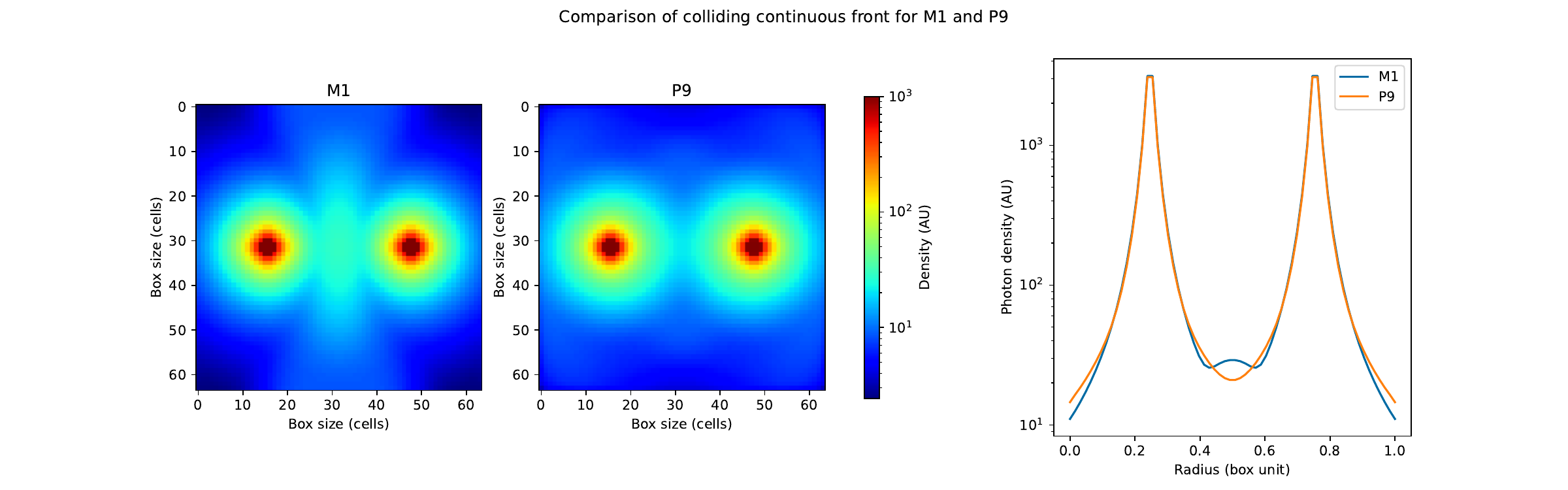}
      \caption{Comparison of colliding fronts of two isotropic continuous sources in $\rm M_1$ and $\rm P_9$ at 400 time steps.}
         \label{fig:Continuous_Comparison}
   \end{figure*}

\begin{figure*} 
    \centering
    \includegraphics[width=17cm,trim={1.5cm 1.0cm 0.5cm 2cm},clip]{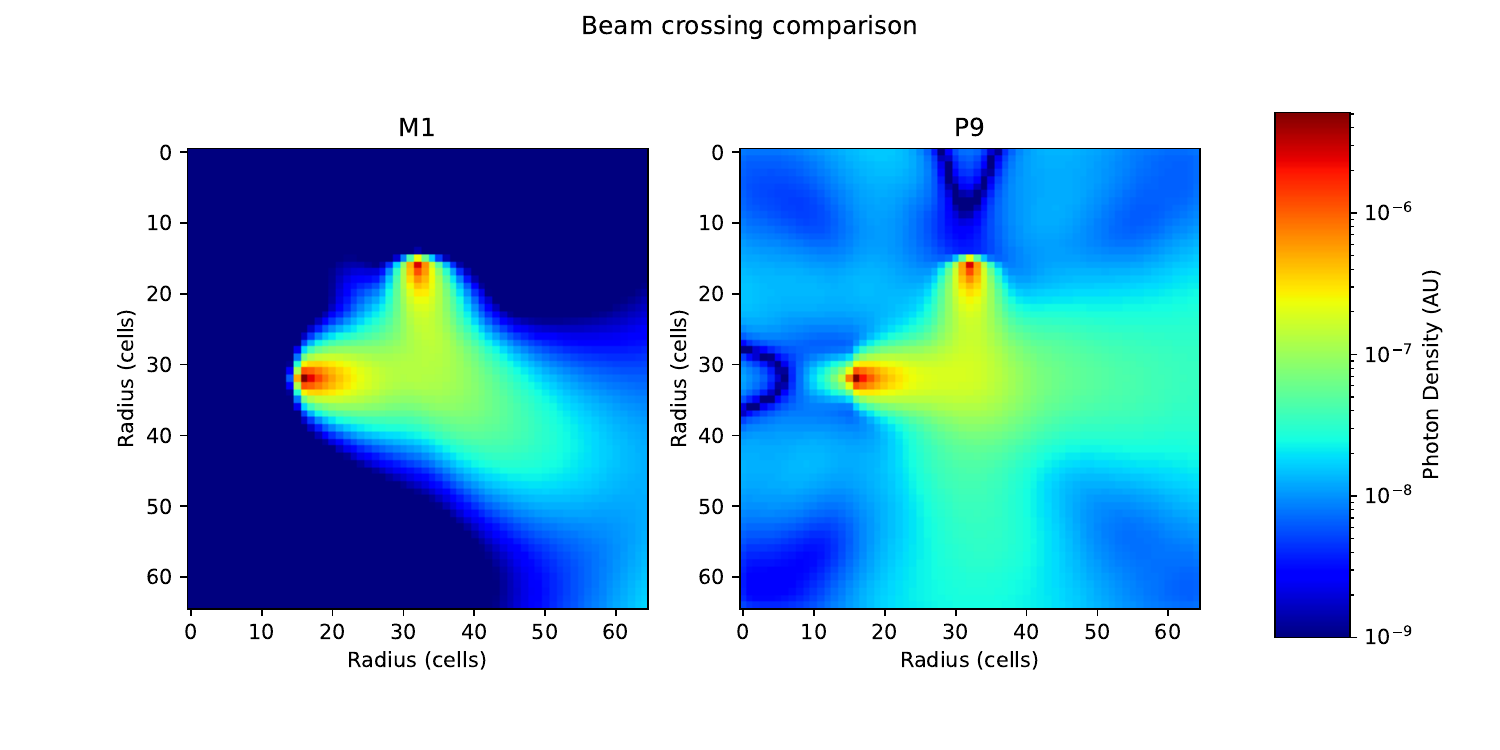}
    \caption{Beam crossing comparison between $\rm M_1$ (left) and $\rm P_9$ (right) at 500 time steps. We can observe the difference between how the beams cross each other in $\rm P_9$ and average out in $\rm M_1$.}
    \label{fig:beam_compare}
\end{figure*}

\section{Results: Radiation-only test cases}\label{section:test_cases}

We start by highlighting the ways in which $\rm P_n$ solves $\rm M_1$'s issues in a series of simple tests. All test cases in this section are purely adimensional radiative transfer models with no chemistry coupling or hydrodynamics and gravity, run on a $\rm 64^3$ grid. Their aim is to showcase how $\rm P_n$ compares to $\rm M_1$ in the specific cases in which it tends to have issues. In this section, considering that the photon density is adimensional, its unit is written as ‘arbitrary units’ (AU). 

\subsection{Continuous isotropic sources}

Our first and most straightforward test case serves to illustrate the main improvement that $\rm P_n$ brings to the table compared to $\rm M_1$. Indeed, as was talked about previously, $\rm M_1$ is a fluid-like approximation of photons resulting in a collisional behaviour. As such, two wave fronts colliding end up adding up instead of just crossing each other as would be physically expected. This tends to create pseudo-sources and outflows of photons that have no physical reality, as was shown in 1D in \citet{Dubroca_2002} and \citet{turpault}. This mainly appears when the photon fluxes coming from several sources interact with each other. To highlight this flaw in $\rm M_1$ and see if $\rm P_n$ can correct it, we put two continuous isotropic sources  in a $\rm 64^3$ box during 400 time steps. Both sources are adimensional of intensity equal to 1 per time step. They were placed at, respectively, [0.5,0.5,0.25] and [0.5,0.5,0.75] in adimensional box co-ordinates. Since these sources are isotropic, we can write them by defining their moments of order 0 to be equal to the intensity, while the moments of higher orders are all equal to 0. Throughout this paper, isotropic sources are always defined this way in the code. \\ 

Results of this test for $\rm M_1$ and $\rm P_9$ are showcased in Fig. \ref{fig:Continuous_Comparison}, showing the photon density map of a slab containing the two sources as well as a photon density profile of the two sources. The pseudo source created by the coalescence of two wave fronts becomes evident in $\rm M_1$, which emits vertically and changes the directionality of the photon flux. It is also visible in the photon density profile, where a bump appears in the photon distribution between the sources, highlighting their interaction. This behaviour is completely absent in $\rm P_9$, where the bump is absent in the photon density profile, and no coalescence is apparent in the slice shown in the same figure, on the right. The directionality of the radiation field is conserved, and the end result in the continuous regime matches what we should physically expect from two isotropic sources next to each other. As such, we can confidently say that $\rm P_n$ corrects this issue of $\rm M_1$ in the isotropic case.

\subsection{Continuous non-isotropic sources}\label{subsection:Non-iso}

The collisional behaviour of $\rm M_1$ can be highlighted even further in a more extreme test case that shows how it can affect the physics of a simulation. In the next test case, we put two adimensional non-isotropic continuous sources in a $\rm 64^3$ box. Those directional sources each emit a beam of light, and are placed at, respectively, [0.25,0.5,0.5] and [0.5,0.5,0.25] in adimensional box units, so that the beams cross at the centre of the box. The simulation runs for 500 time steps. Our sources each emit a Gaussian beam defined as follows:

\begin{equation}
    S(\theta,\phi) = S_0e^{\frac{-(\theta-\theta_0)^2 + (\phi-\phi_0)^2}{\sigma_0^2}},
\end{equation}

with $\rm S_0$ intensity of the source, $\rm \theta_0$ and $\rm \phi_0$ angles for the direction of the beam, and $\rm \sigma_0$ the half width of the beam itself. Our sources are in the plane $\rm y=0.5$. As such, the values of these angles were chosen as follows, with $\rm \theta_0$, $\rm \phi_0$ for source 1 and $\rm \theta_0'$, $\rm \phi_0'$ for source 2:

\begin{subequations}
    \begin{equation}
        \theta_0 = 0,\: \phi_0 = \pi,
    \end{equation}
    \begin{equation}
        \theta_0' = 0,\: \phi_0' = 2\pi.
    \end{equation}
\end{subequations} 

    By choosing $\rm \theta_0=\theta_0'$, we ensure our sources emit in the same plane $\rm y=0.5$, and modify the direction of the beam through the other angle parameters. Finally, we chose $\rm S_0 = 1$ to have unit non-dimensional sources, and $\rm \sigma_0 = 0.05$. These beams were then projected on the respective bases of the two models we were comparing to obtain the value of each moment for the two sources.\\ 

 Physically, the two light beams should just cross each other without interacting, since light is non-collisional. However, considering $\rm M_1$'s properties, we expect the two beams to coalesce and their resultant directional vector to be the sum of the two initial direction vectors. This experiment was already done in \citet{Rosdahl_2013} in 2D, with striking results.\\

Results for this test case are shown in Fig. \ref{fig:beam_compare} and highlight very well the collisional properties of $\rm M_1$. Here, the two beams merge and the resulting new beam changes direction, in an unphysical way. However, we can see that $\rm P_9$ manages to maintain the crossing behaviour that should be modelled here, and corrects once again the issues of $\rm M_1$. On the other hand, we can also see a lot of oscillations around the sources in $\rm P_9$ that are not expected from a test case with directional sources. It is due to the strong anisotropy of the sources, which creates low-intensity ripples in all directions around the sources. We also mention the case of strong oscillations due to sharp time-varying sources (Dirac) in Appendix \ref{Appendix:order_negativity}.

\begin{table*}
\caption{Initial condition parameters of the background medium for the benchmarked tests.}             
\label{table:parameters}      
\centering          
\begin{tabular}{c c c c l l l }    
\hline\hline       
Test Name & Duration (Myr) & Cross section ($\rm m^{-2}$) & T($\rm t_0$) (K) & $\rm n_H$ ($\rm H.m^{-3}$)& $\rm x_{HII}$($\rm t_0$) & Source \\ 
\hline                    
   Isothermal Strömgren Sphere & 500 & $\rm 6.3\times 10^{-22}$ & $\rm 1\times 10^4$ & $\rm 1 \times 10^3$ & $\rm 1.2\times 10^{-3}$ & $5\times 10^{48}$ ph/s\\  
   Adiabatic Strömgren Sphere & 100 & $1.63\times 10^{-22}$ & 100 & $1\times 10^3$ & $1.2\times 10^{-3}$ & $5\times 10^{48}$ ph/s\\
   Shadowing by a dense clump & 3 & $1.63\times 10^{-22}$ & $8\times 10^4$ & $2\times 10^2$ & $1.2\times 10^{-3}$ & $10^{10}$ ph/$\rm m^2$\\
   Cosmological map & 4 & $1.63\times 10^{-22}$ & 100 & Cube\tablefootmark{a} & $1.2\times 10^{-3}$ & List\tablefootmark{b} \\

\hline                  
\end{tabular}
\tablefoot{ All values are set at the initial time, $\rm t_0$.\\
\tablefoottext{a}{Tabulated cube of hydrogen densities taken from \citet{Iliev_2006}.}
\tablefoottext{b}{Tabulated list of sources with their co-ordinates in the box and photon densities taken from \citet{Iliev_2006}.}
}

\end{table*}

\begin{figure}
    \centering
    \includegraphics[width=0.99\linewidth,trim={0 0 0 1.4cm},clip]{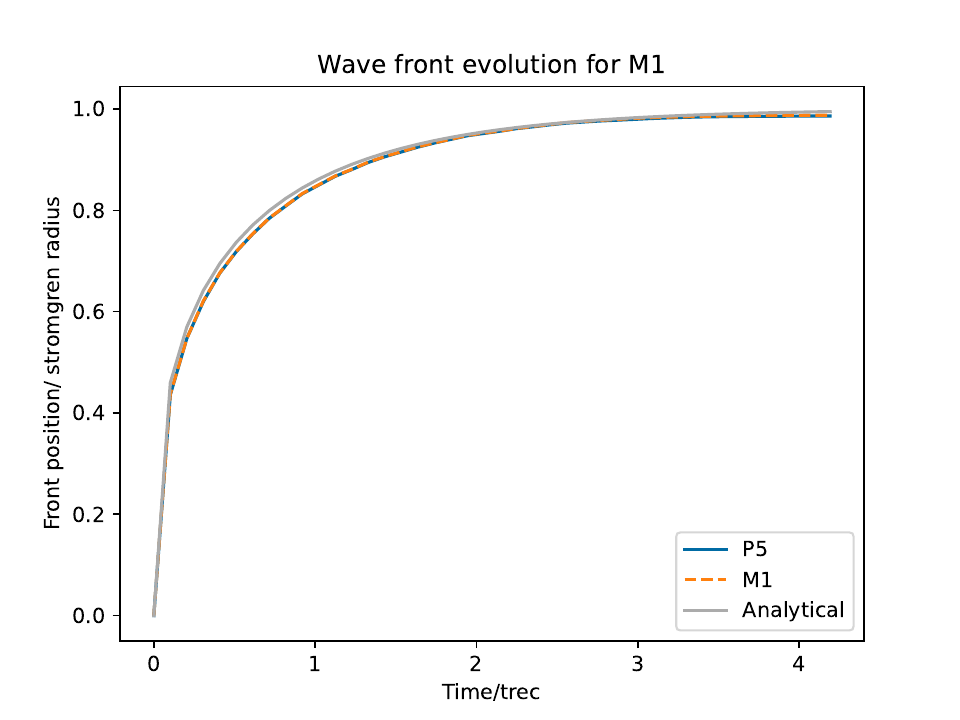}
    \caption{Front radius evolution of $\rm P_5$ and $\rm M_1$ compared to the analytical solution.}
    \label{fig:iostrom_front}
\end{figure}

\begin{figure*}
    \centering
    \resizebox{16cm}{!}
    {\includegraphics[width=12cm,trim={0.2cm 0.3cm 1.6cm 1.4cm},clip]{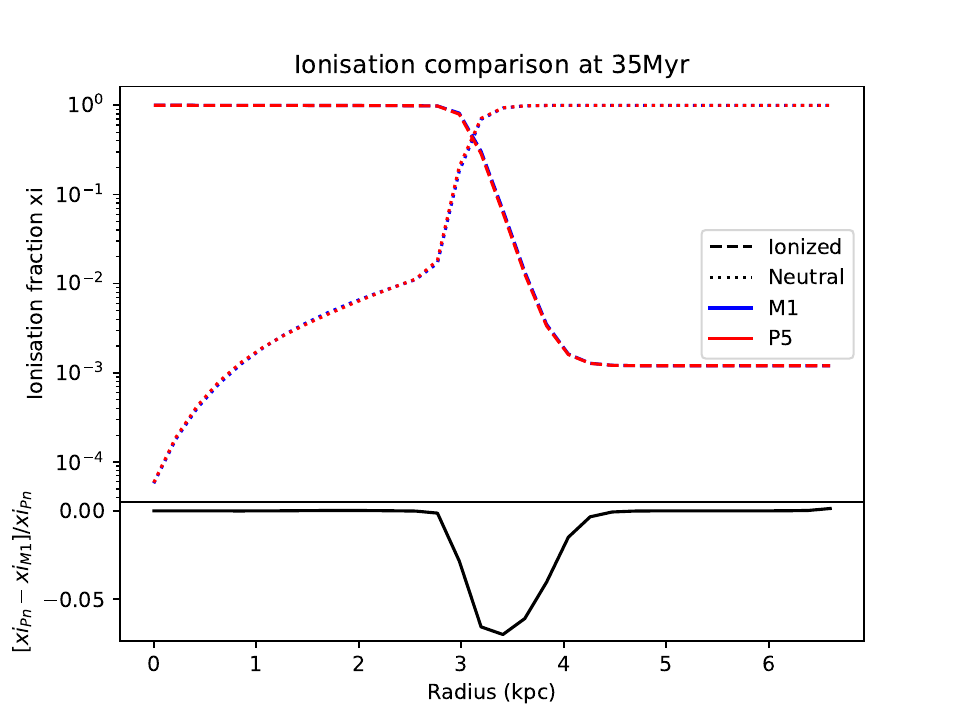}
    \includegraphics[width=12cm,trim={0.2cm 0.3cm 1.6cm 1.4cm},clip]{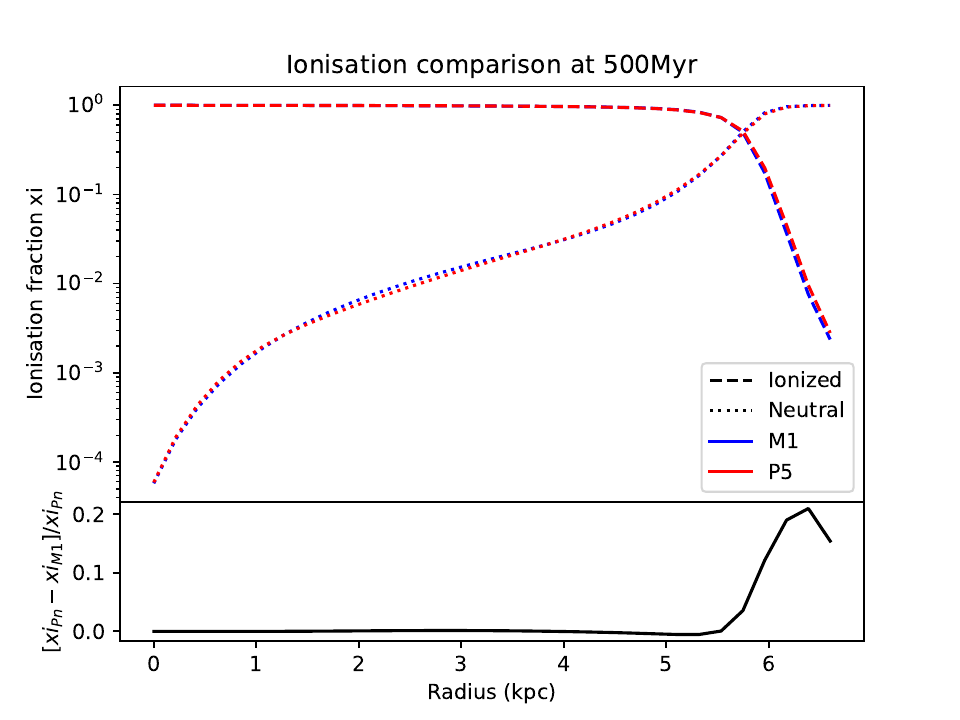}
    }
    \caption{Ionised and neutral hydrogen profile comparison of $\rm M_1$ and $\rm P_5$ at around 35 Myr (left) and 500 Myr (right) in the isothermal case.}
    \label{fig:isostrom_profile}
\end{figure*}

\begin{figure*}
    \centering
    \resizebox{16cm}{!}
    {\includegraphics[width=12cm,trim={0.2cm 0.3cm 1.6cm 1.4cm},clip]{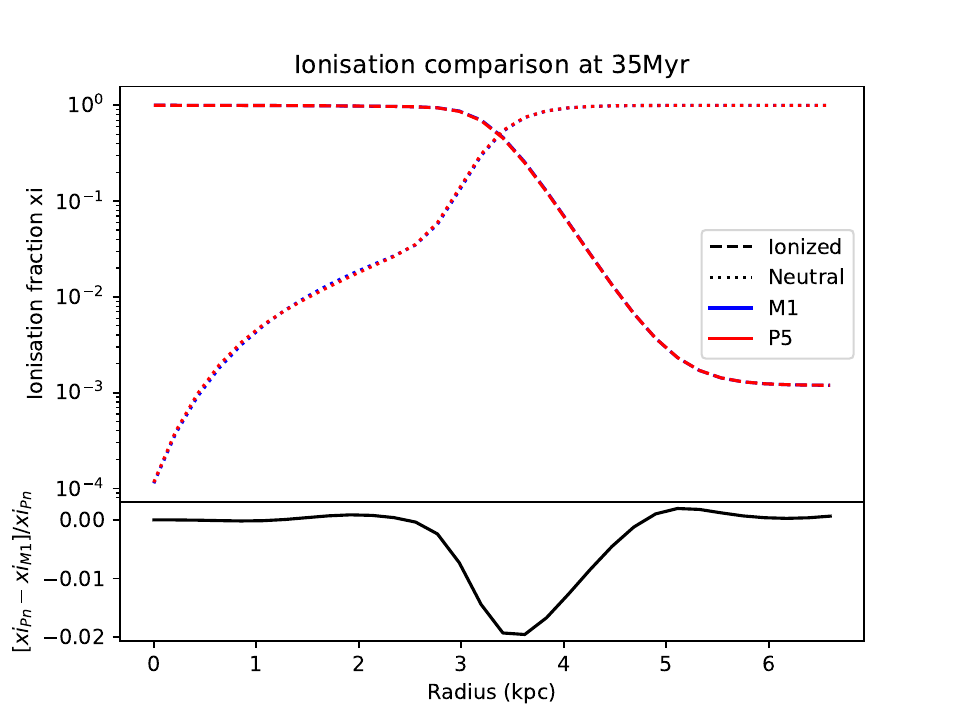}
    \includegraphics[width=12cm,trim={0.2cm 0.3cm 1.6cm 1.4cm},clip]{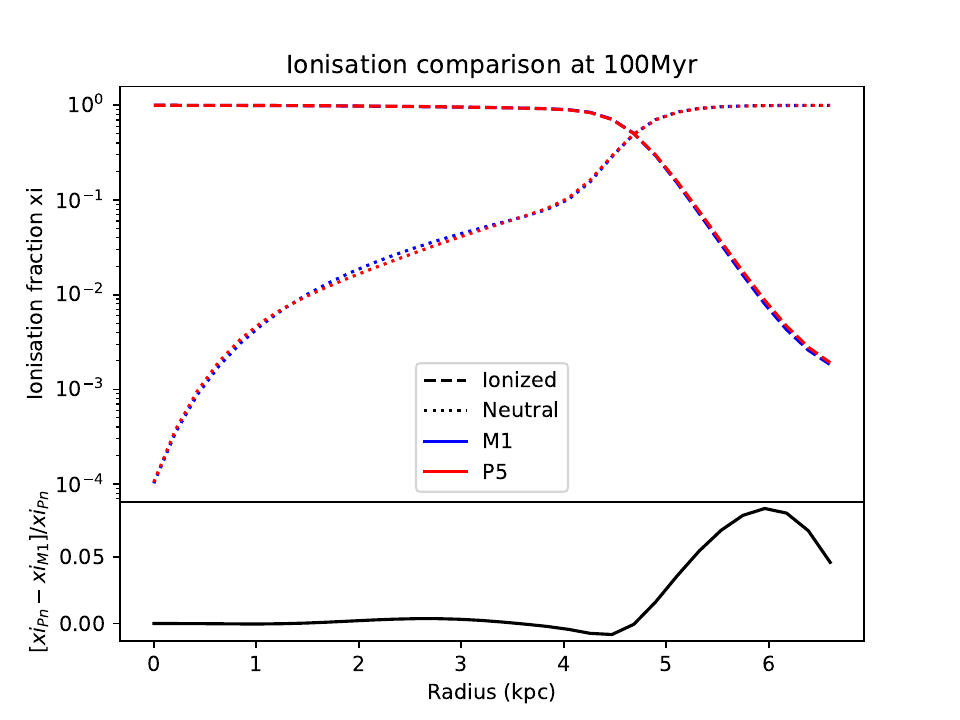}
    }
    \caption{Ionised and neutral hydrogen profile comparison of $\rm M_1$ and $\rm P_5$ at around 35 Myr (left) and 100 Myr (right) with coupled temperature evolution.}
    \label{fig:strom_profile}
\end{figure*}

\begin{figure}
    \centering
    \includegraphics[width=0.99\linewidth,trim={0 0 0 1.4cm},clip]{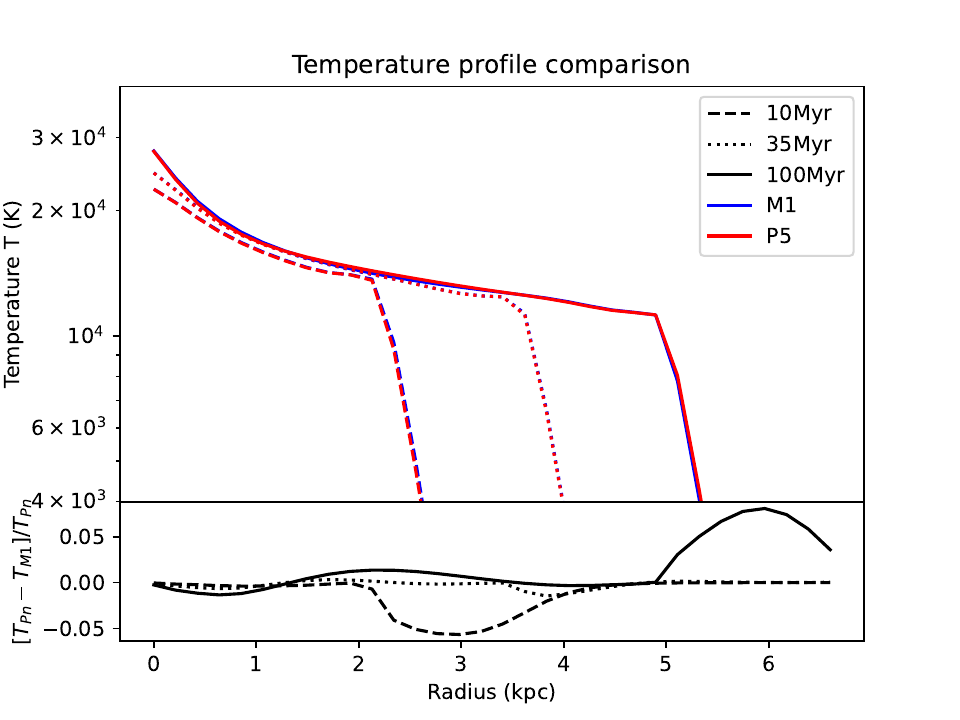}
    \caption{Comparison of the temperature profiles in the Strömgren sphere at three times: 10, 35, and 100 Myr.}
    \label{fig:strom_temp}
\end{figure}

\begin{figure}[!ht]
\centering
\resizebox{9cm}{6cm}{
\begin{tikzpicture}

\draw[gray, ->] (2,7) -- (2,6.5);
\draw[gray, ->] (4,7) -- (4,6.5);
\draw[gray, ->] (6,7) -- (6,6.5);
\draw[gray, ->] (8,7) -- (8,6.5);

\draw[gray, ->] (2,9) -- (2,8.5);
\draw[gray, ->] (4,9) -- (4,8.5);
\draw[gray, ->] (6,9) -- (6,8.5);
\draw[gray, ->] (8,9) -- (8,8.5);

\draw[gray, ->] (2,5) -- (2,4.5);
\draw[gray, ->] (4,5) -- (4,4.5);
\draw[gray, ->] (6,5) -- (6,4.5);
\draw[gray, ->] (8,5) -- (8,4.5);

\draw[gray, ->] (2,3) -- (2,2.5);
\draw[gray, ->] (4,3) -- (4,2.5);
\draw[gray, ->] (6,3) -- (6,2.5);
\draw[gray, ->] (8,3) -- (8,2.5);
    \draw[gray, ->] (2,1) -- (2,0.5);
\draw[gray, ->] (4,1) -- (4,0.5);
\draw[gray, ->] (6,1) -- (6,0.5);
\draw[gray, ->] (8,1) -- (8,0.5);

\draw[black] (0,0) rectangle (10,10);
\filldraw[color=black, fill=black!5, dashed] (0,9.5) rectangle (10,10);
    
\filldraw[color=black, fill=gray!5] (3.79,0) rectangle (6.21,2.5);
\filldraw[color=black!60, fill=red!5, very thick](5,2.5) circle (1.21);
\begin{scope}
    \clip (3,2.5) rectangle (7,4);
    \filldraw[color=red!60, fill=red!60, very thick](5,2.6) circle (1.23);
\end{scope}

\begin{scope}
    \clip (3,2.4) rectangle (7,4);
    \filldraw[color=red!60, fill=red!5, very thick](5,2.4) circle (1.24);
\end{scope}

\draw[black, ->] (-1,3.5) -- (4.3,3.5);
\draw[black, ->] (-1,9.75) -- (1,9.75);
\draw[black, ->] (-1,1) -- (4,1);

\node[text width=3cm] at (6,2.5) {clump};
\node[text width=5cm] at (-2,3.5) {shell of ionised gas};
\node[text width=3cm] at (-1,9.75) {sources};
\node[text width=4cm] at (-2.5,1) {'shadow' of neutral background gas};

\end{tikzpicture}}
    \caption{Schematic representation of the shadowing by a dense clump test. Sources emit in the increasing z direction.}
    \label{fig:clump_schematic}
\end{figure}

\begin{figure*} 
    \centering
    \includegraphics[width=16cm,trim={0.5cm 0.5cm 0 2cm},clip]{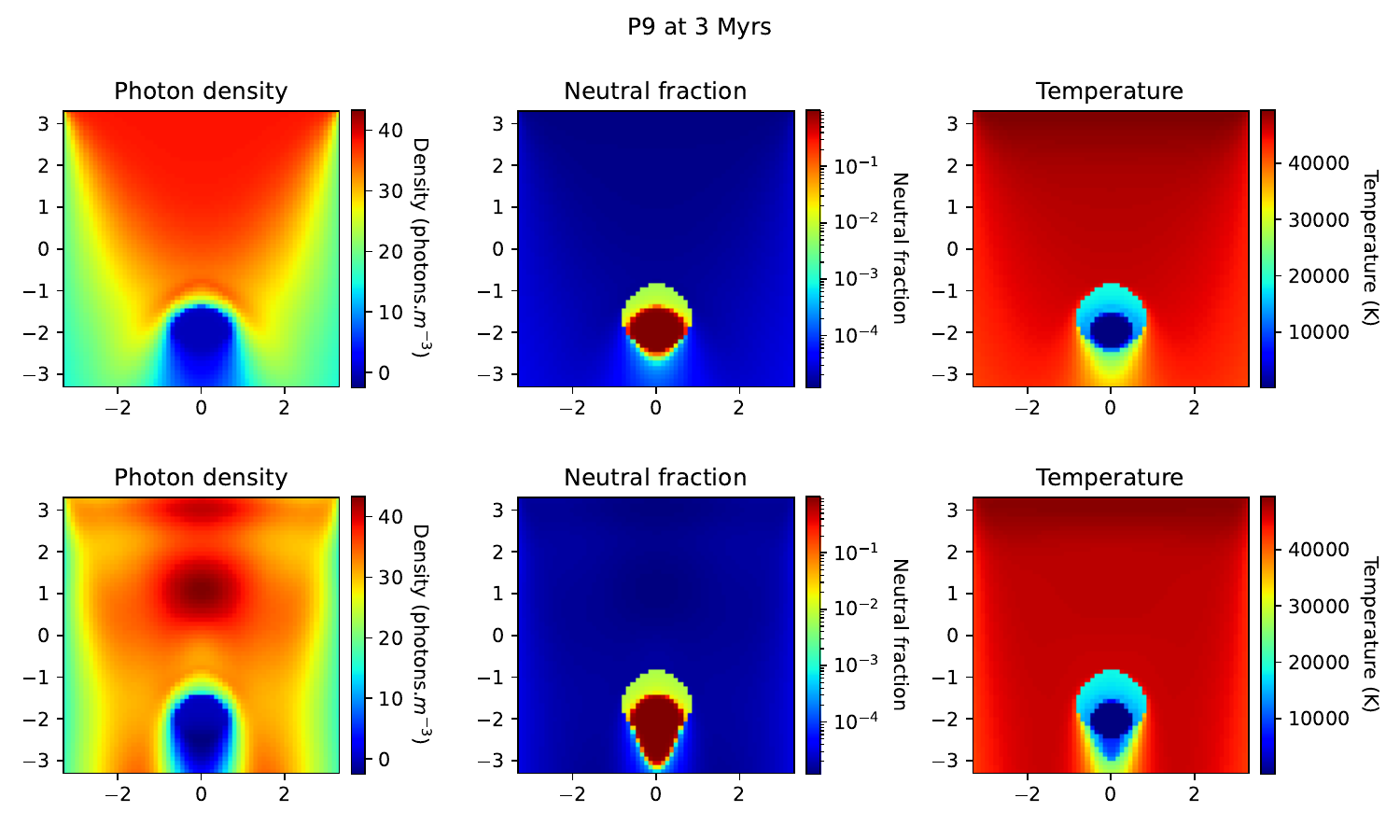}
    \caption{Dense clump of hydrogen in the path of a flux of ionising radiation using $\rm M_1$ (top) and $\rm P_9$ (bottom) at 3 Myr.}
    \label{fig:clump_comp}
\end{figure*}

\begin{figure}
    \centering
    \includegraphics[width=0.99\linewidth,trim={0 0 0 1.4cm},clip]{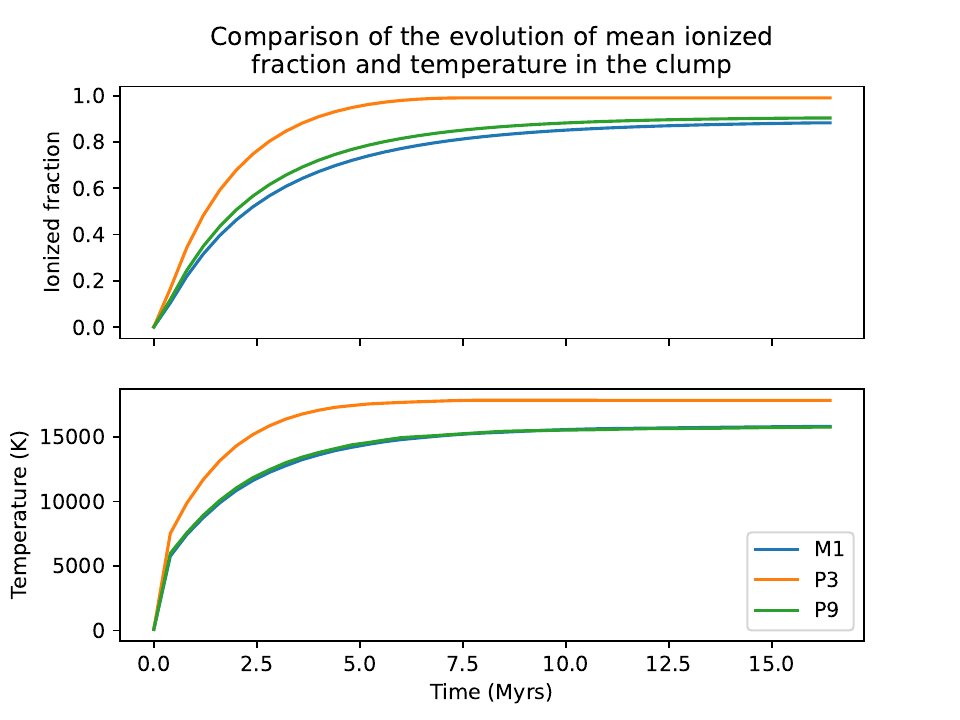}
    \caption{Evolution of the mean ionised fraction (top) and temperature (bottom) depending on the model used at 16 Myr.}
    \label{fig:clump_mean}
\end{figure}

\begin{figure}
    \centering
    \includegraphics[width=0.99\linewidth,trim={0 0 0 1.4cm},clip]{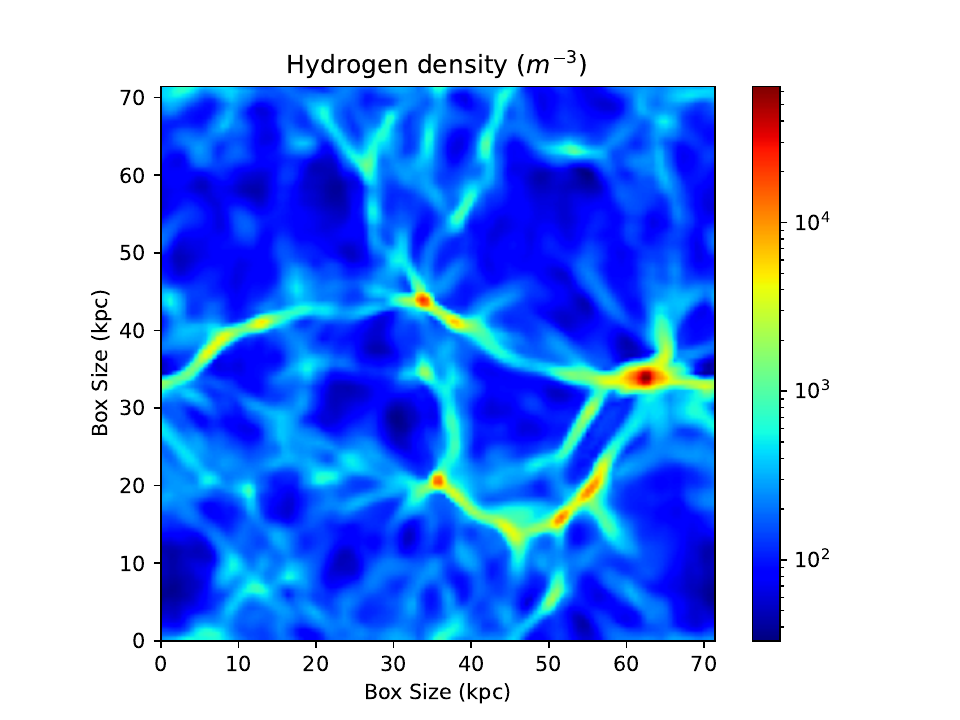}
    \caption{Example slice of hydrogen density map from \citet{Iliev_2006} used in this test.}
    \label{fig:map_nh_dens}
\end{figure}

\begin{figure*}
    \centering
    \includegraphics[width=15cm,trim={1.9cm 1.3cm 1.1cm 1.4cm},clip]{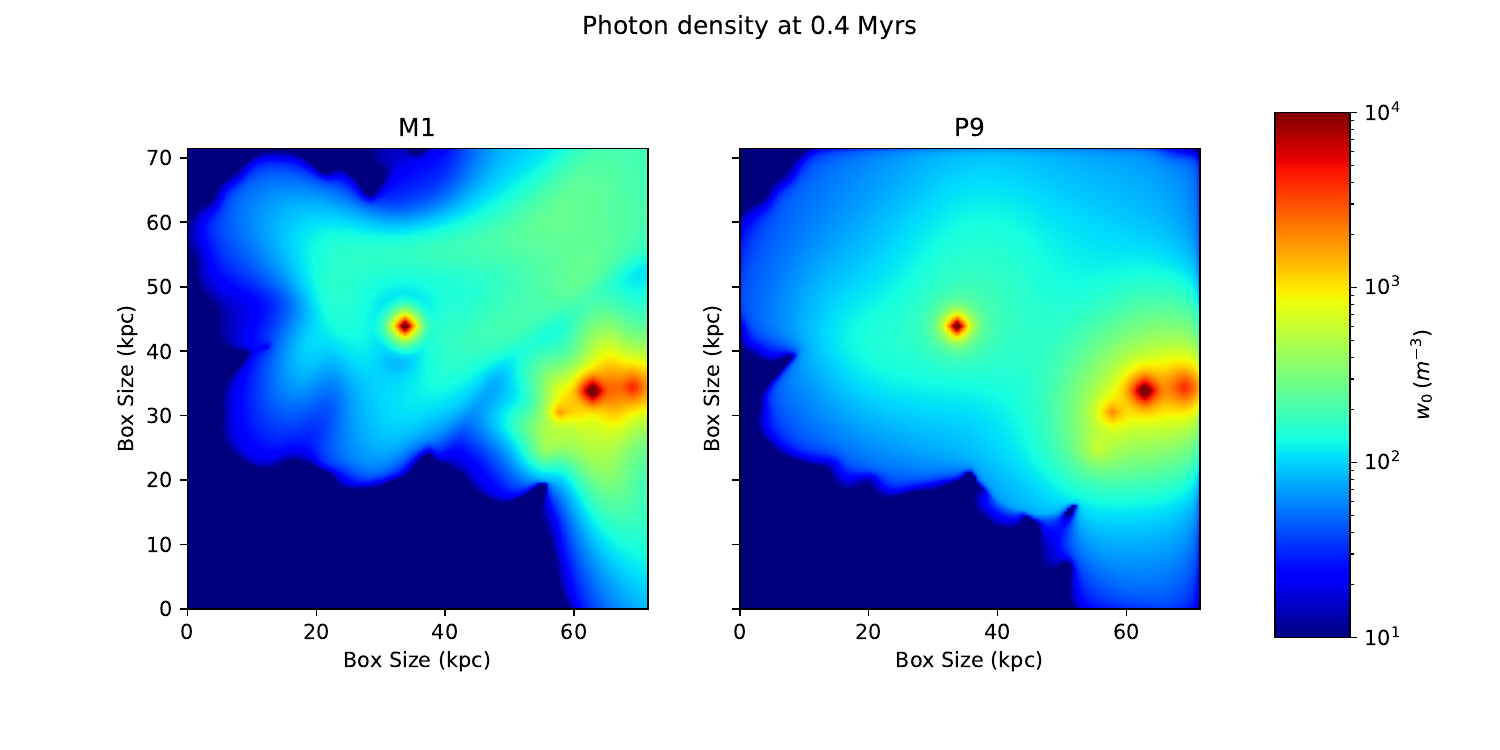}
    \caption{Maps of the photon density at 0.4 Myr in $\rm M_1$ (left) and $\rm P_9$ (right) with a limited dynamic.}
    \label{fig:map_photodens_04_4dec}
\end{figure*}

\begin{figure*}
    \centering
    \includegraphics[width=15cm,trim={1.9cm 1.3cm 1.1cm 1.4cm},clip]{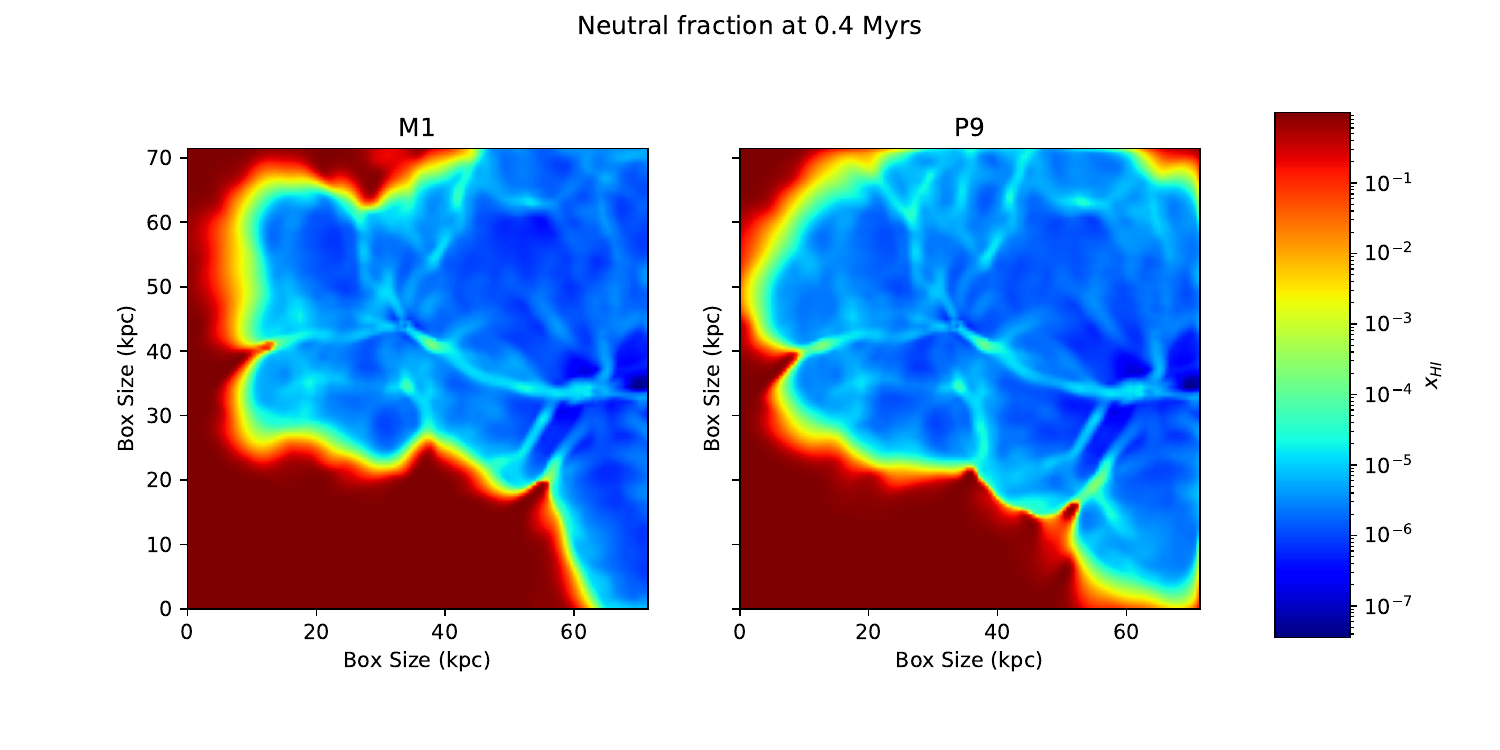}
    \caption{Maps of the neutral fraction of hydrogen at 0.4 Myr in $\rm M_1$ (left) and $\rm P_9$ (right).}
    \label{fig:map_xn_04}
\end{figure*}

\begin{figure}
    \centering
    \includegraphics[width=0.99\linewidth,trim={0 0 0 1.4cm},clip]{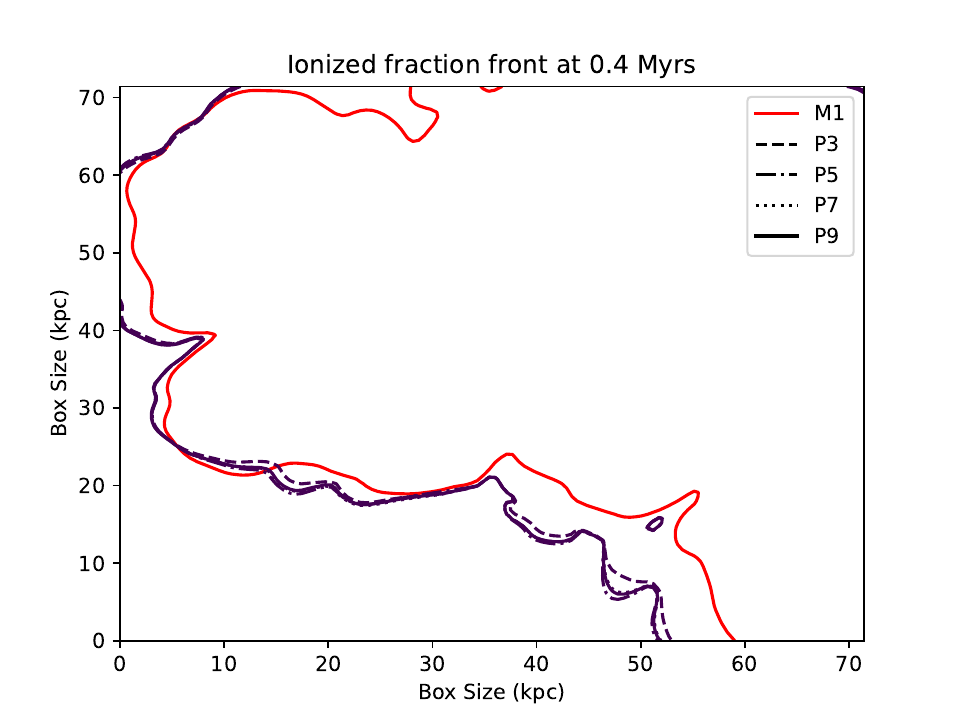}
    \caption{Position of the 50\% ionisation front at 0.4 Myr.}
    \label{fig:map_front_pos}
\end{figure}

\begin{figure*}
    \centering
    \includegraphics[width=15cm,trim={1.1cm 0cm 2.2cm 1.2cm},clip]{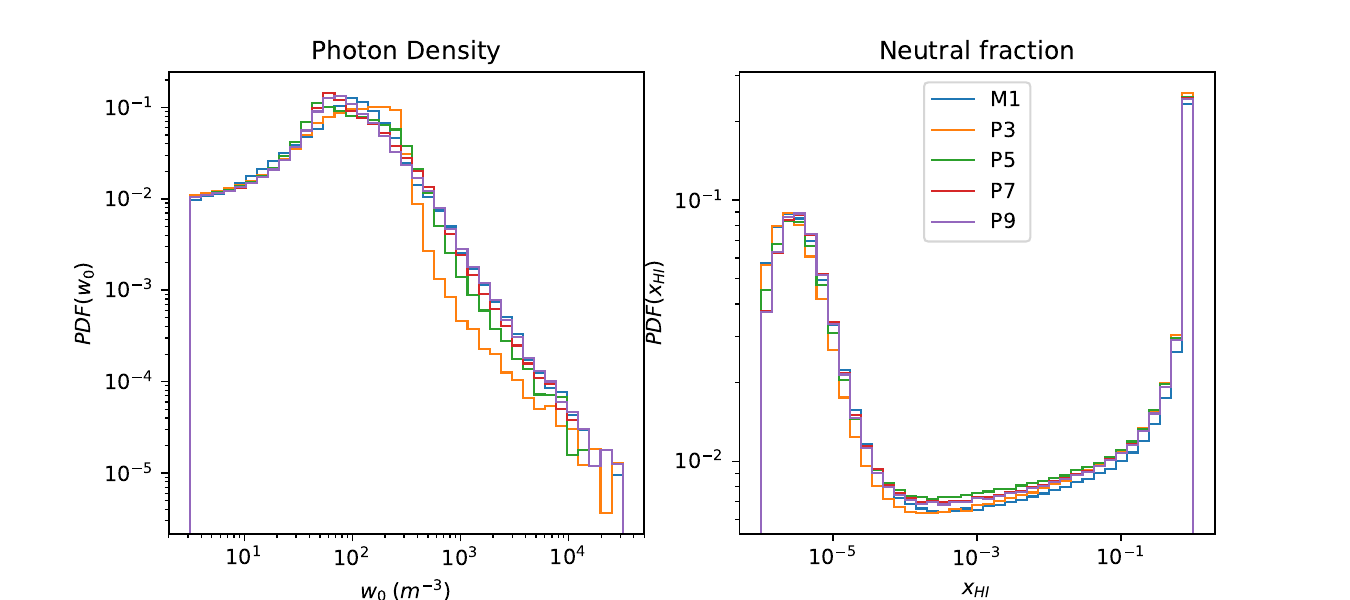}
    \caption{Probability distribution function of photon density at 0.4 Myr (left) and of neutral fraction at 0.4 Myr (right).}
    \label{fig:map_pdf_04}
\end{figure*}

\begin{figure*}
    \centering
    \includegraphics[width=15cm,trim={3.5cm 0cm 3.5cm 1.4cm},clip]{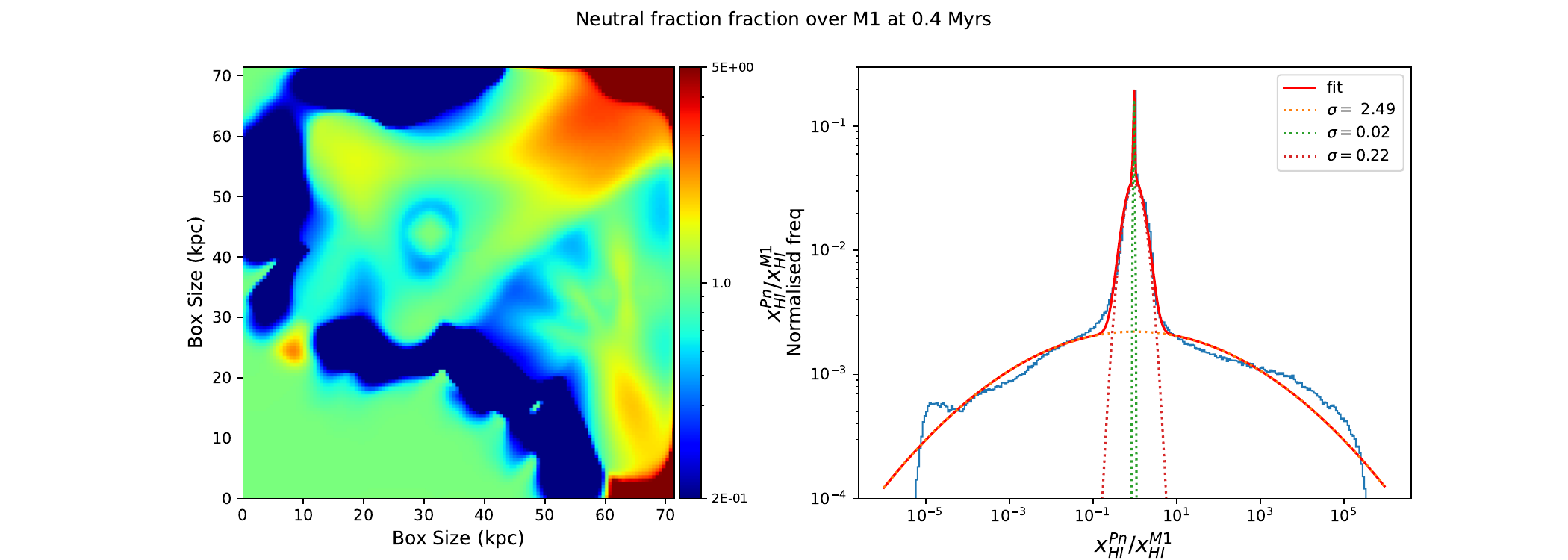}
    \caption{ Map of neutral fraction of $\rm P_n$ over $\rm M_1$ (left) and histogram of this ratio for the whole box, fitted with a sum of 3 log-Gaussian curve, at 0.4 Myr.}
    \label{fig:map_neutraldiff04}
\end{figure*}

\begin{figure*}
    \centering
    \includegraphics[width=15cm,trim={1.9cm 1.3cm 1cm 1.4cm},clip]{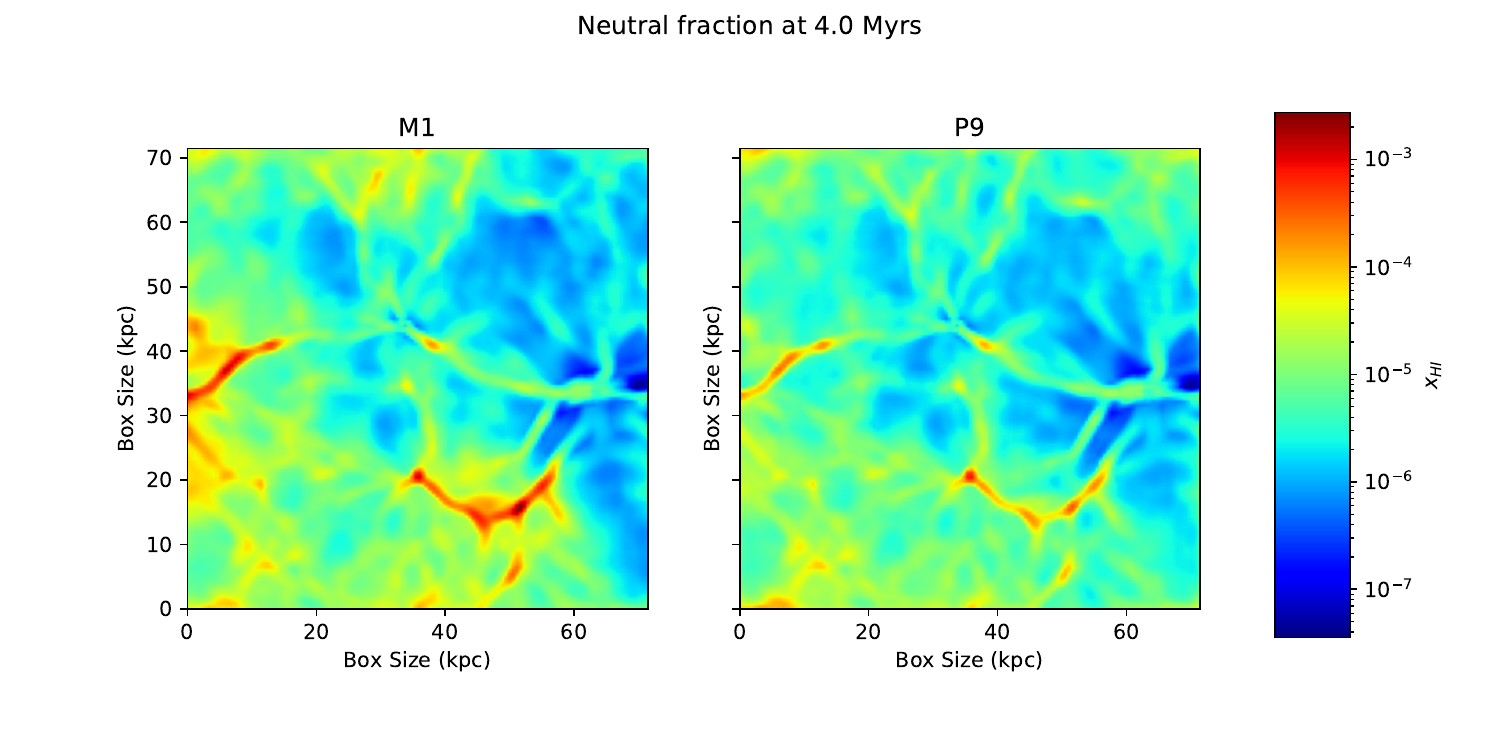}
    \caption{Maps of the neutral fraction of hydrogen at 4.0 Myr in $\rm M_1$ (left) and $\rm P_9$ (right).}
    \label{fig:map_xn_4}
\end{figure*}

\begin{figure*}
    \centering
    \includegraphics[width=15cm,trim={1.5cm 0 2.7cm 1.4cm},clip]{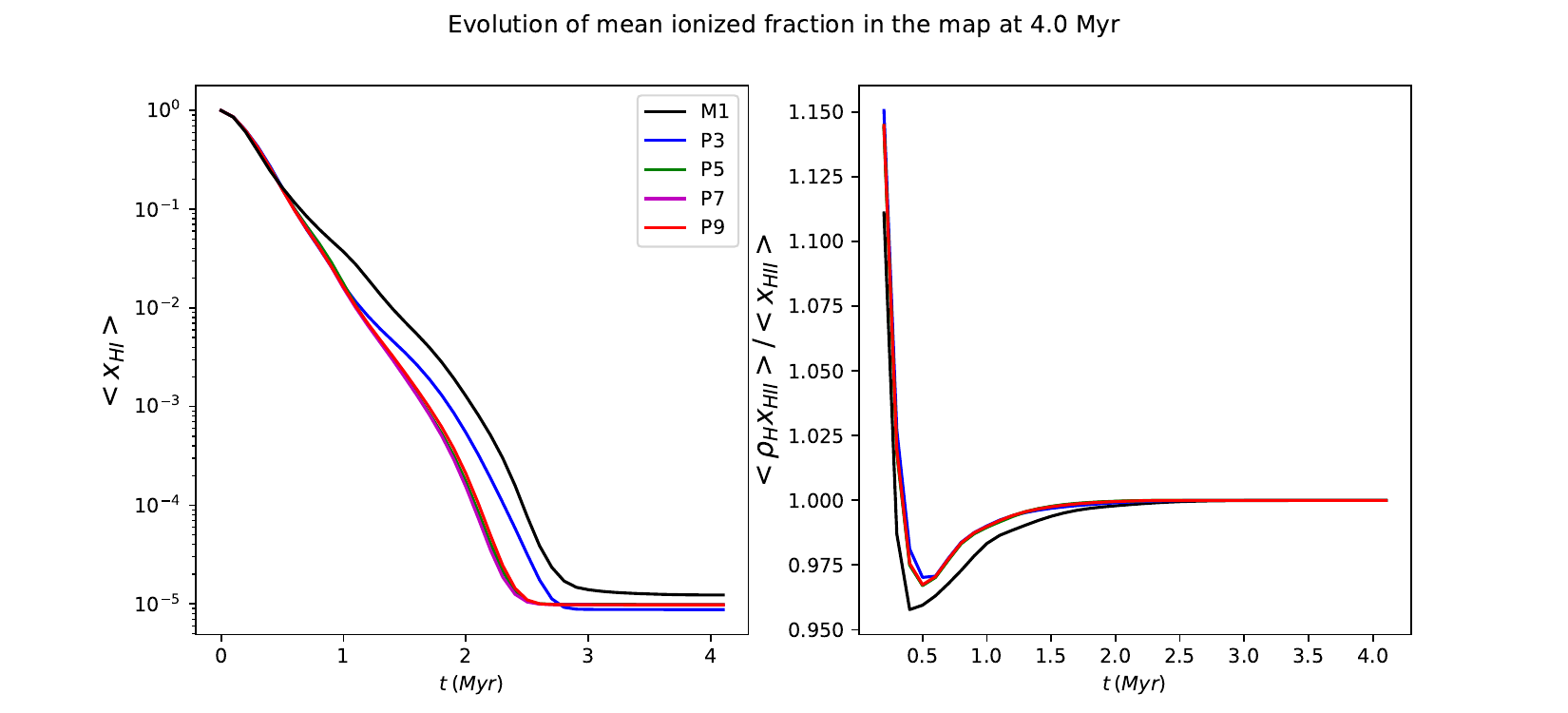}
    \caption{Mean neutral fraction evolution over time.}
    \label{fig:map_nh_mean}
\end{figure*}

\begin{figure*}
    \centering
    \includegraphics[width=15cm,trim={1.2cm 0 2.2cm 1.1cm},clip]{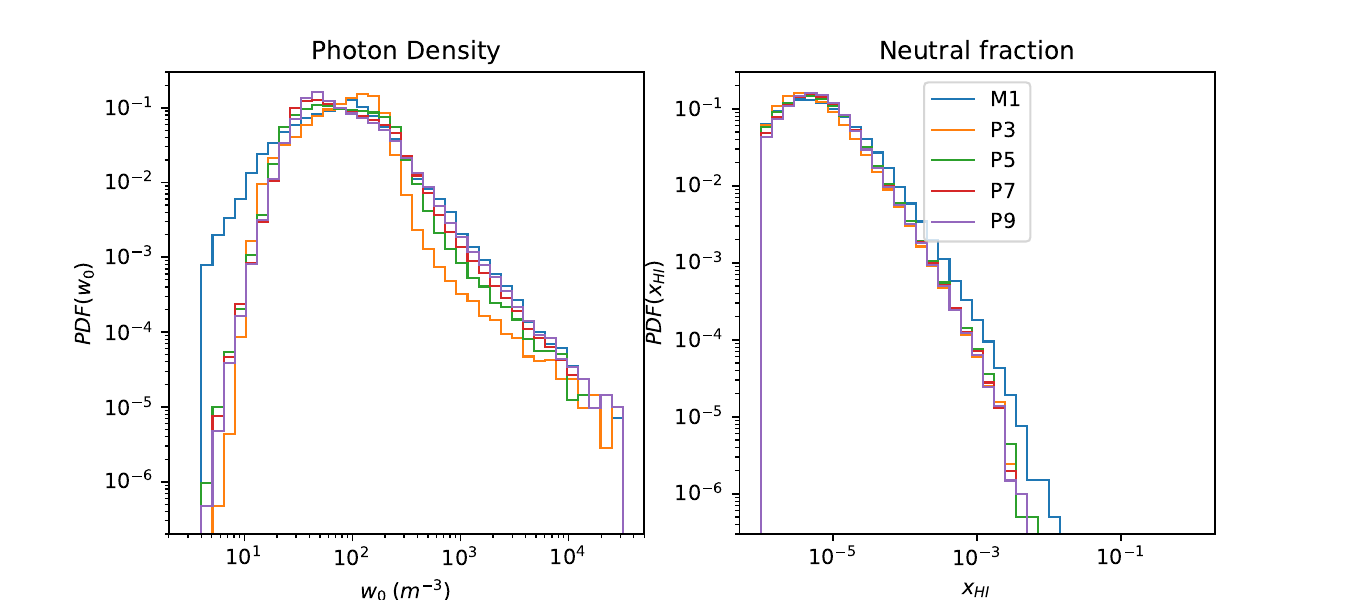}
    \caption{Probability distribution function of photon density at 4.0 Myr (left) and of neutral fraction at 4.0 Myr (right).}
    \label{fig:map_pdf_4}
\end{figure*}

\begin{figure*}
    \centering
    \includegraphics[width=15cm,trim={2cm 1.3cm 1.1cm 1.4cm},clip]{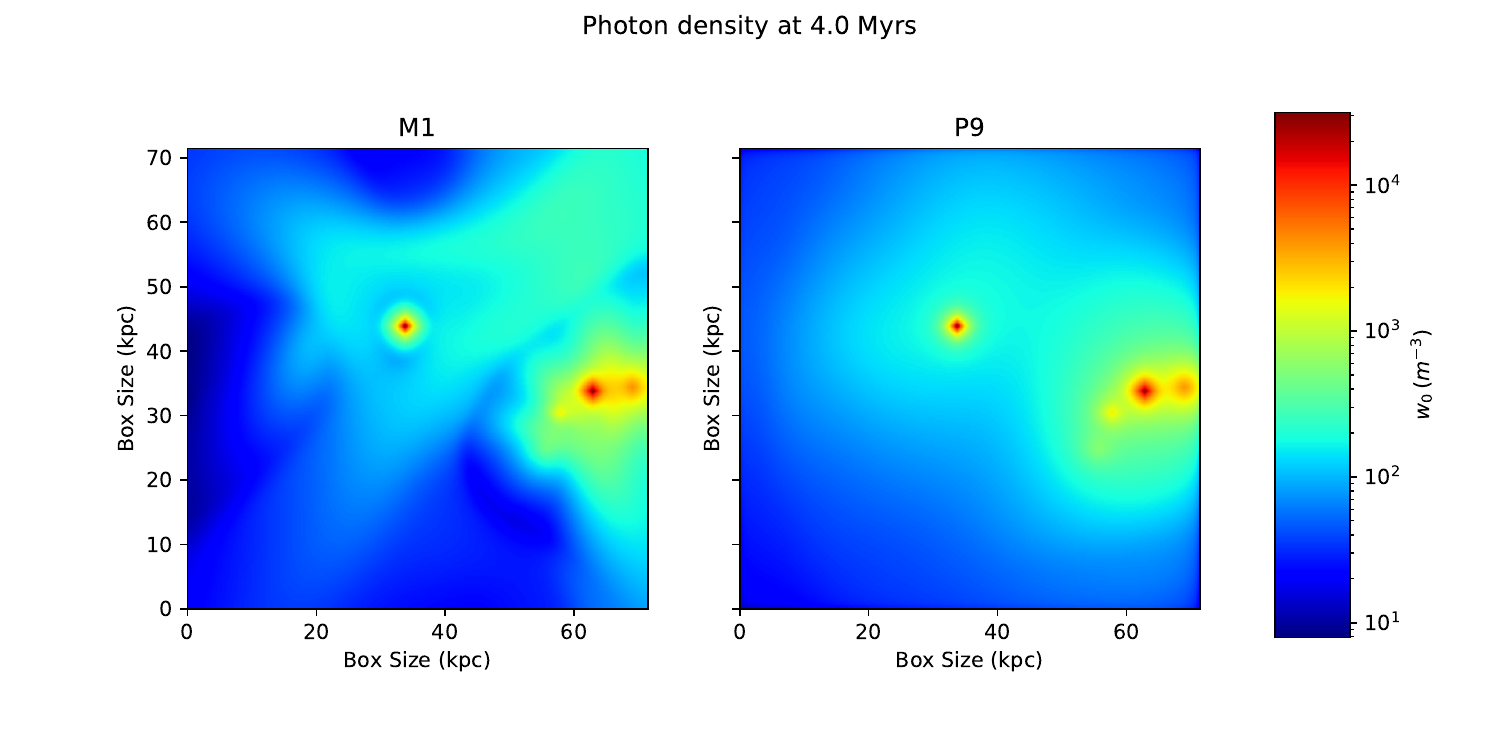}
    \caption{Maps of the photon density at 4.0 Myr in $\rm M_1$ (left) and $\rm P_9$ (right).}
    \label{fig:map_photodens_4}
\end{figure*}

\begin{figure}
    \centering
    \includegraphics[width=0.99\linewidth,trim={0 0 0 1.4cm},clip]{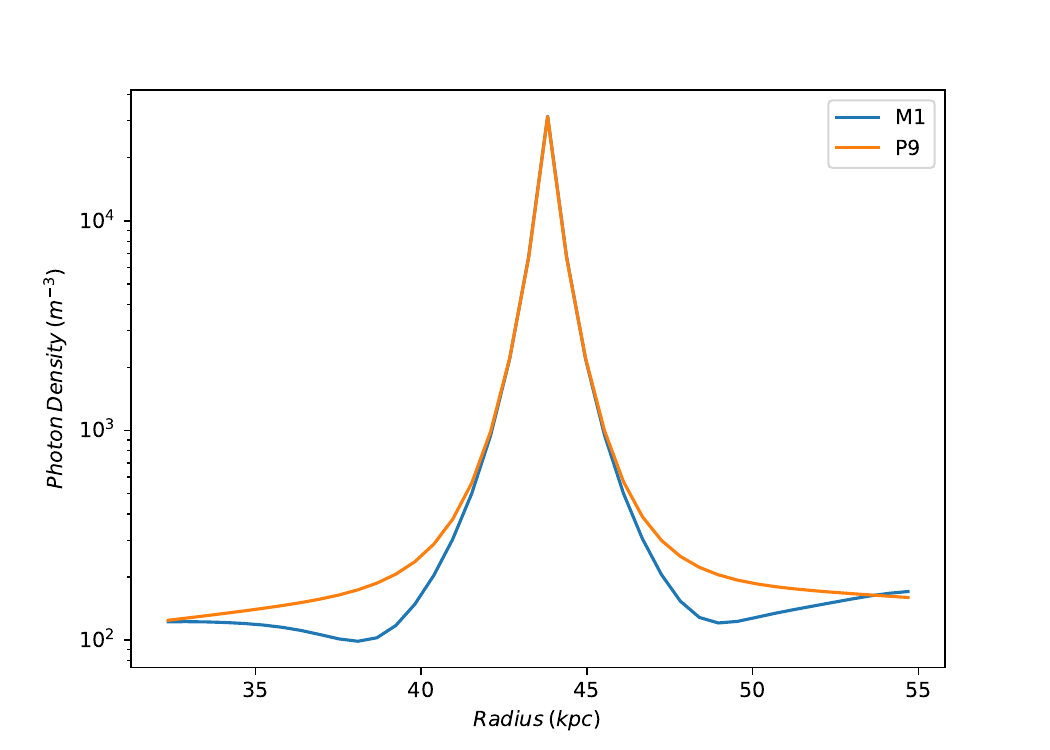}
    \caption{Photon density profile around source 7 at 4.0 Myr.}
    \label{fig:map_source_profile}
\end{figure}

\begin{figure*}
    \centering
    \includegraphics[width=16cm,trim={3.3cm 1.4cm 3.5cm 2cm},clip]{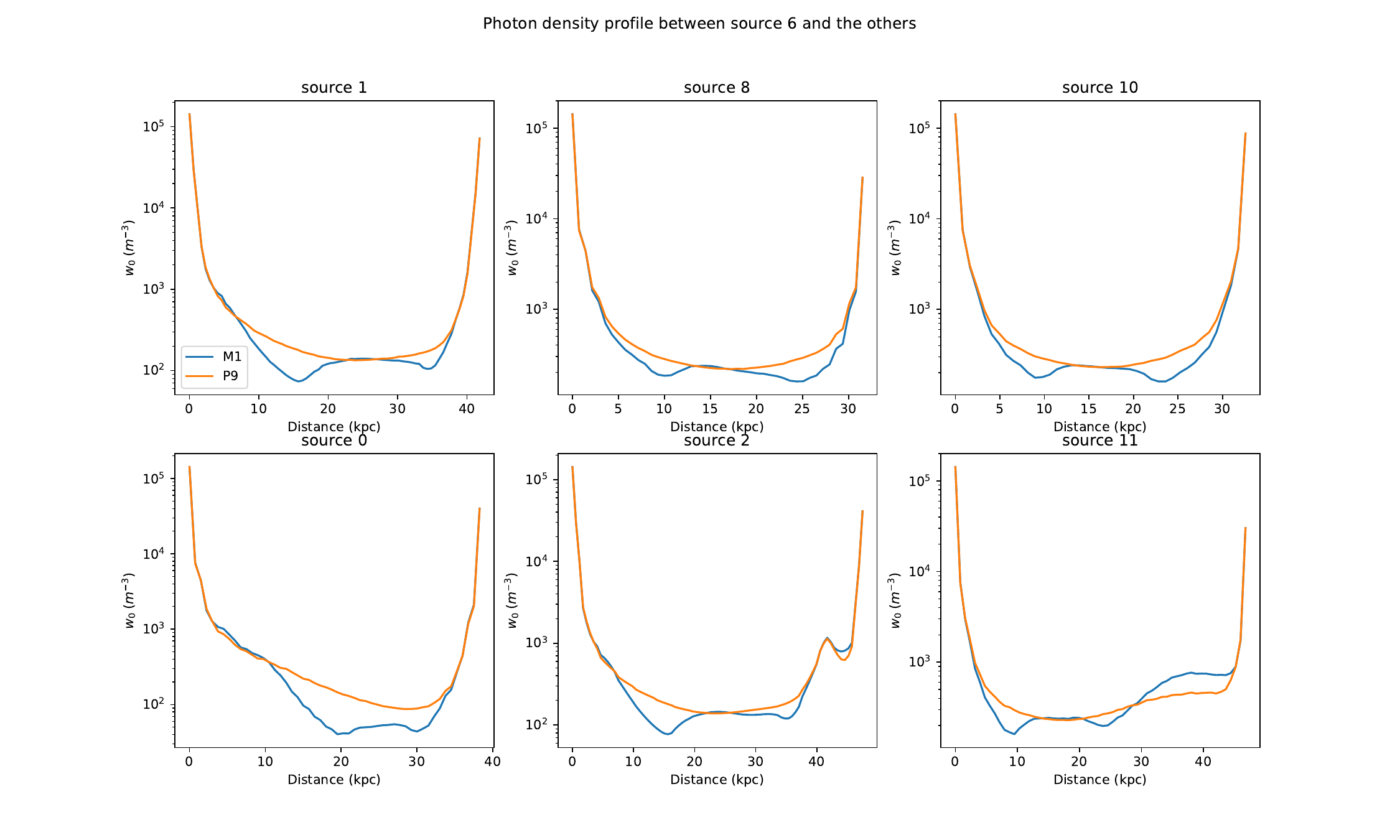}
    \caption{Photon density profiles between source 6 and six other sources of the simulation at 4.0 Myr.}
    \label{fig:map_all_sources_profiles}
\end{figure*}

\begin{figure*}
    \centering
    \includegraphics[width=15cm,trim={5.8cm 0cm 4.0cm 1.4cm},clip]{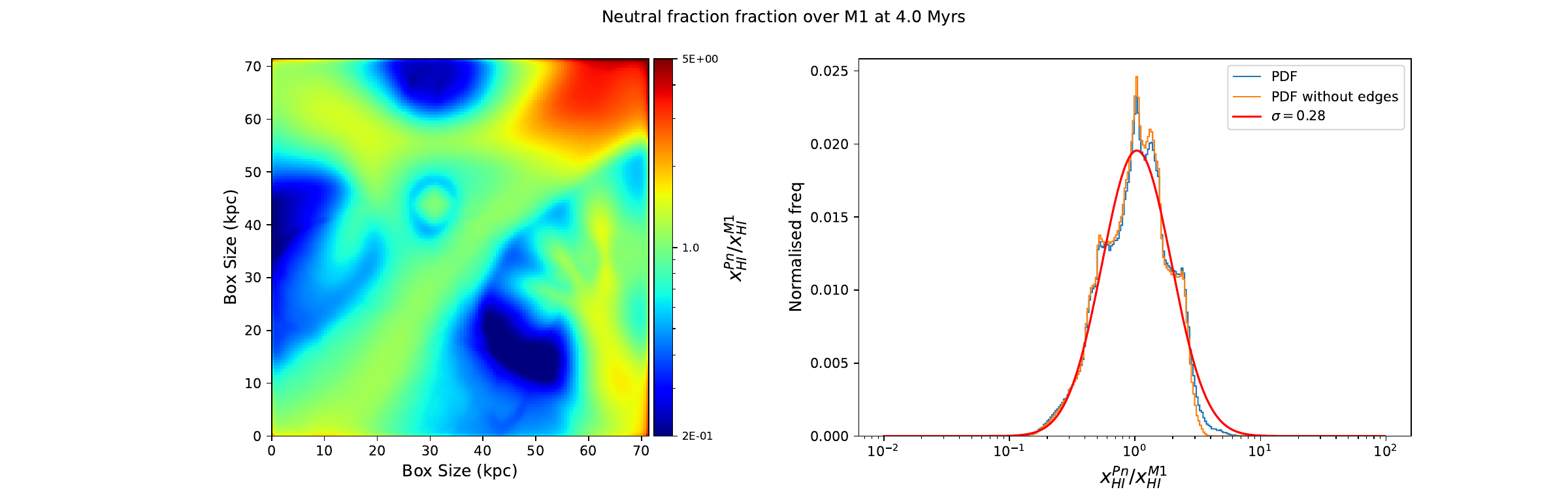}
    \caption{Map of neutral fraction of $\rm P_n$ over $\rm M_1$ (left) and histogram of this ratio fitted with a log-Gaussian curve, at 4.0 Myr for the whole box and for the box excluding the layers of edge cells to avoid boundary condition effects (right).}
    \label{fig:map_neutraldiff}
\end{figure*}

\begin{figure*}
    \centering
    \includegraphics[width=15cm,trim={3cm 1.cm 3cm 1.8cm},clip]{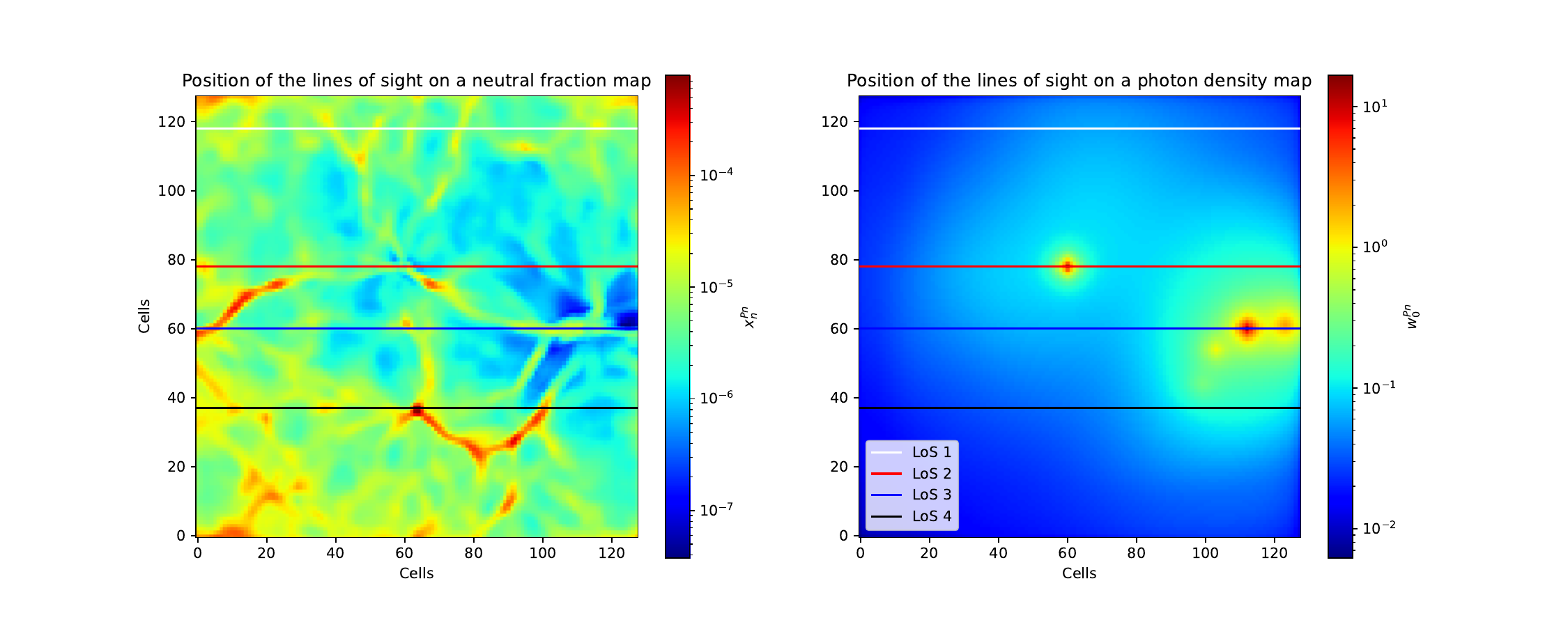}
    \caption{Positions of the LoS used for the transmission spectra.}
    \label{fig:transmit_position}
\end{figure*}

\begin{figure*}
    \centering
    \includegraphics[width=17cm,trim={8cm 0cm 7cm 1cm},clip]{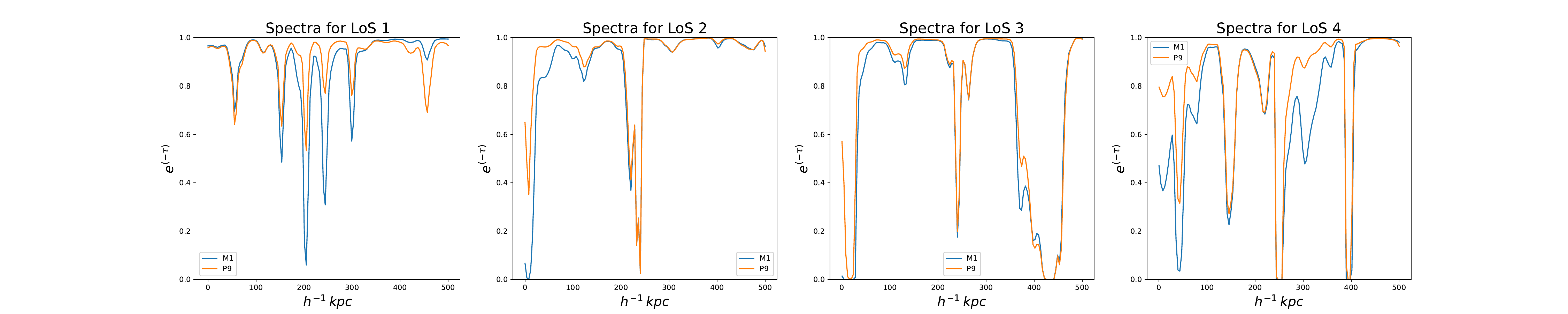}
    \caption{Transmission spectra along several LoSs for $\rm M_1$ and $\rm P_9$.}
    \label{fig:transmit_spectra}
\end{figure*}

\section{Results: Coupled radiation and chemistry - Cosmological radiative benchmark tests}\label{section:comparisonproject}

In this section, we show how $\rm P_n$ fits the requirements for a cosmological radiative transfer model by making it undergo four separate test cases described in \citet{Iliev_2006}. These standardised tests allow for an easy comparison with other radiative transfer models in the literature. A more in-depth analysis of $\rm M_1$'s results for the same tests can be found in \citet{Aubert_2008}. All test cases in this section are dimensional, i.e. use physical box sizes and emissivities, with coupled chemistry as described in \ref{subsection:chemistry} and no hydrodynamics or gravity. 

\subsection{Isothermal Strömgren sphere}\label{subsection:isostrom}

A single ionising source can only ionise a given radius of gas around it before equilibrium is reached and it enters the continuous regime. The sphere formed by that ionised gas is referred to as the Strömgren sphere \citep{Stromgren_1993,Spitzer_1978}, whose radius is given by the following equation:\\

\begin{equation}
        R_s  =  \displaystyle \left[ \frac{3S_0}{4\pi\alpha_{B}(T )n_{H}^2}\right]^{1/3},
\end{equation}

with $\rm S_0$ the intensity of the ionising source in photons per second, $\rm n_H$ the density of the gas in unit per cubic metre, $\rm T$ the gas temperature in Kelvin, and $\rm \alpha_B$ the case B recombination rate of hydrogen. The characteristic evolution time of the front can be derived as $\rm t_r=1/\alpha_B(T)n_H$. As such, with a unique non-variable source and fixed temperature and hydrogen density, the theoretical radius of this sphere can easily be computed and compared to the result of our model. With that in mind, we devised a test similar to the one given in \citet{Iliev_2006} consisting of a single continuous isotropic source at [0.5,0.5,0.5] adimensional box co-ordinates, emitting $\rm S_0=5\times 10^{48}\:\rm photons.s^{-1}$. The source being isotropic, all of its moments, $\rm S_l$, with $\rm l>0$ are equal to 0. The surrounding hydrogen has a fixed density, $\rm n_H=1\times 10^3\:\rm H.m^{-3}$, and temperature, $\rm T=1\times 10^4\:\rm K$. The initial value of ionised fraction is $1.2\times 10^{-3}$. The grid uses a resolution of $65^3$, representing a 13.2 kpc side with reflective boundary conditions. \\

Fig. \ref{fig:iostrom_front} shows the evolution of the position of the ionising front normalised by the Strömgren radius over the duration of the simulation, along with a theoretical evolution of the front derived as $\rm [1-exp(-t/t_r)]^{1/3}$ \citep{Iliev_2006}. Here, the simulation runs for just over four times the characteristic evolution time, for about 500 Myrs. Fig. \ref{fig:isostrom_profile} shows side-by-side profiles of the ionised and neutral fractions at 35 Myr on the left, and 500 Myr on the right. Both $\rm P_5$ and $\rm M_1$ are compared here, with their relative difference being showcased in the lower panels. \\

We can observe that $\rm P_5$ converges on the theoretical value of the Strömgren sphere in Fig. \ref{fig:iostrom_front}, and is in very good agreement with the evolution of the sphere in $\rm M_1$. All in all, the results here show that the coupling of $\rm P_n$ with chemistry is able to reproduce the results of $\rm M_1$ very closely, as shown in Fig. \ref{fig:isostrom_profile}, as the difference between the two models is always smaller than 1\%.

\subsection{Non-isothermal Strömgren sphere}\label{subsection:strom}

A similar experiment can be made, but this time with temperature variation being taken into account. The conditions of the test are similar to the one previously described, but with an initial temperature condition of $100\:\rm K$ and the source being a $10^5\:\rm K$ black body. This time, the computation of the theoretical Strömgren sphere radius is less straightforward, but we use a comparison with $\rm M_1$ as was done previously, as well as with the previous test, to compare the results of the two models. \\

In Fig. \ref{fig:strom_profile} we show side by side the profiles of the ionised and neutral fractions at 35 Myr on the left and 100 Myr on the right. As was done previously, we compare both models, with their relative difference being showcased in the underplot. Fig. \ref{fig:strom_temp} shows the same comparison but on the temperature profile around our source, and at three different time steps instead of two: 10, 35, and 100 Myr.\\

Once again, $\rm P_n$ matches $\rm M_1$ very well, in terms of both the ionised fraction, with a maximum difference of around 2\%, and the temperature, with a maximum difference of 5\%. Surprisingly, we observe that this variation in ionised fraction has been divided by a factor of two compared to the same test without temperature variation at the same time step. The temperature profile also fits what is physically expected of this test, with a sudden drop when the ionising front is reached, since the radiation has not reached the medium beyond that point, and thus has not started photo-heating it.\\

\subsection{Shadowing by a dense clump}\label{subsection:clump}

Previous test cases in this section \ref{section:comparisonproject} took place in a context in which $\rm P_n$ is known to fare pretty well, i.e. with continuous, isotropic sources that tend to limit the oscillations of the model. We have confirmed that, in such cases, $\rm P_n$ fares as well as $\rm M_1$. However, the third test in \citet{Iliev_2006} makes use of non-isotropic sources that may cause more issues for $\rm P_n$, such as oscillations and negativity.\\

The shadowing by a dense clump test aims to mimic how resistant to ionisation a clump of dense, cold gas would be using our radiative transfer model. It consists of a cold, high-density sphere of hydrogen placed on the path of a constant directional flux of photons. The sphere -- hereby referred to as a ‘clump’ -- should resist full ionisation, as it is too dense for the radiation to properly penetrate it, instead creating a slim shell of ionised gas on its exposed surface. On top of that, as the clump blocks the path of the ionising flux, it should create a ‘shadow’ of non-ionised, non-heated gas in its wake. We show a simplified description of the model in Fig. \ref{fig:clump_schematic}.\\

For our experiment, we defined the background hydrogen density as $0.2\:\rm h^{-1}\:H.m^{-3}$ comoving\footnote{h is the dimensionless Hubble parameter, $\rm h= H_0/100$, with $\rm H_0$ the Hubble constant taken as $70 \: \rm km.s^{-1}.Mpc^{-1}$ \citep{Iliev_2006}}, which corresponds to the mean background density of the Universe. In the same way, the mean density hydrogen in a dark matter clump is 200 times greater at $40\:\rm h^{-1}\:m^{-3}$. Considering a redshift of $\rm z=9$ for our test, this turns into a high-density hydrogen sphere of $4\times 10^4\:\rm H.m^{-3}$ plunged in a background gas density of $2\times 10^2\:\rm H.m^{-3}$. We used a $6.6\:\rm kpc$ box with a clump of radius $0.8\:\rm kpc$ that runs for $3\:\rm Myr$. The initial temperature is homogeneous at $8000\:\rm K$, unless in the clump, where it was set at $80\:\rm K$. The ionising flux was emitted in the z direction by a plane of Gaussian beam directional source cells similar to the one used in section \ref{section:test_cases}, at a value of $1\times 10^{10}\: \rm ph.s^{-1}.m^{-2}$. The centre of the clump was placed at co-ordinates [0.5,0.5,0.25] in adimensional box co-ordinates. A further description of the setup and the method used to deal with the oscillations of the directional sources can be found in Appendix \ref{Appendix:nested_boxes}.\\

We can observe that the results of $\rm P_9$ compare well to $\rm M_1$ in Fig. \ref{fig:clump_comp}, where both models show a similar ionisation of the clump. On top of that, we can see that $\rm P_9$ creates a sharper shadow behind said clump than $\rm M_1$, which, in this case, is closer to what should be physically expected. 
Indeed, \citet{Aubert_2008} showed using ray tracing that the physical expectation in this test (and in a vacuum) is a sharp, linear shadow. 
\\

However, we have to point out that other results in the literature show $\rm M_1$ resulting in a far sharper shadow than in our test \citep{Aubert_2008,Gonzalez_2007}. The reason for this discrepancy is the use of a different numerical scheme for this specific test in these other publications, namely the HLL scheme. Our Rusanov scheme is indeed more diffusive than the HLL one in this situation; however, we opted to keep using it as, in practice, in large simulation codes such as RAMSES or DYABLO, $\rm M_1$ uses the similar GLF numerical scheme. As such, we considered observing the difference between $\rm M_1$ and $\rm P_9$ with this numerical scheme to be more in line with the expected results of these simulation codes. 
\\

One can, however, observe some local maxima in the photon density of $\rm P_9$. This is the consequence of the oscillatory out-of-beam components of the source model in $\rm P_n$ due to the truncation of the spherical harmonics development to a finite order. However, this difference compared to $\rm M_1$ has little to no impact on the temperature and neutral fraction maps, and is specific to this situation adverse to $\rm P_n$ that would not be found in most EoR simulations.\\

To get a better quantitative view of the behaviour of the gas inside the clump, we can take a look at Fig. \ref{fig:clump_mean}, where we plot the evolution of the mean ionised fraction and mean temperature inside the clump for a longer period of time, 16 Myr. We can first point out that $\rm M_1$ and $\rm P_9$ converge on similar values, with $\rm M_1$ plateauing around a value of 0.80 mean ionised fraction. The discrepancy is even less obvious with the temperature.\\

We also plot in Appendix \ref{Appendix:order_negativity} the same results for $\rm P_3$, for illustration purposes. In both ionised fraction and temperature, it overshoots compared to high-order $\rm P_n$ and $\rm M_1$, going as far as to fully ionise the clump by 5 Myr of run time. This shows that lower orders of $\rm P_n$ cannot properly pass this test.\\

\subsection{Cosmological density map}\label{subection:cosmomap}
\subsubsection{Setup}
The fourth and final test is also the one that comes closest to the future applications this model could have in cosmological simulations of the EoR. It consists of testing the behaviour of our radiative transfer model on a fixed $128^3$ cell grid of hydrogen density provided by the authors of \citet{Iliev_2006}, along with 16 isotropic continuous sources of varying intensities placed in that cosmological field. This situation mimics a real simulation cube (without hydro or gravity) at a smaller scale to get a better idea of how the model will react in situations more practical than the previously mentioned test cases. We expect to observe strong differences between $\rm M_1$ and $\rm P_n$ due to the interaction between several sources and the pseudo-sources we have shown in section \ref{section:test_cases}. Indeed, since these unwanted interactions tend to change the directionality of part of the radiation field, the photon flux brought to the main ionising front might be diminished or increased compared to $\rm P_n$, where this problem does not arise. We also aim to observe the differences between several orders of $\rm P_n$, especially $\rm P_3$, $\rm P_5$, $\rm P_7$, and $\rm P_9$, to observe at which point the solution seems to converge. As for $\rm P_n$ potential negativity, we expect it to have a minor impact on the test, as this test does not hold any sharp time-discontinuities or anisotropies that might cause the model to strongly oscillate, apart from potential boundary condition effects. Finally, we shall also push the test further than was done in the initial paper to reach the optically thin regime where $\rm M_1$ is known to struggle, to see how much better $\rm P_n$ fares compared to it. \\

The simulation box represents a 0.5/h Mpc sized cube comoving, which, with $\rm h=0.7$ in this specific case, and considering a redshift of $\rm z=9$, translates to a 71.4 kpc sized box. The initial temperature was set to be $100\:\rm K$, and the initial ionised fraction $1.2\times10^{-3}$. The 16 sources were tabulated, and the intensity ranges from $0.64\times10^{52}\:\rm ph.s^{-1} $ to $7.97\times10^{52}\:\rm ph.s^{-1}$. For our analysis, we observe specifically one slice of this box, at co-ordinate $\rm z=64$, of which we show the hydrogen density map in Fig. \ref{fig:map_nh_dens}.\\ 

The hydrogen density map is the output of a gravity-hydrodynamics cosmological simulation, which was performed using periodic boundary conditions, as is customary in this type of simulation \citep{Ryu_1993}. On the other hand, the radiative transfer test uses open boundary conditions. While seemingly inconsistent, this is the setup reported and used in \citet{Iliev_2006}, which is why we stuck to it. \\
In our study, we ran the test further in time than usually reported in the literature, showing results not only at $\rm t=0.4\:\rm Myr$ but also at $\rm t=4.0\:\rm Myr$. Doing that, we discovered an unreported, seemingly minor issue of this test: source 4 is located near the edge of the grid, at position x=124. While the source itself lives in a single cell, the gas cloud it belongs to is extended, and shows up as an over-density at x=124-128 (the grid size is $128^3$) but also at x=1-5 because of periodicity. However, because of the non-periodicity of the RT test, source 4, while successfully ionising its direct surroundings, is unable to ionise the isolated counterpart of its host gas cloud in the x=1-5 region, as it is much further away. Remarkably, we found that this happened with M1 only, while P9 for instance did not show such an issue.\\
Having identified this flaw in the test setup, we decided to solve it by shifting the hydrogen density map in x by minus 5 cells, so as to reunite the source with the bulk of its host gas cloud, and avoid having any cloud cut in 2 by the boundaries. Doing this successfully solves the issue and allows for a cleaner comparison between the RT methods.\\

\subsubsection{Results}

At $0.4\:\rm Myr$, we can already spot some strong differences between the two models. Qualitatively, it appears in Fig. \ref{fig:map_photodens_04_4dec} that photons in $\rm P_9$ and $\rm M_1$ seem to favour different propagation directions, with $\rm M_1$ showing photon ejectas on the top right and bottom right of the slice we observe, while photons in $\rm P_n$ seem to propagate more evenly in all directions. This translates in the neutral fraction too as shown in Fig. \ref{fig:map_xn_04}, where we observe a discrepancy in ionised regions between the two models at the same time. 
We can quantify this difference, by observing the position of the ionising front at $0.4\:\rm Myr$, defined as the point where the gas is 50\% ionised, or $\rm x = 0.5$. This is shown in Fig. \ref{fig:map_front_pos}, where this difference appears even more strikingly. This may be the result of the collisional behaviour of photons and the interactions between sources in $\rm M_1$, resulting in photons contributing to the ionising of neutral hydrogen in specific directions rather than in an isotropic way. We can also point out that all instances of $\rm P_n$ seem to converge on a single solution as the order increases, the highest tested order being $\rm P_9$. It looks as if, based on this metric, $\rm P_7$ and even $\rm P_5$ could be potentially less expensive alternatives to $\rm P_9$ with similar results.\\
To better compare the homogeneity of $\rm P_9$ and $\rm M_1$ in ionised regions, we can look at the probability distribution function (PDF) of the photon density and the neutral fraction, which are shown in Fig. \ref{fig:map_pdf_04}. We do not observe here a significant difference between $\rm M_1$ and $\rm P_9$ yet, apart from a slightly lower number of cells in $\rm M_1$ between $\rm 10^{-3}<x_{HI}<1$. In this regime, the $\rm M_1$ PDF is about 0.05 dex below the $\rm P_n$ PDF. We can also point out that $\rm P_3$ diverges significantly from $\rm P_5$, $\rm P_7$, and $\rm P_9$ in photon density distribution, with its peak in photon density being at $2\times10^2\:\rm ph.m^{-3}$, while it is around $10^2\:\rm ph.m^{-3}$ for other models, and then experiences a lower number of cells than other orders between $2\times10^2$ and $10^4\:\rm ph.m^{-3}$, with a difference ranging from one to six decades. This difference also shows in the neutral fraction PDF between $10^{-5}$ and $10^{-3}$, with $\rm P_3$ also being lower than all higher orders. This tends to show that it is too low an order to use, as it strays from the converged solution. In Fig. \ref{fig:map_photodens_04_4dec}, it already appears that, in the ionised regions, there is a significant discrepancy between $\rm M_1$ and $\rm P_9$, with a strong anisotropy in the radiation field in $\rm M_1$, for instance in the top right corner, which likely results from the interaction between the emissions from the two main sources (in the centre and the cluster of sources on the right), since it does not appear in $\rm P_9$. As such, we consider this configuration of the radiation field to be unphysical. This difference is highlighted further in Fig. \ref{fig:map_neutraldiff04}, where we plot the ratio of neutral fraction of $\rm P_n$ over $\rm M_1$ ($\rm x_{HI}^{P_9}/x_{HI}^{M_1}$) along with a histogram of this ratio in the whole simulation box. The resulting histogram can be approximated as the sum of three distinct distributions centred at $\rm x_{HI}^{P_9}/x_{HI}^{M_1} =1$:
\begin{itemize}
    \item  A narrow peak with $\rm \sigma = 0.02$ corresponding to the regions where the gas is neutral in both $\rm M_1$ and $\rm P_9$. Those are the areas not yet reached by radiation in both models.
    \item  A broad peak with $\rm \sigma = 2.49$ corresponding to regions where the gas is neutral in either $\rm M_1$ or $\rm P_9$. Those areas correspond to the propagation discrepancy between the two models.
    \item  A medium regime with $\rm \sigma = 0.22$ corresponding to regions where the gas is ionised in both models. Those areas have reached the optically thin regime.
\end{itemize}
The third regime is especially relevant as it quantifies the theoretical error on $\rm x_{HI}$ due to the radiative transfer model, as would be observed in the Lyman-$\rm \alpha$ forest. In conclusion, we can say that, at $\rm 0.4\:Myr$, directionality loss in $\rm M_1$ causes a difference in ionisation directions between the two models, and that the inhomogeneities appearing in the ionised region in $\rm M_1$ are already significant, despite being outmatched by the difference in the position of the ionisation fronts between the two models.  \\

We then observe the simulation at $\rm 4.0\:Myr$, in the optically thin regime. Here, there are no ionising fronts to observe any more as the whole box should be ionised already. Yet, when we observe neutral fraction maps in Fig. \ref{fig:map_xn_4}, we can very quickly see that $\rm M_1$ is less ionised than $\rm P_9$, with bigger patches of more neutral hydrogen in the higher-density regions. These patches are not neutral per se, as their neutral fraction is quite low ($\sim 10^{-3}$), but are far less ionised than $\rm P_9$ even though the simulation has converged by now, as shown by the mean neutral fraction evolution in the left plot of Fig. \ref{fig:map_nh_mean}, since both $\rm P_9$ and $\rm M_1$'s neutral fractions have reached a plateau or a near-plateau. The apparent difference in ionisation can be found in this plot too, as the final mean neutral fraction in $\rm M_1$ is slightly higher than all the mean values for $\rm P_n$ models, which seem to converge on a value of $10^{-5}$. $\rm M_1$ is also a bit late compared to most $\rm P_n$ models, converging around 0.5 Myr later, which could also have an impact on simulations as the timing of reionisation is an important open question \citep{Kulkarni_2019, Cain_2021}. This contrast is conveyed in the PDF of the neutral fraction distribution (right plot of Fig. \ref{fig:map_pdf_4}), where it appears that $\rm M_1$ under-ionises cells even as it stands above all $\rm P_n$ orders, which have all converged on a lower proportion of neutral cells. Indeed, going from around $10^{-4}$ until $10^{-2}$, the PDF of $\rm M_1$ is higher than all $\rm P_n$ orders by at least a factor of 10, even showing a tail of the distribution with cells more ionised than all of $\rm P_n$ at a value of around $2\times10^{-2}$, while the largest neutral fraction for $\rm P_n$ models is around $8\times10^{-3}$. This difference in the neutral fraction can be explained by the loss of directionality in the $\rm M_1$ model, causing some of the photon budget to exit the simulation box through the photon ejecta that were already visible at 0.4 Myr in Fig.\ref{fig:map_photodens_04_4dec}. As a consequence of this, it is likely that $\rm M_1$ has a lower photon budget for ionising the box compared to $\rm P_9$, and thus ends up more neutral once the simulation has converged.\\

We can verify this hypothesis by looking at the photon density distribution at 4.0 Myr in Fig. \ref{fig:map_photodens_4}. At this point in the simulation, we expect a $\rm 1/r^2$ photon density profile around sources, as the fully ionised medium is now transparent to radiation. However, it appears in this figure that, while $\rm P_9$ seems to fit that description, it is not the case for $\rm M_1$. We can observe large photon outflows probably causing a lack of a photon budget for ionising the box as much as $\rm P_9$.

\subsubsection{Dark sombreros}

We also observe unexpected rings of photon under-densities surrounding the sources in $\rm M_1$, most prominently around the central source of the slab we study. We refer to this structure as the ‘dark doughnut’ or the ‘dark sombrero’. This specific artefact is highlighted in Fig. \ref{fig:map_source_profile}, where we can see how the photon density dips isotropically around sources in $\rm M_1$, forming the said dark sombrero shape that is unphysical. This phenomenon is ubiquitous in the simulation, as shown in Fig. \ref{fig:map_all_sources_profiles}, where we plot the profiles between source 6 and the six other sources in the simulation. We observe that, as long as two sources are distant enough, the same kind of dips appear in the photon density of $\rm M_1$, whereas the density of $\rm P_9$ is smooth and monotonically decreasing around each source, as is physically expected in this optically thin regime. We also observe this structure in all directions surrounding the sources, confirming that it is indeed a three-dimensional shell of surrounding lower-photon-density sources. This hints at the fact that this unphysical phenomenon, as well as the photon ejecta previously mentioned, could also appear in more physics-rich simulations \citep{Lewis_2022}, where it could impact the final neutral fraction. This is also highlighted in the PDF on the left plot of Fig. \ref{fig:map_pdf_4}, where a tail appears in the $\rm M_1$ distribution that is absent from all $\rm P_n$ ones between $3\times10^0$ and $3\times10^1\:\rm ph.m^{-3}$, showcasing far more cells at low density, and, as such, a far more diffuse distribution of photons in the box. This difference can be as big as 4 orders of magnitude compared to $\rm P_n$. We can quantify this difference in ionisation by observing the plot and histogram in Fig. \ref{fig:map_neutraldiff}, showcasing the ratio of neutral fractions, $\rm x_{HI}(P_9)/x_{HI}(M_1)$, at 4.0 Myr. The sigma of the log-Gaussian is $\rm \sigma=0.28$ ($\rm \sigma=0.26$ if we disregard three layers of boundary cells), which is very similar to the transitional regime. This error is bound to have an impact on the global output of the neutral fraction, which is the observable that is used to probe the reionisation epoch. We checked real cosmological simulations and found examples of the so-called dark sombrero artefact as well (see Appendix \ref{Appendix:Dark_sombrero}). 

We would also like to point out that the $\rm x_{HI}$ ratios tend to be abnormally high at the edges of the box, as can be seen on the left plot of Fig. \ref{fig:map_neutraldiff}. These result from edge effects in the photon density of $\rm P_n$. Indeed, as is explained in Appendix \ref{Appendix:boundary_conditions}, we must use different boundary conditions for $\rm M_1$ (transmissive) and $\rm P_n$ (forced to a very low value outside in the ghost region outside of the box). In $\rm P_n$, this results in an artificially elevated photon density close to the edges, which in turns lead to an enhanced deficit of $\rm x_{HI}$ in $\rm M_1$ as compared to Pn in edge regions.
We checked that this does not affect our main conclusions by removing the three outermost layers of cells from our statistics and observing the differences with the whole box. The right plot of Fig. \ref{fig:map_neutraldiff} shows that this removes a small tail of elevated $\rm x_{HI}^{P_n}/x_{HI}^{M_1}$ cells, but the general shape and width of the PDF is maintained.\\

\subsubsection{Lyman-$\rm \alpha$ transmission spectra}

To get a better insight into the observational equivalent of what is presented in this simplified version of a cosmological simulation, we can observe the transmission spectra of our cells, i.e. the fraction of radiation that they allow to pass through depending on the wavelength of light, as a makeshift mock Lyman-$\rm \alpha$ forest spectra. The transmission in a cell was computed by taking the inverse exponential of its optical depth, $\rm \tau$, which is defined as follows \citep{Dijkstra_2019}:

\begin{equation}
    \tau = n_H(1-x)dx\sigma_{HI}(T),
\end{equation}

with

\begin{equation}
    \sigma_{HI}(T) = 5.9\times10^{-14}\times\left(\frac{T}{10^4}\right)^{-\frac{1}{2}}\times10^{-4}.
\end{equation}

However, to ensure that we got a representative sample of how the transmission can evolve depending on the medium it is going through, we took four separate lines of sight (LoSs) in our box to extract the spectra from. They are shown in Fig. \ref{fig:transmit_position}, and were chosen to investigate different regimes:
\begin{itemize}
    \item LoS 1 goes through two regions with a large contrast in neutral fraction when comparing the two models, as seen in Fig. \ref{fig:map_neutraldiff}.
    \item LoS 2 goes through source 7, whose dark sombrero artefact is very visible in most of our images and in Fig. \ref{fig:map_source_profile}.
    \item LoS 3 goes through the high-photon-density region created by the cluster of sources 3 to 6, whose interaction with source 7 is shown in most of our images.
    \item LoS 4 goes through a region with a high neutral fraction, $\rm x_{HI}$, and is thus less ionised.
\end{itemize}

The results are shown in Fig. \ref{fig:transmit_spectra} for all four LoSs. We observe less transmission in $\rm M_1$ when the LoS goes through a less ionised medium as expected from its globally lower ionisation. We also highlight the fact that this difference is strongest in areas with a large difference in neutral fractions between the two models, especially with LoS 1. Comparing it to Fig. \ref{fig:map_neutraldiff}, we can see regions where $\rm P_9$ ionises more, where $\rm M_1$'s transmission is lower, and where $\rm M_1$ over-ionises (LoS 1, 400-500 $\rm h^{-1} \rm kpc$), i.e. where $\rm M_1$'s transmission is higher than $\rm P_9$'s. This last area has been shown to be the result of non-physical interactions in $\rm M_1$, which in turn has an observable effect in these spectra. What we also observe is the presence of cells with a transmission value of 0 for both models in three of the four LoSs, showing that there are still areas that are opaque to radiation at this epoch. 
   
\section{Conclusions}

$\rm P_n$ is an alternative moment-based radiative transfer model to the usual $\rm M_1$ model, based on a projection of the photon density on the spherical harmonics basis truncated to a chosen order leading to a quite simple closure equation. After showcasing how it can correct some of the issues in $\rm M_1$, such as the collisionality of photons, but also pointing at its weaknesses, namely its sensitivity to temporal and angular discontinuities and its ability to output negative photon densities, we compared the two models using benchmarked tests created for this purpose. We find that even in a physical context reminiscent of the EoR, $\rm P_n$ fares equally as good as or better than $\rm M_1$. Test cases with a single source such as the Strömgren sphere tests show no significant difference between the two models, while test cases with several sources in interaction such as the cosmological map test show a tendency of $\rm P_n$ to better describe the physically expected distribution of photons. As for the clump test, $\rm P_9$ seems to give closer results to the MC-RT ones, with a sharper shadow behind the clump. 
As such, $\rm P_n$ appears to be a viable replacement for $\rm M_1$ from a purely physical standpoint. \\

By comparing $\rm P_9$ and $\rm M_1$ radiation fields in an idealised and cosmological test case, we highlight a new, thus far unreported artefact of $\rm M_1$, the dark sombrero. A dark sombrero appears in $\rm M_1$ solutions as a spherical photon-deficit shell around the source. The photon density in dark sombreros can be underestimated by a factor of up to 2-3. 
Dark sombreros occur in regions where a source's radiation field connects with that of another source or group of sources. These basic properties (position and amplitude) of the dark sombreros may depend on the sources' relative intensities, positions, and spatial resolution, although we have not been able to test this in detail in this study.
Understanding the root numerical causes of this artefact requires a dedicated investigation, beyond the scope of this paper\footnote{Prior to this study, such profiles have been seen in e.g. the Cosmic Dawn II simulation \citep{Ocvirk_2020} (p.c. P. Ocvirk)}. Moreover, the larger-scale $\rm M_1$ photon density also exhibits spurious features, enhancing or reducing photon density in various regions. We used a small reionisation-like test simulation to characterise the  relative error in hydrogen neutral fractions between $\rm M_1$ and $\rm P_9$. 
We find that there is a small difference in the timing of the reionisation of the test box. Also, in its final state, the $\rm M_1$ solution is slightly more neutral than the $\rm P_9$ realization. In regions where reionisation is finished in both models, the relative error is well represented by a Gaussian with a dispersion of between 0.22 and 0.28 dex in $\rm \log_{10}(\rm x_{HI})$.  Both aspects are likely related to the photons' collisional behaviour in $\rm M_1$.

We also computed, as an illustration, the Lyman-$\rm \alpha$ transmission on 4 LoS of this mini-reionisation box, to produce mock Lyman-$\rm \alpha$ forests. The main trend we observe is $\rm M_1$ being slightly more opaque, consistent with it also being slightly more neutral. However, the trend is reversed in a few regions. 
We highlight that this is mostly an experiment. Ideally proper, large-scale, well-resolved, Cosmic-Dawn-like simulations would be performed to quantify these transmission differences in a more realistic scenario.

Even though $\rm P_n$ may display a form of oscillatory behaviour in the presence of either quickly time-varying or angle-discontinuous sources (such as strongly collimated or focused emission), this can be mitigated by increasing $\rm P_n$'s order, at the cost of a larger computational cost (complexity in O($\rm (N+1)^2$), see Appendix \ref{Appendix:order_negativity}), or filtering. We have shown that models such as $\rm P_3$ or $\rm P_5$ might have too low orders to be used in simulations, but that $\rm P_7$ and $\rm P_9$ already seem close to convergence in the tests we have showcased. We consider there to be probably no need to push the order further for EoR simulations, which, in general, fit the model's needs quite well, with their isotropic sources and simple geometry. However, further tests could be done comparing $\rm P_9$, $\rm P_{11}$, $\rm P_{13}$, and $\rm P_{15}$ to ensure this convergence in more complex environments. \\

All the tests presented in this paper were done in a simplified context, with only hydrogen, simplified chemical reactions, and no gravity or hydrodynamics added. Knowing that RT already takes up a large part of the computing power and time in current cosmological simulations with the use of $\rm M_1$ as a radiative transfer model, the question of the cost $\rm P_n$ could add to these simulations has to be asked. Furthermore, even if we have highlighted some of the starkest differences between $\rm M_1$ and $\rm P_n$, we still need to study these differences and their impact in larger, more complex simulations. As such, a comparison should be made between fully fledged cosmological runs with both models to ensure that this increased computational costs translates into an actual numerical convergence in our final results.\\

Ideally, the results presented in this thesis should motivate and support an implementation of $\rm P_n$ in state-of-the-art and exascale astrophysical codes such as RAMSES, miniRAMSES\footnote{\url{https://bitbucket.org/rteyssie/mini-ramses/src/develop/}}, and DYABLO. Due to $\rm P_n$'s simplicity and explicit formulation, we foresee a fairly straightforward implementation in such legacy codes, unlocking massive MPI, OpenMP, and GPU parallelism, which should be the path pursued in the short term. Such parallelism is required in order to run, for example, $\rm P_9$ in realistic setups, due to its large computational and memory cost. Alternatively, other methods should be researched to overcome $\rm M_1$'s shortcomings, such as the use of a neural network to correct its closure or to facilitate the computation of the closure of higher orders of $\rm M_n$.\\

\begin{acknowledgements}

      The authors would like to acknowledge the High Performance Computing Center of the University of Strasbourg for supporting this work by providing scientific support and access to computing resources. Part of the computing resources were funded by the Equipex Equip@Meso project (Programme Investissements d'Avenir) and the CPER Alsacalcul/Big Data.\\
      
      This work of the Interdisciplinary Thematic Institute IRMIA++, as part of the ITI 2021-2028 program of the University of Strasbourg, CNRS and Inserm, was supported by IdEx Unistra (ANR-10-IDEX-0002), and by SFRI-STRAT’US project (ANR-20-SFRI-0012) under the framework of the French Investments for the Future Program.
\end{acknowledgements}

\bibliographystyle{aa} 
\bibliography{pn.bib}

@article{alldredge_2012,
    adsurl = {https://ui.adsabs.harvard.edu/abs/2012SJSC...34B.361A},
    author = {{Alldredge}, Graham W. and {Hauck}, Cory D. and {Tits}, Andr{\'e} L.},
    doi = {10.1137/11084772X},
    journal = {SIAM Journal on Scientific Computing},
    month = {January},
    number = {4},
    pages = {B361-B391},
    title = {{High-Order Entropy-Based Closures for Linear Transport in Slab Geometry II: A Computational Study of the Optimization Problem}},
    volume = {34},
    year = {2012},
}

@article{AREPO-IDORT,
    adsurl = {https://ui.adsabs.harvard.edu/abs/2025arXiv250316627M},
    archiveprefix = {arXiv},
    author = {{Ma}, Jing-Ze and {Pakmor}, R\textbackslash''udiger and {Justham}, Stephen and {de Mink}, Selma E.},
    doi = {10.48550/arXiv.2503.16627},
    eid = {arXiv:2503.16627},
    eprint = {2503.16627},
    journal = {arXiv e-prints},
    month = {March},
    pages = {arXiv:2503.16627},
    primaryclass = {astro-ph.IM},
    title = {{AREPO-IDORT: Implicit Discrete Ordinates Radiation Transport for Radiation Magnetohydrodynamics on an Unstructured Moving Mesh}},
    year = {2025},
}

@article{Aubert_2008,
    adsurl = {https://ui.adsabs.harvard.edu/abs/2008MNRAS.387..295A},
    archiveprefix = {arXiv},
    author = {{Aubert}, Dominique and {Teyssier}, Romain},
    doi = {10.1111/j.1365-2966.2008.13223.x},
    eprint = {0709.1544},
    journal = {\mnras},
    month = {June},
    number = {1},
    pages = {295-307},
    primaryclass = {astro-ph},
    title = {{A radiative transfer scheme for cosmological reionization based on a local Eddington tensor}},
    volume = {387},
    year = {2008},
}

@article{Barkana&Loeb2001,
    adsurl = {https://ui.adsabs.harvard.edu/abs/2001PhR...349..125B},
    archiveprefix = {arXiv},
    author = {{Barkana}, R. and {Loeb}, A.},
    doi = {10.1016/S0370-1573(01)00019-9},
    eprint = {astro-ph/0010468},
    journal = {\physrep},
    month = {July},
    number = {2},
    pages = {125-238},
    primaryclass = {astro-ph},
    title = {{In the beginning: the first sources of light and the reionization of the universe}},
    volume = {349},
    year = {2001},
}

@article{Berthon_2010,
    author = {Berthon, Christophe and Dubois, Joanne and Dubroca, Bruno and Nguyen-Bui, Thanh-Ha and Turpault, Rodolphe},
    doi = {10.4208/aamm.09-m09105},
    journal = {Advances in Applied Mathematics and Mechanics},
    month = {06},
    pages = {259-285},
    title = {A Free Streaming Contact Preserving Scheme for the M1 Model},
    volume = {2},
    year = {2010},
}

@article{C2ray,
    adsurl = {https://ui.adsabs.harvard.edu/abs/2006NewA...11..374M},
    archiveprefix = {arXiv},
    author = {{Mellema}, Garrelt and {Iliev}, Ilian T. and {Alvarez}, Marcelo A. and {Shapiro}, Paul R.},
    doi = {10.1016/j.newast.2005.09.004},
    eprint = {astro-ph/0508416},
    journal = {\na},
    month = {March},
    number = {5},
    pages = {374-395},
    primaryclass = {astro-ph},
    title = {{C $^{2}$-ray: A new method for photon-conserving transport of ionizing radiation}},
    volume = {11},
    year = {2006},
}

@article{Cain_2021,
    adsurl = {https://ui.adsabs.harvard.edu/abs/2021ApJ...917L..37C},
    archiveprefix = {arXiv},
    author = {{Cain}, Christopher and {D'Aloisio}, Anson and {Gangolli}, Nakul and {Becker}, George D.},
    doi = {10.3847/2041-8213/ac1ace},
    eid = {L37},
    eprint = {2105.10511},
    journal = {\apjl},
    month = {August},
    number = {2},
    pages = {L37},
    primaryclass = {astro-ph.CO},
    title = {{A Short Mean Free Path at z = 6 Favors Late and Rapid Reionization by Faint Galaxies}},
    volume = {917},
    year = {2021},
}

@article{Dijkstra_2019,
    adsurl = {https://ui.adsabs.harvard.edu/abs/2019SAAS...46....1D},
    author = {{Dijkstra}, Mark},
    doi = {10.1007/978-3-662-59623-4_1},
    journal = {Saas-Fee Advanced Course},
    month = {January},
    pages = {1},
    title = {{Physics of Ly{\ensuremath{\alpha}} Radiative Transfer}},
    volume = {46},
    year = {2019},
}

@article{Dodorico_2023,
    adsurl = {https://ui.adsabs.harvard.edu/abs/2023MNRAS.523.1399D},
    archiveprefix = {arXiv},
    author = {{D'Odorico}, Valentina and {Ba{\~n}ados}, E. and {Becker}, G.~D. and {Bischetti}, M. and {Bosman}, S.~E.~I. and {Cupani}, G. and {Davies}, R. and {Farina}, E.~P. and {Ferrara}, A. and {Feruglio}, C. and {Mazzucchelli}, C. and {Ryan-Weber}, E. and {Schindler}, J.-T. and {Sodini}, A. and {Venemans}, B.~P. and {Walter}, F. and {Chen}, H. and {Lai}, S. and {Zhu}, Y. and {Bian}, F. and {Campo}, S. and {Carniani}, S. and {Cristiani}, S. and {Davies}, F. and {Decarli}, R. and {Drake}, A. and {Eilers}, A.-C. and {Fan}, X. and {Gaikwad}, P. and {Gallerani}, S. and {Greig}, B. and {Haehnelt}, M.~G. and {Hennawi}, J. and {Keating}, L. and {Kulkarni}, G. and {Mesinger}, A. and {Meyer}, R.~A. and {Neeleman}, M. and {Onoue}, M. and {Pallottini}, A. and {Qin}, Y. and {Rojas-Ruiz}, S. and {Satyavolu}, S. and {Sebastian}, A. and {Tripodi}, R. and {Wang}, F. and {Wolfson}, M. and {Yang}, J. and {Zanchettin}, M.~V.},
    doi = {10.1093/mnras/stad1468},
    eprint = {2305.05053},
    journal = {\mnras},
    month = {July},
    number = {1},
    pages = {1399-1420},
    primaryclass = {astro-ph.GA},
    title = {{XQR-30: The ultimate XSHOOTER quasar sample at the reionization epoch}},
    volume = {523},
    year = {2023},
}

@article{Dubroca_1999,
    adsurl = {https://ui.adsabs.harvard.edu/abs/1999CRASM.329..915D},
    author = {{Dubroca}, B. and {Feugeas}, J.},
    doi = {10.1016/S0764-4442(00)87499-6},
    journal = {Academie des Sciences Paris Comptes Rendus Serie Sciences Mathematiques},
    month = {November},
    number = {10},
    pages = {915-920},
    title = {{Etude th{\'e}orique et num{\'e}rique d'une hi{\'e}rarchie de mod{\`e}les aux moments pour le transfert radiatif}},
    volume = {329},
    year = {1999},
}

@article{Dubroca_2002,
    adsurl = {https://ui.adsabs.harvard.edu/abs/2002JCoPh.180..584D},
    author = {{Dubroca}, B. and {Klar}, A.},
    doi = {10.1006/jcph.2002.7106},
    journal = {Journal of Computational Physics},
    month = {August},
    number = {2},
    pages = {584-596},
    title = {{Half-Moment Closure for Radiative Transfer Equations}},
    volume = {180},
    year = {2002},
}

@article{Garett_2014,
    adsurl = {https://ui.adsabs.harvard.edu/abs/2013TTSP...42..203G},
    author = {{Garrett}, C. Kristopher and {Hauck}, Cory D.},
    doi = {10.1080/00411450.2014.910226},
    journal = {Transport Theory and Statistical Physics},
    month = {September},
    number = {6-7},
    pages = {203-235},
    title = {{A Comparison of Moment Closures for Linear Kinetic Transport Equations: The Line Source Benchmark}},
    volume = {42},
    year = {2013},
}

@phdthesis{gerhard2020,
    author = {Gerhard, Pierre},
    note = {Thèse de doctorat dirigée par Helluy, Philippe Mathématiques Strasbourg 2020},
    title = {Réduction de modèles cinétiques et applications à l'acoustique du bâtiment},
    url = {http://www.theses.fr/2020STRAD001},
    year = {2020},
}

@article{Gonzalez_2007,
    adsurl = {https://ui.adsabs.harvard.edu/abs/2007A&A...464..429G},
    author = {{Gonz{\'a}lez}, M. and {Audit}, E. and {Huynh}, P.},
    doi = {10.1051/0004-6361:20065486},
    journal = {\aap},
    month = {March},
    number = {2},
    pages = {429-435},
    title = {{HERACLES: a three-dimensional radiation hydrodynamics code}},
    volume = {464},
    year = {2007},
}

@article{Harikane_2025,
    adsurl = {https://ui.adsabs.harvard.edu/abs/2025ApJ...991..222O},
    archiveprefix = {arXiv},
    author = {{Ono}, Yoshiaki and {Ouchi}, Masami and {Harikane}, Yuichi and {Yajima}, Hidenobu and {Nakajima}, Kimihiko and {Fujimoto}, Seiji and {Nakane}, Minami and {Xu}, Yi},
    doi = {10.3847/1538-4357/adfc4d},
    eid = {222},
    eprint = {2502.08885},
    journal = {\apj},
    month = {October},
    number = {2},
    pages = {222},
    primaryclass = {astro-ph.GA},
    title = {{Morphological Demographics of Galaxies at z {\ensuremath{\sim}} 10─16: Log-normal Size Distribution and Exponential Profiles Consistent with the Disk Formation Scenario}},
    volume = {991},
    year = {2025},
}

@article{HC_method,
    adsurl = {https://ui.adsabs.harvard.edu/abs/2006A&A...452..907R},
    archiveprefix = {arXiv},
    author = {{Rijkhorst}, E.-J. and {Plewa}, T. and {Dubey}, A. and {Mellema}, G.},
    doi = {10.1051/0004-6361:20053401},
    eprint = {astro-ph/0505213},
    journal = {\aap},
    month = {June},
    number = {3},
    pages = {907-920},
    primaryclass = {astro-ph},
    title = {{Hybrid characteristics: 3D radiative transfer for parallel adaptive mesh refinement hydrodynamics}},
    volume = {452},
    year = {2006},
}

@article{Hui_Gnedin_97,
    adsurl = {https://ui.adsabs.harvard.edu/abs/1997MNRAS.292...27H},
    archiveprefix = {arXiv},
    author = {{Hui}, Lam and {Gnedin}, Nickolay Y.},
    doi = {10.1093/mnras/292.1.27},
    eprint = {astro-ph/9612232},
    journal = {\mnras},
    month = {November},
    number = {1},
    pages = {27-42},
    primaryclass = {astro-ph},
    title = {{Equation of state of the photoionized intergalactic medium}},
    volume = {292},
    year = {1997},
}

@article{Iliev_2006,
    adsurl = {https://ui.adsabs.harvard.edu/abs/2006MNRAS.371.1057I},
    archiveprefix = {arXiv},
    author = {{Iliev}, Ilian T. and {Ciardi}, Benedetta and {Alvarez}, Marcelo A. and {Maselli}, Antonella and {Ferrara}, Andrea and {Gnedin}, Nickolay Y. and {Mellema}, Garrelt and {Nakamoto}, Taishi and {Norman}, Michael L. and {Razoumov}, Alexei O. and {Rijkhorst}, Erik-Jan and {Ritzerveld}, Jelle and {Shapiro}, Paul R. and {Susa}, Hajime and {Umemura}, Masayuki and {Whalen}, Daniel J.},
    doi = {10.1111/j.1365-2966.2006.10775.x},
    eprint = {astro-ph/0603199},
    journal = {\mnras},
    month = {September},
    number = {3},
    pages = {1057-1086},
    primaryclass = {astro-ph},
    title = {{Cosmological radiative transfer codes comparison project - I. The static density field tests}},
    volume = {371},
    year = {2006},
}

@article{Katz_2019,
    adsurl = {https://ui.adsabs.harvard.edu/abs/2019MNRAS.483.1029K},
    archiveprefix = {arXiv},
    author = {{Katz}, Harley and {Kimm}, Taysun and {Haehnelt}, Martin G. and {Sijacki}, Debora and {Rosdahl}, Joakim and {Blaizot}, Jeremy},
    doi = {10.1093/mnras/sty3154},
    eprint = {1806.07474},
    journal = {\mnras},
    month = {February},
    number = {1},
    pages = {1029-1041},
    primaryclass = {astro-ph.CO},
    title = {{Tracing the sources of reionization in cosmological radiation hydrodynamics simulations}},
    volume = {483},
    year = {2019},
}

@article{Kulkarni_2019,
    adsurl = {https://ui.adsabs.harvard.edu/abs/2019MNRAS.485L..24K},
    archiveprefix = {arXiv},
    author = {{Kulkarni}, Girish and {Keating}, Laura C. and {Haehnelt}, Martin G. and {Bosman}, Sarah E.~I. and {Puchwein}, Ewald and {Chardin}, Jonathan and {Aubert}, Dominique},
    doi = {10.1093/mnrasl/slz025},
    eprint = {1809.06374},
    journal = {\mnras},
    month = {May},
    number = {1},
    pages = {L24-L28},
    primaryclass = {astro-ph.CO},
    title = {{Large Ly {\ensuremath{\alpha}} opacity fluctuations and low CMB {\ensuremath{\tau}} in models of late reionization with large islands of neutral hydrogen extending to z < 5.5}},
    volume = {485},
    year = {2019},
}

@inbook{Larsen_2010,
    author = {Larsen, Edward and Morel, Jim and Azmy, Yousry and Sartori, Enrico},
    doi = {10.1007/978-90-481-3411-3_1},
    isbn = {978-90-481-3410-6},
    journal = {Nuclear Computational Science: A Century in Review},
    month = {04},
    pages = {1-84},
    title = {Advances in Discrete-Ordinates Methodology},
    year = {2010},
}

@article{LEVERMORE_1984,
    adsurl = {https://ui.adsabs.harvard.edu/abs/1984JQSRT..31..149L},
    author = {{Levermore}, C.~D.},
    doi = {10.1016/0022-4073(84)90112-2},
    journal = {\jqsrt},
    month = {February},
    number = {2},
    pages = {149-160},
    title = {{Relating Eddington factors to flux limiters.}},
    volume = {31},
    year = {1984},
}

@article{Lewis_2020,
    adsurl = {https://ui.adsabs.harvard.edu/abs/2020MNRAS.496.4342L},
    archiveprefix = {arXiv},
    author = {{Lewis}, Joseph S.~W. and {Ocvirk}, Pierre and {Aubert}, Dominique and {Sorce}, Jenny G. and {Shapiro}, Paul R. and {Deparis}, Nicolas and {Dawoodbhoy}, Taha and {Teyssier}, Romain and {Yepes}, Gustavo and {Gottl{\"o}ber}, Stefan and {Ahn}, Kyungjin and {Iliev}, Ilian T. and {Chardin}, Jonathan},
    doi = {10.1093/mnras/staa1748},
    eprint = {2001.07785},
    journal = {\mnras},
    month = {August},
    number = {4},
    pages = {4342-4357},
    primaryclass = {astro-ph.GA},
    title = {{Galactic ionizing photon budget during the epoch of reionization in the Cosmic Dawn II simulation}},
    volume = {496},
    year = {2020},
}

@article{Lewis_2022,
    adsurl = {https://ui.adsabs.harvard.edu/abs/2022MNRAS.516.3389L},
    archiveprefix = {arXiv},
    author = {{Lewis}, Joseph S.~W. and {Ocvirk}, Pierre and {Sorce}, Jenny G. and {Dubois}, Yohan and {Aubert}, Dominique and {Conaboy}, Luke and {Shapiro}, Paul R. and {Dawoodbhoy}, Taha and {Teyssier}, Romain and {Yepes}, Gustavo and {Gottl{\"o}ber}, Stefan and {Rasera}, Yann and {Ahn}, Kyungjin and {Iliev}, Ilian T. and {Park}, Hyunbae and {Th{\'e}lie}, {\'E}milie},
    doi = {10.1093/mnras/stac2383},
    eprint = {2202.05869},
    journal = {\mnras},
    month = {November},
    number = {3},
    pages = {3389-3397},
    primaryclass = {astro-ph.CO},
    title = {{The short ionizing photon mean free path at z = 6 in Cosmic Dawn III, a new fully coupled radiation-hydrodynamical simulation of the Epoch of Reionization}},
    volume = {516},
    year = {2022},
}

@article{Lymanalpha_opacities,
    adsurl = {https://ui.adsabs.harvard.edu/abs/2021MNRAS.507.6108O},
    archiveprefix = {arXiv},
    author = {{Ocvirk}, Pierre and {Lewis}, Joseph S.~W. and {Gillet}, Nicolas and {Chardin}, Jonathan and {Aubert}, Dominique and {Deparis}, Nicolas and {Th{\'e}lie}, {\'E}milie},
    doi = {10.1093/mnras/stab2502},
    eprint = {2105.01663},
    journal = {\mnras},
    month = {November},
    number = {4},
    pages = {6108-6117},
    primaryclass = {astro-ph.CO},
    title = {{Lyman-alpha opacities at z = 4-6 require low mass, radiatively-suppressed galaxies to drive cosmic reionization}},
    volume = {507},
    year = {2021},
}

@article{Maselli_2003,
    adsurl = {https://ui.adsabs.harvard.edu/abs/2003MNRAS.345..379M},
    archiveprefix = {arXiv},
    author = {{Maselli}, A. and {Ferrara}, A. and {Ciardi}, B.},
    doi = {10.1046/j.1365-8711.2003.06979.x},
    eprint = {astro-ph/0307117},
    journal = {\mnras},
    month = {October},
    number = {2},
    pages = {379-394},
    primaryclass = {astro-ph},
    title = {{CRASH: a radiative transfer scheme}},
    volume = {345},
    year = {2003},
}

@article{MCCLARREN_2010,
    adsurl = {https://ui.adsabs.harvard.edu/abs/2010JCoPh.229.5597M},
    author = {{McClarren}, Ryan G. and {Hauck}, Cory D.},
    doi = {10.1016/j.jcp.2010.03.043},
    journal = {Journal of Computational Physics},
    month = {August},
    number = {16},
    pages = {5597-5614},
    title = {{Robust and accurate filtered spherical harmonics expansions for radiative transfer}},
    volume = {229},
    year = {2010},
}

@phdthesis{Meltz_2015,
    author = {B. Meltz},
    school = {University of Paris-Saclay, France},
    title = {Analyse math{\'{e}}matique et num{\'{e}}rique de syst{\`{e}}mes d'hydrodynamique compressible et de photonique en coordonn{\'{e}}es polaires. (Mathematical and Numerical Analysis of Systems of Compressible Hydrodynamics and Photonics with Polar Coordinates)},
    url = {https://tel.archives-ouvertes.fr/tel-01280237},
    year = {2015},
}

@article{Naidu_2020,
    adsurl = {https://ui.adsabs.harvard.edu/abs/2020ApJ...892..109N},
    archiveprefix = {arXiv},
    author = {{Naidu}, Rohan P. and {Tacchella}, Sandro and {Mason}, Charlotte A. and {Bose}, Sownak and {Oesch}, Pascal A. and {Conroy}, Charlie},
    doi = {10.3847/1538-4357/ab7cc9},
    eid = {109},
    eprint = {1907.13130},
    journal = {\apj},
    month = {April},
    number = {2},
    pages = {109},
    primaryclass = {astro-ph.GA},
    title = {{Rapid Reionization by the Oligarchs: The Case for Massive, UV-bright, Star-forming Galaxies with High Escape Fractions}},
    volume = {892},
    year = {2020},
}

@article{Ocvirk_2020,
    adsurl = {https://ui.adsabs.harvard.edu/abs/2020MNRAS.496.4087O},
    archiveprefix = {arXiv},
    author = {{Ocvirk}, Pierre and {Aubert}, Dominique and {Sorce}, Jenny G. and {Shapiro}, Paul R. and {Deparis}, Nicolas and {Dawoodbhoy}, Taha and {Lewis}, Joseph and {Teyssier}, Romain and {Yepes}, Gustavo and {Gottl{\"o}ber}, Stefan and {Ahn}, Kyungjin and {Iliev}, Ilian T. and {Hoffman}, Yehuda},
    doi = {10.1093/mnras/staa1266},
    eprint = {1811.11192},
    journal = {\mnras},
    month = {August},
    number = {4},
    pages = {4087-4107},
    primaryclass = {astro-ph.GA},
    title = {{Cosmic Dawn II (CoDa II): a new radiation-hydrodynamics simulation of the self-consistent coupling of galaxy formation and reionization}},
    volume = {496},
    year = {2020},
}

@book{Osterbrock_1974,
    adsurl = {https://ui.adsabs.harvard.edu/abs/1974agn..book.....O},
    author = {{Osterbrock}, Donald E.},
    title = {{Astrophysics of gaseous nebulae}},
    year = {1974},
}

@techreport{osti,
    author = {Brunner, Thomas A},
    doi = {10.2172/800993},
    institution = {Sandia National Lab. (SNL-NM), Albuquerque, NM (United States); Sandia National Lab. (SNL-CA), Livermore, CA (United States)},
    month = {06},
    place = {United States},
    title = {Forms of Approximate Radiation Transport},
    url = {https://www.osti.gov/biblio/800993},
    year = {2002},
}

@article{papovich_2025,
    adsurl = {https://ui.adsabs.harvard.edu/abs/2026ApJ..1000..111P},
    archiveprefix = {arXiv},
    author = {{Papovich}, Casey and {Cole}, Justin W. and {Hu}, Weida and {Finkelstein}, Steven L. and {Shen}, Lu and {Arrabal Haro}, Pablo and {Amor{\'\i}n}, Ricardo O. and {Backhaus}, Bren E. and {Bagley}, Micaela B. and {Bhatawdekar}, Rachana and {Calabr{\`o}}, Antonello and {Carnall}, Adam C. and {Cleri}, Nikko J. and {Daddi}, Emanuele and {Dickinson}, Mark and {Grogin}, Norman A. and {Holwerda}, Benne W. and {Jaskot}, Anne E. and {Koekemoer}, Anton M. and {Llerena}, Mario and {Lucas}, Ray A. and {Mascia}, Sara and {Pacucci}, Fabio and {Pentericci}, Laura and {P{\'e}rez-Gonz{\'a}lez}, Pablo G. and {Pirzkal}, Nor and {Raghunathan}, Srinivasan and {Seill{\'e}}, Lise-Marie and {Somerville}, Rachel S. and {Yung}, L.~Y. Aaron},
    doi = {10.3847/1538-4357/ae3b25},
    eid = {111},
    eprint = {2505.08870},
    journal = {\apj},
    month = {March},
    number = {1},
    pages = {111},
    primaryclass = {astro-ph.GA},
    title = {{Galaxies in the Epoch of Reionization Are All Bark and No Bite{\textemdash}Plenty of Ionizing Photons, Low Escape Fractions}},
    volume = {1000},
    year = {2026},
}

@article{Pawlik_2008,
    adsurl = {https://ui.adsabs.harvard.edu/abs/2008MNRAS.389..651P},
    archiveprefix = {arXiv},
    author = {{Pawlik}, Andreas H. and {Schaye}, Joop},
    doi = {10.1111/j.1365-2966.2008.13601.x},
    eprint = {0802.1715},
    journal = {\mnras},
    month = {September},
    number = {2},
    pages = {651-677},
    primaryclass = {astro-ph},
    title = {{TRAPHIC - radiative transfer for smoothed particle hydrodynamics simulations}},
    volume = {389},
    year = {2008},
}

@article{Radice_2013,
    adsurl = {https://ui.adsabs.harvard.edu/abs/2013JCoPh.242..648R},
    archiveprefix = {arXiv},
    author = {{Radice}, David and {Abdikamalov}, Ernazar and {Rezzolla}, Luciano and {Ott}, Christian D.},
    doi = {10.1016/j.jcp.2013.01.048},
    eprint = {1209.1634},
    journal = {Journal of Computational Physics},
    month = {June},
    pages = {648-669},
    primaryclass = {astro-ph.HE},
    title = {{A new spherical harmonics scheme for multi-dimensional radiation transport I. Static matter configurations}},
    volume = {242},
    year = {2013},
}

@article{Rosdahl_2013,
    adsurl = {https://ui.adsabs.harvard.edu/abs/2013MNRAS.436.2188R},
    archiveprefix = {arXiv},
    author = {{Rosdahl}, J. and {Blaizot}, J. and {Aubert}, D. and {Stranex}, T. and {Teyssier}, R.},
    doi = {10.1093/mnras/stt1722},
    eprint = {1304.7126},
    journal = {\mnras},
    month = {December},
    number = {3},
    pages = {2188-2231},
    primaryclass = {astro-ph.CO},
    title = {{RAMSES-RT: radiation hydrodynamics in the cosmological context}},
    volume = {436},
    year = {2013},
}

@article{RUSANOV1962,
    author = {V.V Rusanov},
    doi = {https://doi.org/10.1016/0041-5553(62)90062-9},
    issn = {0041-5553},
    journal = {USSR Computational Mathematics and Mathematical Physics},
    number = {2},
    pages = {304-320},
    title = {The calculation of the interaction of non-stationary shock waves and obstacles},
    url = {https://www.sciencedirect.com/science/article/pii/0041555362900629},
    volume = {1},
    year = {1962},
}

@article{Ryu_1993,
    adsurl = {https://ui.adsabs.harvard.edu/abs/1993ApJ...414....1R},
    author = {{Ryu}, Dongsu and {Ostriker}, Jeremiah P. and {Kang}, Hyesung and {Cen}, Renyue},
    doi = {10.1086/173051},
    journal = {\apj},
    month = {September},
    pages = {1},
    title = {{A Cosmological Hydrodynamic Code Based on the Total Variation Diminishing Scheme}},
    volume = {414},
    year = {1993},
}

@phdthesis{sahmim_2005,
    author = {Sahmim, Slah},
    hal_id = {tel-00010000},
    hal_version = {v1},
    month = {June},
    note = {Mme Anela Kumbaro, Mme Laure Quivy, M. Fran{\c c}ois Alouges (rapporteur), M. Claude Basdevant (pr{\'e}sident), M. Fayssal Benkhaldoun (Directeur), M. Herv{\'e} Guillard (Rapporteur)},
    school = {{Universit{\'e} Paris-Nord - Paris XIII}},
    title = {{Un sch{\'e}ma aux volumes finis avec matrice signe pour les syst{\`e}mes non homog{\`e}nes}},
    type = {Theses},
    url = {https://theses.hal.science/tel-00010000},
    year = {2005},
}

@article{SKA,
    adsurl = {https://ui.adsabs.harvard.edu/abs/2009IEEEP..97.1482D},
    author = {{Dewdney}, P.~E. and {Hall}, P.~J. and {Schilizzi}, R.~T. and {Lazio}, T.~J.~L.~W.},
    doi = {10.1109/JPROC.2009.2021005},
    journal = {IEEE Proceedings},
    month = {August},
    number = {8},
    pages = {1482-1496},
    title = {{The Square Kilometre Array}},
    volume = {97},
    year = {2009},
}

@book{Spitzer_1978,
    adsurl = {https://ui.adsabs.harvard.edu/abs/1978ppim.book.....S},
    author = {{Spitzer}, Lyman},
    doi = {10.1002/9783527617722},
    title = {{Physical processes in the interstellar medium}},
    year = {1978},
}

@article{Stromgren_1993,
    adsurl = {https://ui.adsabs.harvard.edu/abs/1939ApJ....89..526S},
    author = {{Str{\"o}mgren}, Bengt},
    doi = {10.1086/144074},
    journal = {\apj},
    month = {May},
    pages = {526},
    title = {{The Physical State of Interstellar Hydrogen.}},
    volume = {89},
    year = {1939},
}

@phdthesis{turpault,
    author = {Turpault, Rodolphe},
    hal_id = {tel-00004620},
    hal_version = {v1},
    month = {December},
    school = {{Universit{\'e} Sciences et Technologies - Bordeaux I}},
    title = {{Modelisation, approximation numerique et applications du transfert radiatif en desequilibre spectral couple avec l'hydrodynamique}},
    type = {Theses},
    url = {https://theses.hal.science/tel-00004620},
    year = {2003},
}

@article{Wu_2021,
    adsurl = {https://ui.adsabs.harvard.edu/abs/2021JCAP...02..042W},
    archiveprefix = {arXiv},
    author = {{Wu}, Xiaohan and {McQuinn}, Matthew and {Eisenstein}, Daniel},
    doi = {10.1088/1475-7516/2021/02/042},
    eid = {042},
    eprint = {2009.07278},
    journal = {\jcap},
    month = {February},
    number = {2},
    pages = {042},
    primaryclass = {astro-ph.CO},
    title = {{On the accuracy of common moment-based radiative transfer methods for simulating reionization}},
    volume = {2021},
    year = {2021},
}

@inproceedings{Zaroubi_2012,
    adsurl = {https://ui.adsabs.harvard.edu/abs/2013ASSL..396...45Z},
    archiveprefix = {arXiv},
    author = {{Zaroubi}, Saleem},
    booktitle = {The First Galaxies},
    doi = {10.1007/978-3-642-32362-1_2},
    editor = {{Wiklind}, Tommy and {Mobasher}, Bahram and {Bromm}, Volker},
    eprint = {1206.0267},
    month = {January},
    pages = {45},
    primaryclass = {astro-ph.CO},
    series = {Astrophysics and Space Science Library},
    title = {{The Epoch of Reionization}},
    volume = {396},
    year = {2013},
}

\begin{appendix}

    \section{Spherical Harmonics}\label{Appendix:SH}
    The spherical harmonics are a set of functions defining an infinite basis. With $\rm P$ the Legendre polynomials, they are written as follows

    \begin{align}\label{equ:SphericalHarmonics}
    Y_{l}^{m}(\theta,\phi)=\left\{\begin{aligned} 
    &\sqrt{\frac{1}{\pi}\frac{(l-m)!}{(l+m)!}\frac{2l+1}{2} } P_l^m(\cos(\phi))\cos(m\theta), &\quad m>0 \\
    &\sqrt{\frac{1}{2\pi}\frac{2l+1}{2}}P_l(\cos(\phi)), &\quad m=0 \\
    &-\sqrt{\frac{1}{\pi}\frac{(l+m)!}{(l-m)!}\frac{2l+1}{2} } P_l^{-m}(\cos(\phi))\sin(m\theta), &\quad m<0
    \end{aligned}
    \right.
    \end{align}
    with
    \begin{equation}\label{equ:LegendrePolynomials}
        P_{l}^{m}(\mu)=(-1)^{m}\sqrt((1-\mu^2)^{m})\frac{d^{m}P_{l}}{d\mu^{m}}.
    \end{equation}

    This infinite basis was then truncated to project the $\rm P_n$ model in our case.\\
    
    \section{$\rm P_n$ matrices}\label{appendixA}

    In this appendix, we write the definitions of the matrices $\rm J_x, J_y, J_z$ as defined in the thesis of Bertrand Meltz  \citep{Meltz_2015}. We however do not go through the extensive proof and recurrence developed in said paper. 
    Let us define $\rm A_l^m$ and $\rm B_l^m$ in the context of spherical harmonics with $\rm l<n$, $\rm m\in [\![-l;l]\!]$, as follows

    \begin{equation}
        A_l^m = \sqrt{\frac{(l-m)(l+m)}{(2l+1)(2l-1)}}\\
        B_l^m = \sqrt{\frac{(l+m-1)(l+m)}{(2l+1)(2l-1)}}.
    \end{equation}

    Then, to each orbital $\rm (l,m)$ we associated a positional index defined as:
    \begin{equation}\label{equ:orbital}
        i(l_i,m_i)=\sum\limits_{l=0}^{l_i-1}\sum\limits_{m=-l}^l 1 + \sum\limits_{m=-l_i}^{m_i} 1 = l_i^2 + l_i + m_i + 1.
    \end{equation}
    
    Counting all possible orbitals for a model of order n gave us $\rm (n+1)^2$ coefficients, which was the size of our vector and also, consequently the size of our matrices. With $\rm (i,j)\in[\![1,(n+1)^2]\!]^2$ indices in the matrices each corresponding to an orbital as defined in Eq. \ref{equ:orbital}, we could define the matrices as

    \begin{equation}
        \begin{aligned}
            J_{i,j}^x&=\frac{sgn(m_i)}{2}(1+(\sqrt{2}-1)\delta_{m_i,1})\\
            &\times[-\delta_{l_j,l_i-1}\delta_{m_j,m_i-1}B_{l_i}^{m_i}+\delta_{l_j,l_i+1}\delta_{m_j,m_i-1}B_{l_j}^{-m_j}]\\
            &+\frac{sgn(m_i+1)}{2}(1+(\sqrt{2}-1)\delta_{m_i,0})\\ &\times[\delta_{l_j,l_i-1}\delta_{m_j,m_i+1}B_{l_i}^{-m_i}-\delta_{l_j,l_i+1}\delta_{m_j,m_i+1}B_{l_j}^{m_j}],
        \end{aligned}
    \end{equation}

    \begin{equation}
        \begin{aligned}
            J_{i,j}^y&=\frac{sgn(m_i)}{2}(1-\delta_{m_i,1})\\ &\times[\delta_{l_j,l_i-1}\delta_{m_j,-(m_i-1)}B_{l_i}^{m_i}-\delta_{l_j,l_i+1}\delta_{m_j,-(m_i-1)}B_{l_j}^{m_j}]\\
            &+\frac{sgn(m_i+1/2)}{2}(1+(\sqrt{2}-1)(\delta_{m_i,0}+\delta_{m_i,-1}))\\ &\times[\delta_{l_j,l_i-1}\delta_{m_j,-(m_i+1)}B_{l_i}^{-m_i}-\delta_{l_j,l_i+1}\delta_{m_j,-(m_i+1)}B_{l_j}^{-m_j}],
        \end{aligned}
    \end{equation}

    \begin{equation}
        J_{i,j}^z=\delta_{l_j,l_i-1}\delta_{mj,m_i}A_{l_i}^{m_i}+\delta_{l_j,l_i+1}\delta_{m_j,m_i}A_{l_j}^{m_j},
    \end{equation}

    where $\rm sgn$ is a sign function defined as
    
    \begin{align}
        \forall x \in \mathbb{R},\\
        sgn(x)=\left\{\begin{aligned} 
        &1, &\quad x>0, \\
        &0, &\quad x=0, \\
        &-1, &\quad x<0.
        \end{aligned}
        \right.
    \end{align}

    \section{Photo-chemistry}\label{Appendix:chemistry}

    We could write the coupled equation of the first moment in 1D and the ionising equation as

    \begin{equation}\label{equ:CoupledTransport}
        \frac{dw_0}{dt} + \frac{dw_1}{dr} = S + \dot{w_0}^{rec} - n_H \sigma c w_0 (1-x),
    \end{equation}
    
    where $\rm S$ and $\rm \dot{w_0}^{rec}$ corresponded respectively to ionising sources and ionising photons from recombination. Considering that $\rm \alpha_A n_H^2 x^2 = \dot{w_0}^{rec} + \alpha_B n_H^2 x^2$, and using Eq. \ref{equ:ionisingEqu}, we rewrote Eq. \ref{equ:CoupledTransport} as
    
    \begin{equation}\label{equ:ModifCoupled}
        \frac{dw_0}{dt} + \frac{dw_1}{dr} = S - \alpha_B n_H^2 x^2 + \beta n_H^2 x(1-x) - n_H \frac{dx}{dt}.
    \end{equation}
    
    Then, we approximated $\rm \frac{dw_0}{dt} = \frac{w_0^{p+1} - w_0^p}{\Delta t}$. We also write $\rm x^p=x^{p+1/2}=x$ and $\rm x^{p+1}=X$ for sake of comprehension. Doing this, and knowing that the terms $\rm w_0^p + S - \frac{dw_1}{dr}$ were already integrated in the transport step and are thus equal to $\rm w_0^{p+1/2}$, we rewrote Eq. \ref{equ:ModifCoupled} as
    
    \begin{equation}\label{equ:CoupledSolver}
        w_0^{p+1} = w_0^{p+1/2} + \beta n_H^2 (1-X)X \Delta t - \alpha_B n_H^2 X^2 \Delta t - n_H(X-x).
    \end{equation}
    
    Replacing $\rm w_0^{p+1}$ in Eq. \ref{equ:CoupledTransport} and using the truncated version of the photon density $\rm \Bar{w_0}^{p+1/2}$ instead of the real value $\rm w_0^{p+1/2}$ to ensure positivity, we obtained a third degree polynomial in X defined as
    
    \begin{subequations}\label{equ:x_system}
    \begin{equation}
        m X^3 + n X^2 + p X + q = 0,
    \end{equation}
    \begin{equation}
        m = (\alpha_B + \beta)n_H^2 \Delta t,
    \end{equation}
    \begin{equation}
        n = n_H - \frac{(\alpha_B + \beta)n_H}{\sigma c} - \alpha_B n_H^2 \Delta t - 2 \beta n_H^2 \Delta t,
    \end{equation}
    \begin{equation}
        p = \beta n_H^2 \Delta t -n_H(1+x)-\Bar{w_0}^{p+1/2}- \frac{1}{\sigma c \Delta t} + \frac{\beta n_H}{\sigma c},
    \end{equation}
    \begin{equation}
        q = \Bar{w_0}^{p+1/2} + n_H x + \frac{x}{\sigma c \Delta t}.
    \end{equation}
    \end{subequations}
    
    Solving this system by finding the only real root between 0 and 1 gives the updated value of the ionised fraction $\rm x^{p+1}$ in the cell. Given our knowledge of this updated ionised fraction and Eq. \ref{equ:CoupledSolver}, we defined the updated photon density as
    \begin{equation}\label{equ:updated_w0}
        \begin{aligned}
        &w_0^{p+1}=w_0^{p+1/2}+dw_0^{p+1/2},\\
        &with\:dw_0^{p+1/2} = \beta n_H^2 (1-X)X \Delta t - \alpha_B n_H^2 X^2 \Delta t - n_H(X-x).
        \end{aligned}
    \end{equation}
    
    This way, we added the photon density variation $\rm dw_0^{p+1/2}$ derived from the truncated photon density $\rm \Bar{w_0}^{p+1/2}$ to the real photon density $\rm w_0^{p+1/2}$. As such, even if $\rm dw_0^{p+1/2}$ is strictly positive, $\rm w_0^{p+1}$ can be negative if $\rm w_0^{p+1/2}$ already was. \\
    
    In the same way as done in Eq. \ref{equ:CoupledTransport}, we wrote all coefficients of order superior to 0 as follows
    \begin{equation}
        \frac{dw_l}{\partial t} + \frac{dw_{l+1}}{dr} = - n_H \sigma w_l (1-x).
    \end{equation}
    As such, using an implicit scheme and the updated $\rm X$ previously computed, the updated value of all coefficients of order superior to 0 is written
    \begin{equation}\label{equ:updated_wk}
        w_l^{p+1} = \frac{w_l^{p+1/2}}{1+ n_H \sigma c \Delta t (1-X)},
    \end{equation}
    with $\rm w_l^{p+1/2}= w_l^p - \frac{dw_{l+1}^p}{dr} \Delta t$ value of the coefficient at the end of a transport step. This way, all of our coefficients were updated with photon variation due to absorption.\\
    
    The last parameter to be computed is the temperature of the hydrogen gas. Our code followed its variations through heating and cooling processes. The heating rate $\rm \mathcal{H}$ mainly involves photoionisation and is given by
    
    \begin{equation}
        \mathcal{H}=n_H (1-X) \Bar{w_0}^{p+1/2} \sigma c \Delta e,
    \end{equation}
    
    where $\rm \Delta e =  e_{HI} - e_{HII}$ where delta e is the energy leftover in the unbound electron and proton system after a photo-ionisation, available as thermal energy. In our case, this difference was equal to $\rm \Delta e = 29.61 - 13.6\:\rm eV$, except in test 1 of section \ref{section:comparisonproject} which is isothermal, meaning that the photon energy was equal to the ionisation energy, $\rm e_{HI}= 13.6\: eV$. For stability reasons, we chose to use the updated value of the ionised fraction $\rm X$, but the non-updated value of the photon density $\rm w_0^{p+1/2}$.\\
    The cooling rate $\rm \mathcal{L}$ is the result of collisional cooling due to case A and B recombination, collisional ionisation and excitation, and bremsstrahlung effect. We used fits from \citet{Hui_Gnedin_97} and \citet{Maselli_2003} to compute these processes. 
    
    Our internal gas energy $\rm E=\frac{3}{2}n_{tot}k_B T$ is linked to these two rates by Eq. \ref{equ:energy}. Knowing that $\rm n_{tot}=n_{e^-}+n_H=x n_H + n_H = n_H (1+x)$, we wrote it as
    
    \begin{equation}
        \frac{3}{2}k_B n_H \left[T \frac{\partial(x+1)}{\partial t} + (x+1) \frac{\partial T}{\partial t}\right] = \mathcal{H} - \mathcal{L},
    \end{equation}
    
    which, when discretising, gave the following explicit equation for the evolution of the temperature, remembering that $\rm T^p=T^{p+1/2}$, $\rm x^p=x^{p+1/2}=x$ and $\rm x^{p+1}=X$
    
    \begin{equation}\label{equ:updated_T}
        \begin{aligned}
        \frac{T^{p+1}-T^{p+\frac{1}{2}}}{\Delta t} &= \frac{1}{X+1} \times \\
            &\left[\frac{2[\mathcal{H}(x,X,\Bar{w_0}^{p+1/2})-\mathcal{L}(X,T^{p+\frac{1}{2}})]}{3k_B n_H} - \frac{(X-x)T^{p+1}}{\Delta t}\right].
        \end{aligned}
    \end{equation}

    \section{Boundary conditions}\label{Appendix:boundary_conditions}

    So far, we have defined our physical processes and implemented them in the simulation without mentioning the issue of boundary conditions. Indeed, since our simulation box were not infinite, we needed to define the behaviour of our transport models at the edge of the limited space we had defined.\\

    Transport between cells using our numerical scheme consisted in computing the value at the interface between two neighbouring cells. However, cells situated at the edge of our box lacked one or more neighbours, and thus have sides where the numerical flux could not be computed. One way to circumvent this issue and define our boundary condition was the use of what we will be calling 'ghost cells'. These ghost cells were not physically defined in our model. However, they allowed us to better represent what we wanted our boundary condition to represent, and aimed at defining in which way the numerical flux behaved along the edge of the box. \\

    In this paper, we used two different types of boundary conditions, neither of which are periodic. The first one we worked with was a transparent boundary condition, which consisted in considering the box as if it was in a larger setting that we did not simulate. As such, radiation going through the boundary simply traversed it and exited the box. \\

    A way to implement it is to consider the ghost cell as very similar to the cell it is neighbouring, as if the surrounding medium we do not simulate had similar properties to the boundary cells.\\

\begin{figure}[!ht]
\centering
\begin{tikzpicture}

\draw[black, dashed] (1,0) rectangle (2,1) node[pos=.5] {(0;0)};
\draw[black, dashed] (2,0) rectangle (3,1) node[pos=.5] {(0;1)};
\draw[black, dashed] (3,0) rectangle (4,1) node[pos=.5] {(0;2)};

\draw[black, dashed] (0,1) rectangle (1,2) node[pos=.5] {(0;0)};
\draw[black, very thick] (1,1) rectangle (2,2) node[pos=.5] {(0;0)};
\draw[black, very thick] (2,1) rectangle (3,2) node[pos=.5] {(0;1)};
\draw[black, very thick] (3,1) rectangle (4,2) node[pos=.5] {(0;2)};
\draw[black, dashed] (4,1) rectangle (5,2) node[pos=.5] {(0;2)};

\draw[black, dashed] (0,2) rectangle (1,3) node[pos=.5] {(1;0)};
\draw[black, very thick] (1,2) rectangle (2,3) node[pos=.5] {(1;0)};
\draw[black, very thick] (2,2) rectangle (3,3) node[pos=.5] {(1;1)};
\draw[black, very thick] (3,2) rectangle (4,3) node[pos=.5] {(1;2)};
\draw[black, dashed] (4,2) rectangle (5,3) node[pos=.5] {(1;2)};

\draw[black, dashed] (0,3) rectangle (1,4) node[pos=.5] {(2;0)};
\draw[black, very thick] (1,3) rectangle (2,4) node[pos=.5] {(2;0)};
\draw[black, very thick] (2,3) rectangle (3,4) node[pos=.5] {(2;1)};
\draw[black, very thick] (3,3) rectangle (4,4) node[pos=.5] {(2;2)};
\draw[black, dashed] (4,3) rectangle (5,4) node[pos=.5] {(2;2)};

\draw[black, dashed] (1,4) rectangle (2,5) node[pos=.5] {(2;0)};
\draw[black, dashed] (2,4) rectangle (3,5) node[pos=.5] {(2;1)};
\draw[black, dashed] (3,4) rectangle (4,5) node[pos=.5] {(2;2)};

\end{tikzpicture}
    \caption{Schematic representation of p a $\rm 3^2$ 2D grid with transparent boundary conditions}
    \label{fig:transparent_BC}
\end{figure}

We showed a simplified representation of how this would be implemented in Fig. \ref{fig:transparent_BC}. Here, we just had to copy the value of the boundary cell into the neighbouring ghost cell to compute the numerical Rusanov flux. \\

However, even though this boundary condition worked very well with $\rm M_1$, it completely broke $\rm P_n$. Indeed, this method seemed to create a strong discontinuity that caused the model to break and return NaN values. As such, despite how good of an option this boundary condition seemed to be, we only used it for the $\rm M_1$ model.\\

The boundary condition used for the $\rm P_n$ model was the condition already used in RKMS. It consisted in putting a constant value $\rm \epsilon$ in ghost cells, chosen to be very close to zero. This might seem a bit counter intuitive, as one would expect the sharp gradient in photon densities between boundary cells and ghost cells to make the $\rm P_n$ solution oscillates, but it did not end up being the case. \\

\begin{figure}[!ht]
\centering
\begin{tikzpicture}

\draw[black, dashed] (1,0) rectangle (2,1) node[pos=.5] {$\epsilon$};
\draw[black, dashed] (2,0) rectangle (3,1) node[pos=.5] {$\epsilon$};
\draw[black, dashed] (3,0) rectangle (4,1) node[pos=.5] {$\epsilon$};

\draw[black, dashed] (0,1) rectangle (1,2) node[pos=.5] {$\epsilon$};
\draw[black, very thick] (1,1) rectangle (2,2) node[pos=.5] {(0;0)};
\draw[black, very thick] (2,1) rectangle (3,2) node[pos=.5] {(0;1)};
\draw[black, very thick] (3,1) rectangle (4,2) node[pos=.5] {(0;2)};
\draw[black, dashed] (4,1) rectangle (5,2) node[pos=.5] {$\epsilon$};

\draw[black, dashed] (0,2) rectangle (1,3) node[pos=.5] {$\epsilon$};
\draw[black, very thick] (1,2) rectangle (2,3) node[pos=.5] {(1;0)};
\draw[black, very thick] (2,2) rectangle (3,3) node[pos=.5] {(1;1)};
\draw[black, very thick] (3,2) rectangle (4,3) node[pos=.5] {(1;2)};
\draw[black, dashed] (4,2) rectangle (5,3) node[pos=.5] {$\epsilon$};

\draw[black, dashed] (0,3) rectangle (1,4) node[pos=.5] {$\epsilon$};
\draw[black, very thick] (1,3) rectangle (2,4) node[pos=.5] {(2;0)};
\draw[black, very thick] (2,3) rectangle (3,4) node[pos=.5] {(2;1)};
\draw[black, very thick] (3,3) rectangle (4,4) node[pos=.5] {(2;2)};
\draw[black, dashed] (4,3) rectangle (5,4) node[pos=.5] {$\epsilon$};

\draw[black, dashed] (1,4) rectangle (2,5) node[pos=.5] {$\epsilon$};
\draw[black, dashed] (2,4) rectangle (3,5) node[pos=.5] {$\epsilon$};
\draw[black, dashed] (3,4) rectangle (4,5) node[pos=.5] {$\epsilon$};

\end{tikzpicture}
    \caption{Schematic representation of p a $\rm 3^2$ 2D grid with absorbent boundary conditions}
    \label{fig:absorbent_BC}
\end{figure}

Indeed, having a ghost cell whose value is 0 meant that, when computing the numerical flux, the result turned out to be not zero, but equal to the value of the boundary cell. As such, this boundary condition that we refer to as 'absorbent' could be used with $\rm P_n$ without breaking the closure.\\

However, it brought up another issue when used with $\rm M_1$ this time. Indeed, the $\rm M_1$ closure has a range of authorised reduced fluxes, as it involves a subtraction inside a square root. If $\rm \epsilon$ is chosen badly, then the numerical flux may end up using forbidden values for the $\rm M_1$ closure and thus, return NaN just as the transparent boundary conditions did with $\rm P_n$.\\

This can be mitigated by finding values of $\rm \epsilon$ which do not violate the conditions of the $\rm M_1$ closure. However, since the transparent boundary conditions already worked quite well with $\rm M_1$, we kept it for that model, and used the absorbent condition for $\rm P_n$ with $\rm \epsilon=1e-8$.\\

All of our tests used transparent boundary conditions for $\rm M_1$ and absorbent boundary conditions for $\rm P_n$. Despite different boundary conditions for each model, we considered the final results to be similar, excluding some artefacts close to the edges of the cell that could be removed from the statistics.\\

    \section{Order and negativity}\label{Appendix:order_negativity}

         \begin{figure}
    \centering
    \includegraphics[width=0.99\linewidth,trim={0 0 0 1.2cm},clip]{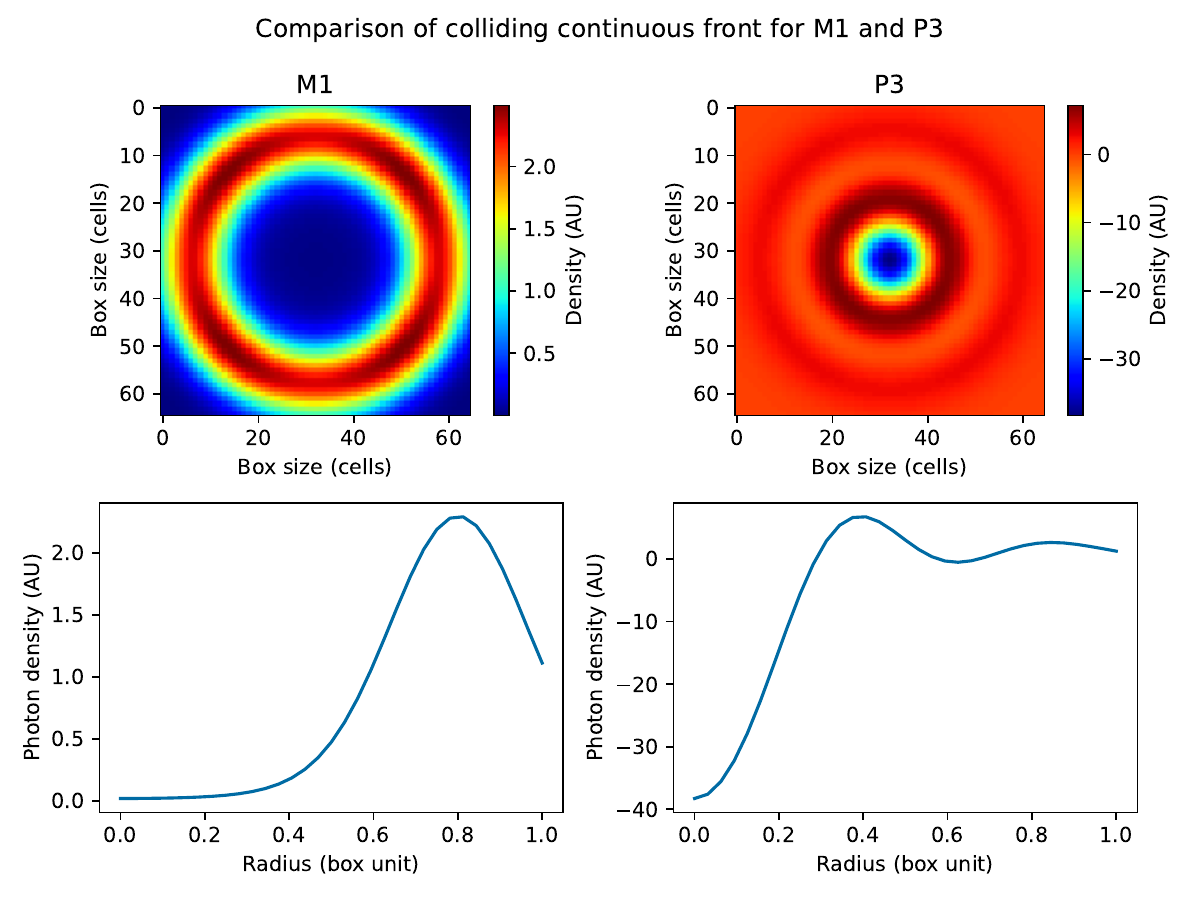}
    \caption{Pulse response of $\rm M_1$ (left) and $\rm P_3$ (right) after 200 time steps. The top plots show a slab of the photon density, while the bottom plots show a profile along the radius of the emission, with 0 being the position of the initial pulse.}
    \label{fig:pulse_comp}
    \end{figure}
  \begin{figure}
    \centering
    \includegraphics[width=0.99\linewidth,trim={0 0 0 1.3cm},clip]{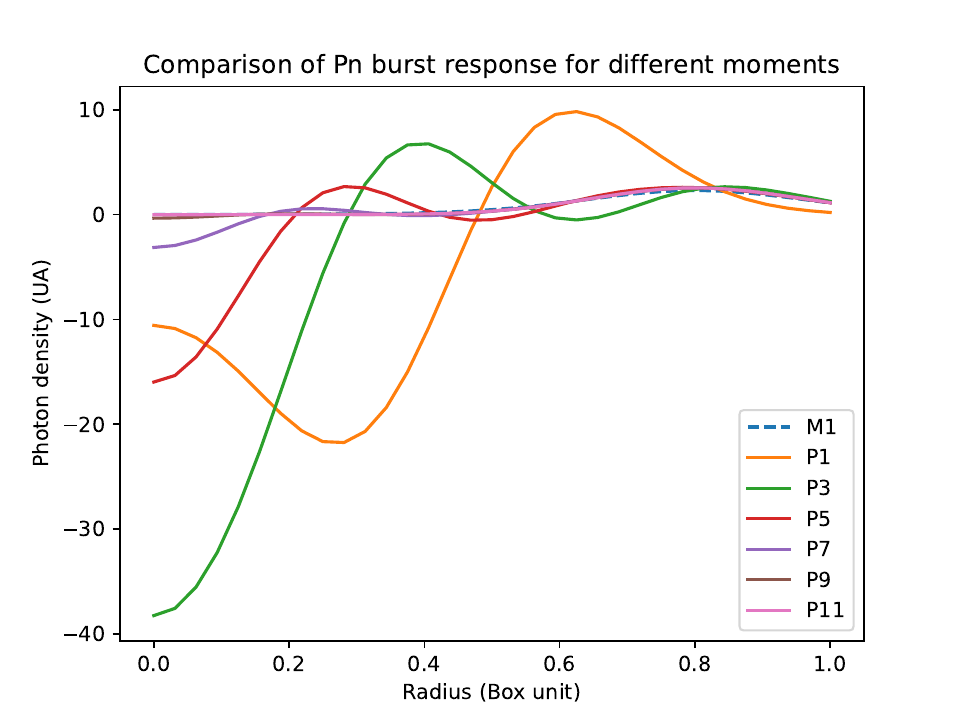}
    \caption{Radial profiles of an impulse response using $\rm M_1$ and various orders of $\rm P_n$ after 200 time steps, with 0 being the position of the initial pulse.}
    \label{fig:pulse_order}
    \end{figure}

    \begin{figure*}
    \resizebox{\hsize}{!}
    {\includegraphics[width=0.99\linewidth,trim={0 0 0 1.2cm},clip]{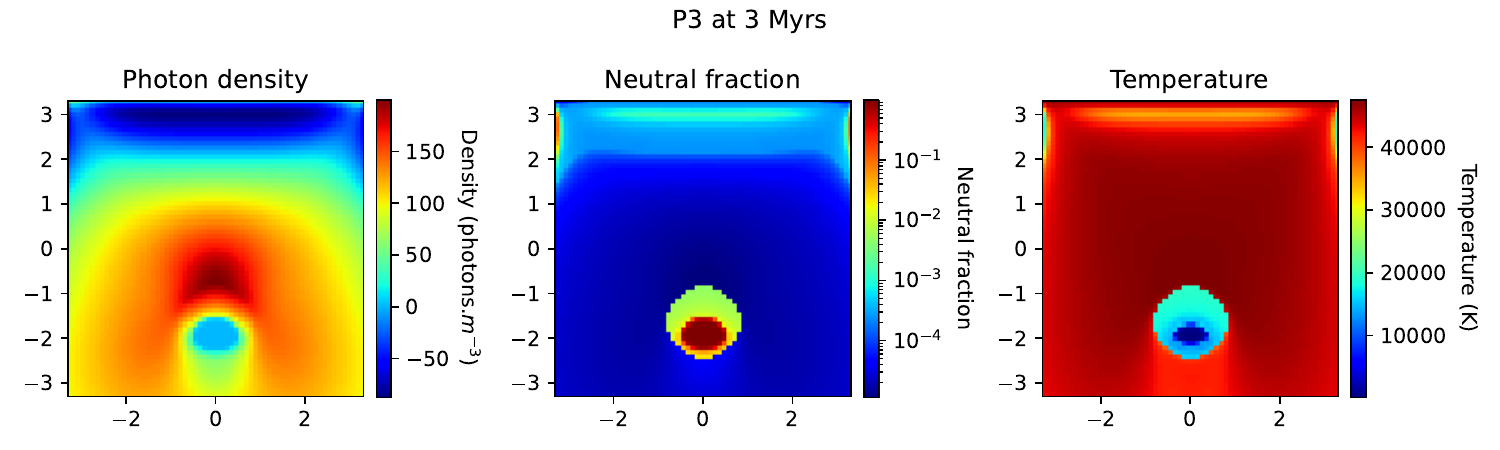}
    }
    \caption{Dense clump of hydrogen in the path of a flux of ionising radiation using $\rm P_3$ at 16 Myr}
    \label{fig:P3_clump_comp}
\end{figure*}

\begin{figure*}
    \resizebox{\hsize}{!}
    {\includegraphics[width=0.99\linewidth,trim={0 0 0 1.4cm},clip]{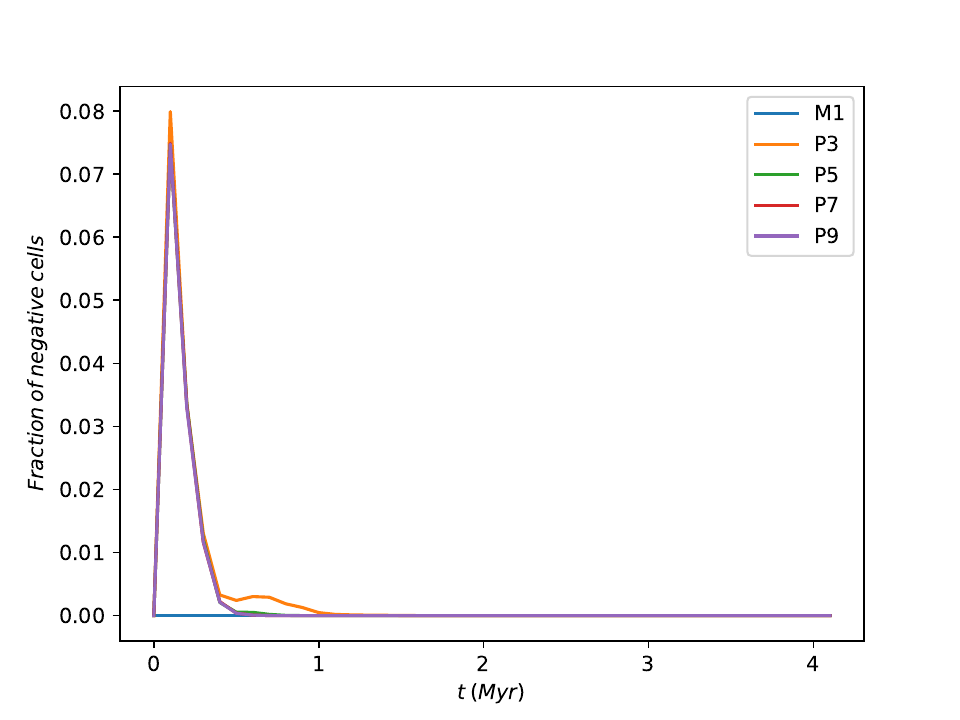}
    \includegraphics[width=1.05\linewidth,trim={0 0 0 1.2cm},clip]{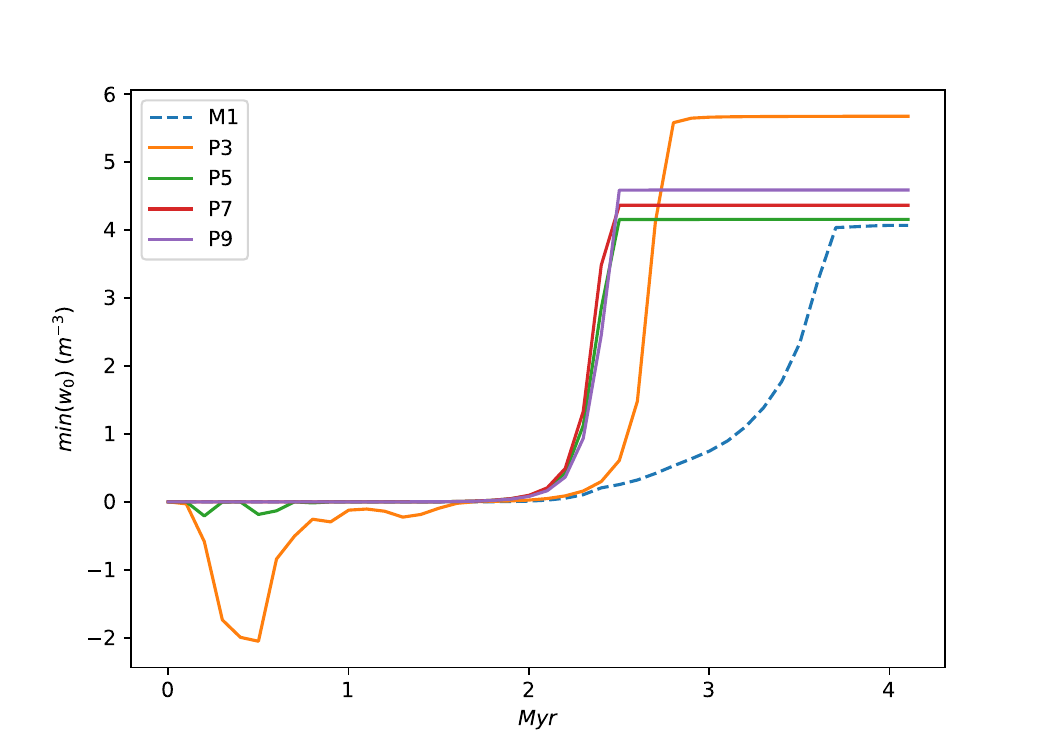}
    }
    \caption{Fraction of negative photon density cells (left) and minimum density (right) over time in the cosmological map test}
    \label{fig:map_negfrac}
\end{figure*}
    As mentioned previously, one of $\rm P_n$'s limitations was its propensity to oscillate in the presence of strong spatial, angular and time discontinuities. Source angular discontinuities are quite unusual in the context of reionisation simulation, since sources are in general isotropic and continuous, and the boundary conditions are generally periodic. Here, we tried to put $\rm P_n$ through the worst case scenario and see how it fared.\\

    This short test consisted of a single adimensional source discontinuous in space and time. There was no chemistry involved here, we just wanted to observe how $\rm P_n$ reacted to a pulse-like source that we defined as a Dirac $\rm \delta$, such that the integral of the photon density over the cell containing the source was equal to 1. As such, our source emitted as follows during one time step
    \begin{equation}
        S_0= \frac{1}{dx dy dz}.
    \end{equation}

    This source was only lit up during a single time step at the beginning of the simulation, in the cell [0.5,0.5,0.5] in adimensional box co-ordinates. We expected the differences between $\rm M_1$ and $\rm P_n$ to be quite sharp, with $\rm M_1$ being strictly positive and non oscillatory, contrary to $\rm P_n$. However, we were interested in the difference between the various orders of $\rm P_n$ and at what point the order was high enough for these oscillations to dampen enough to be comparable to $\rm M_1$. \\

    These results can already be seen quite clearly in Fig \ref{fig:pulse_comp} where we show the comparative results of this pulse response for the two models after 200 time steps. The 'ring' of the wave front for $\rm M_1$ is very visible, but $\rm P_3$ oscillates and outputs negative photon densities, which makes the position of the wave front less obvious. This is even more blatant when looking at the profiles for both models, as $\rm P_3$'s minimum at this time step is $\rm -40$, its negative amplitude being far superior to its positive part. It is important to note that $\rm P_3$ was used here on purpose, as its very low order showcases the limits of $\rm P_n$ the most, and demonstrates why there is a need to go to higher orders for a sensible result, which we did during this paper. \\

    To convince ourselves further, a simple look at Fig. \ref{fig:pulse_order} shows the various profiles of the impulse response for several orders of $\rm P_n$, with $\rm M_1$ in dashed line as a reference. The oscillations of the models are very strong under order 9. For orders above 9 however, the output becomes almost indistinguishable from $\rm M_1$. As such, we can assume that orders above $\rm P_9$ are able to handle situations where $\rm P_9$ already fared well, as their oscillations are dampened enough for them to be negligible. This is the reason why we focused mainly on $\rm P_9$ in this paper, instead of less expensive orders like $\rm P_7$ or $\rm P_5$. We also point out that it is possible to use filters to dampen the oscillations for almost no computational cost \citep{MCCLARREN_2010, Radice_2013, Garett_2014}, but it is beyond the scope of this paper. \\

    The issue of the lower orders of $\rm P_n$ appears in other unfavourable cases, for example the clump test of section \ref{subsection:clump}, where all sources were directional, thus non isotropic, thus introducing a sharp variation into $\rm P_n$. Fig. \ref{fig:P3_clump_comp} shows the result of this test using $\rm P_3$, and we can clearly note that $\rm P_n$, at low order and similar computational cost to $\rm M_1$, cannot reproduce correctly what we physically expect from this test. Specifically, $\rm P_3$, which only requires 16 coefficients, created visible local maxima in the photon density, and tended to over-ionise the sphere. On top of that, its modes output negative photon densities in a large area close to the sources, at the top of the simulation box, which created artefacts in the neutral fraction and temperature that should not exist. As such, we can say that, to mitigate the negativity of the $\rm P_n$ model and to be able to reproduce or outperform $\rm M_1$'s result, a higher order is required, and as such, a higher computational cost. In the case of our test cases, $\rm P_9$ seemed to be an order high enough to match or outperform $\rm M_1$. \\

    However, can we ensure the positivity of our model at all times, even at higher orders and in a favourable test? This can be answered partially by Fig. \ref{fig:map_negfrac}, where we plotted the fraction of negative cells (left) and the minimum photon density (right) in the cosmological map test. One surprising result is that, in this specific test case, the fraction of negative cells seemed to be almost independent of the order of our model, with a peak around 7.5\% of cells around 0.2 Myr. It meant that no matter which model we used, there is always at least some cells outputting a negative photon density, which might pose a problem. However, we can note that in the case of this test, these negative cells are transient and do not last until the end of the simulation, but also that these cells are not as negative as others. Indeed, when looking at the right plot of Fig. \ref{fig:map_negfrac} that shows the minimum photon density value at each time step, it appears obvious that all models above $\rm P_7$ have an almost negligible negative outputs, which in turn will not impact the simulation as much as $\rm P_3$ or $\rm P_5$ would. Indeed, since we approximated the value of a negative cell as 0 (see Eq.\ref{equ:positivation}), the impact of this modification was smaller if the value of our negative cell is already almost a 0. This is yet another reason for favouring higher-order $\rm P_n$ even in simple scenarios where lower-orders models would perform well, such as the Strömgren sphere test.

    \section{Nested Boxes}\label{Appendix:nested_boxes}

    \begin{figure}
    \centering
\resizebox{9cm}{6cm}{
\begin{tikzpicture}

\draw[pattern=north west lines, pattern color=black] (-1.25,-1.25) rectangle (6.25,6.25);
\filldraw[color=black,fill=white,ultra thick] (0,0) rectangle (5,5);
\filldraw[color=black, fill=black!5, dashed] (0,4.75) rectangle (5,5);

\filldraw[color=black, fill=gray!5] (1.9,0) rectangle (3.1,1.25);

\filldraw[color=red!60, fill=red!60, very thick](2.5,1.26) circle (0.607);
\filldraw[color=red!60, fill=red!5, very thick](2.5,1.25) circle (0.605);

\draw[black, ->] (-1.5,4.88) -- (1,4.88);
\draw[black, ->] (-1.5,2.5) -- (0,2.5);
\draw[black, ->] (-1.5,5.5) -- (1,5.5);
\draw[black, ->] (-1.5,-0.5) -- (1,-0.5);

\node[text width=3cm] at (-1.5,4.88) {sources};
\node[text width=4cm] at (-3.5,2.5) {main simulation box};
\node[text width=3cm] at (-1.5,5.5) {buffer};
\node[text width=3cm] at (-1.5,-0.5) {buffer};

\end{tikzpicture}}
    \caption{Schematic representation of the complete simulation box for the shadowing by a dense clump test, using the nested boxes method}
    \label{fig:clump_schematic_nested}
\end{figure}

    For the 'shadowing by a dense clump' test described in section \ref{subsection:clump}, all the sources generating the flux of photons were anisotropic, using directional Gaussian beams. However, as shown in Fig. \ref{fig:beam_compare}, anisotropic sources in $\rm P_n$ create oscillations in the photon density, which means that part of the radiation flux, even if negligible, is not aligned with the main direction of our beams. Worse, some of this photon density goes in the opposite direction, and can create further oscillations if met with the discontinuity of the absorbent boundary conditions. \\
    
    To minimise the impact of such oscillations on the results of the experiment, we tested out several modified setups of this test. The one we ended up using is described in Fig. \ref{fig:clump_schematic_nested}. We created buffers between the cells used in our model and the border of our box by enlarging the simulation domain in all directions. Our $\rm 64^3,\: 6.6\:\rm kpc$ simulation box thus becomes a $\rm 128^3,\: 13.2\:\rm kpc$ one. Indeed, the boundary conditions could interact with the ripples created by non-isotropic $\rm P_n$ sources and create oscillations which prevents a fair comparison between the two models. This bigger box served to minimise the impact of boundary conditions on the simulation box. To ensure a fair comparison, $\rm M_1$ used the same box as $\rm P_n$. However, only the physics happening inside the main $\rm 64^3$ box was of interest to us. Thus, in all plots of this test case shown throughout this paper, we only showcased the main simulation box and omitted the buffers. \\

     \section{Dark Sombrero}\label{Appendix:Dark_sombrero}

\begin{figure}
    \centering
    \includegraphics[width=0.99\linewidth,trim={0 0 0 0cm},clip]{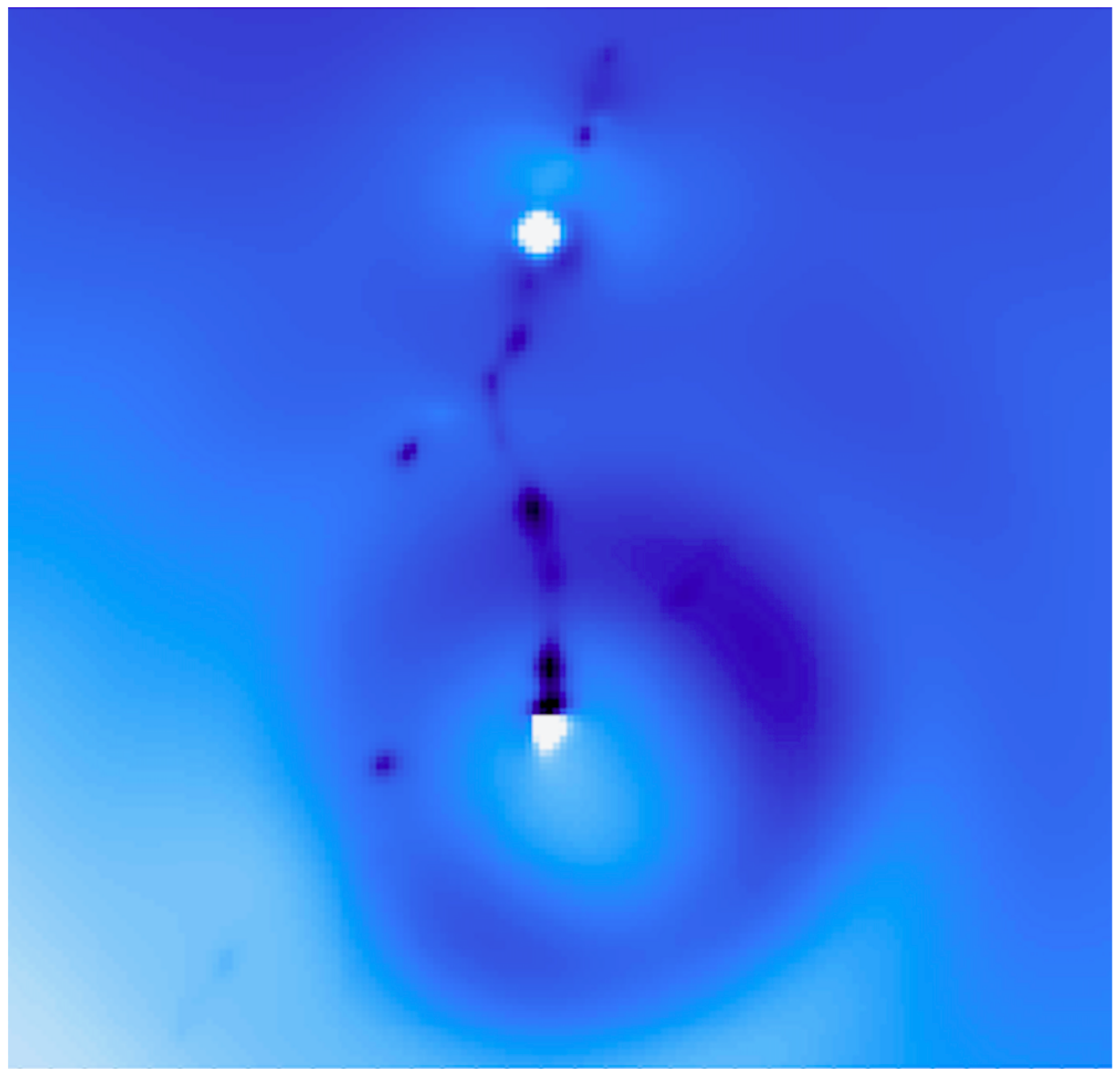}
    \caption{An example of a dark sombrero found in the Cosmic Dawn II simulation, which also uses the M1 radiative transfer model of Aubert et al. 2008. The map shows a photon density slice 2 comoving Mpc across and 23 comoving kpc thick. Two bright galactic sources (in white) can be seen, at both ends of a filament containing several dark clumps. A few dark clumps, away from the filament, can also be seen. The sources and absorbers are bathed in a smooth radiation background. The colour map scaling has been chosen so as to emphasise the dark ring around the bottom source, similar to the dark sombrero or dark doughnut reported in this study.}
    \label{fig:sombrero_coda}
    \end{figure}

    Despite not having been reported on prior to this paper, the Dark Sombrero artefact is a property of the $\rm M_1$ model that can be found in larger published simulations. Fig. \ref{fig:sombrero_coda} shows a photon density map taken from the cosmological simulation CoDa II \citep{Ocvirk_2020} performed with RAMSES and using the CUDATON RT model \citep{Aubert_2008} based on the $\rm M_1$ closure with GLF numerical scheme.\\

    The same dark ring we observed around sources in the tests of this paper can be seen very clearly in the photon density map around the white source at the bottom. The discovery of this artefact in the dataset of CoDa II was one of the reasons that led our team to investigate the potential issues of the $\rm M_1$ closure and to compare it with another state of the art model.\\
    
\end{appendix}

\end{document}